\documentclass[letterpaper,twocolumn,10pt]{article}
\usepackage{usenix-2020-09}
\usepackage{amsmath}
\usepackage{subfigure}
\usepackage[english]{babel}
\usepackage{blindtext}
\usepackage[capitalize]{cleveref}
\usepackage{subcaption} 
\usepackage[labelformat=simple,skip=0pt]{subcaption}
\usepackage{array}
\usepackage{float}
\usepackage{multirow}
\usepackage{listings}
\usepackage{listing}
\usepackage{xcolor} 
\usepackage{circledsteps}
\usepackage{tikz}
\usepackage[title]{appendix}
\usepackage{booktabs}
\usepackage[ruled,vlined]{algorithm2e}
\usepackage{subcaption}
\usepackage{titlesec}
\usepackage{amsfonts}
\usepackage{enumitem}

\titlespacing*{\section}{0pt}{4pt}{2pt}
\titlespacing*{\subsection}{0pt}{4pt}{2pt}
\titlespacing*{\subsubsection}{0pt}{4pt}{2pt}
\titlespacing*{\paragraph}{0pt}{2pt}{2pt}
\setlist{noitemsep, topsep=0pt}

\definecolor{LightGray}{gray}{0.9}

\newcommand{\circnum}[1]{%
  \tikz[baseline=(char.base)]{
    \node[shape=circle,draw,inner sep=1pt] (char) {#1};
  }%
}

\DeclareCaptionLabelSeparator{emdash}{--- }
\newcommand{\systemname}{StreamGuard}
\newcommand{\fig}[1]{Fig.~\ref{#1}}

\newcommand{\ie}{\textit{i.e., }}
\newcommand{\eg}{\textit{e.g., }}

\newcommand{\parahead}[1]{\vspace*{0.5ex plus 0.25ex minus 0.25ex}\noindent %
  {\itshape #1}}
\newcommand{\parabreak}{\vspace*{0.5ex plus 0.5ex}\noindent}

\begin{document}

\title{StreamGuard: Exploring a 5G Architecture for Efficient, \\Quality of Experience-Aware Video Conferencing}

\author{
Xuyang Cao\\
Princeton University\\
\texttt{xyc@princeton.edu}
\and
Oliver Michel\\
Illinois Institute of Technology\\
\texttt{omichel@illinoistech.edu}
\and
Kyle Jamieson\\
Princeton University\\
\texttt{kylej@princeton.edu}
}

\maketitle

\begin{abstract}
\noindent
Video conferencing over 5G is increasingly prevalent, yet its Quality of Experience (QoE) often degrades under limited radio resources. This has two causes: 5G networks must serve many users, while interactive traffic requires careful handling. Motivated by the insight that different subflows within an interactive session have a disproportionate effect on QoE, we present the design and implementation of \systemname{}, a practical 5G architecture for subflow-level, QoE-aware prioritization. \systemname{} forms a closed control loop with three components: (1) a monitor in the Radio Access Network (RAN) that uses deep packet inspection to infer QoE and RAN state, (2) a controller that selects prioritization actions to balance QoE and fairness, and (3) a marking module that applies these decisions by marking packets to steer subflows into appropriate priority queues. \systemname{} further shapes application behaviors via mechanisms including selective subflow dropping and probe-based rate control, to align application behavior with radio constraints. Implemented in a real 5G testbed, \systemname{} achieves a superior QoE–fairness tradeoff compared to vanilla 5G and prior state-of-the-art approaches, improving QoE by up to 70\% at comparable background throughput or preserving up to 2× higher background throughput at similar QoE.

\end{abstract}

\setlength{\abovedisplayskip}{2pt}
\setlength{\belowdisplayskip}{2pt}
\setlength{\abovedisplayshortskip}{2pt}
\setlength{\belowdisplayshortskip}{2pt}

\pagestyle{plain}

\section{Introduction}
\label{sec:intro}



\begin{figure}
    \centering
        \includegraphics[width=0.99\linewidth]{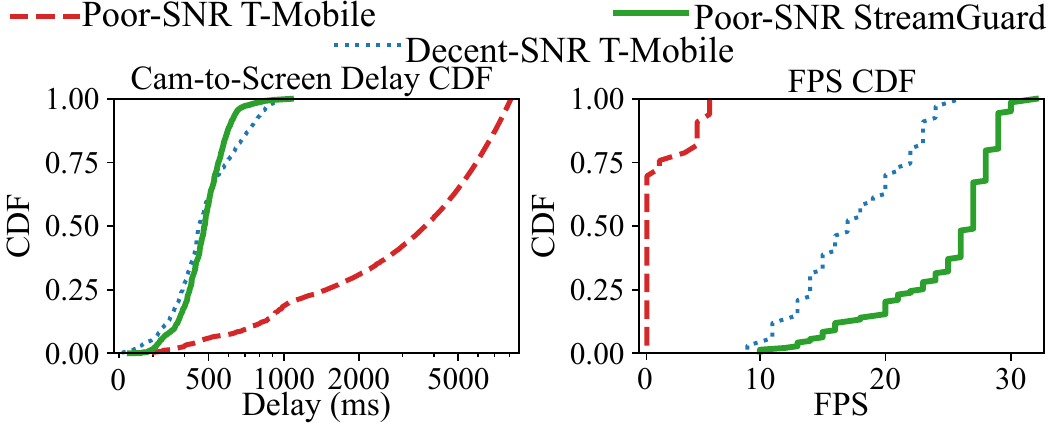}
        \caption{Zoom delay and FPS in a congested 15 MHz T-Mobile cell
        at low ($\approx$3 dB) and moderate ($\approx$15 dB) SNR, compared to a 15 MHz \systemname{} cell under low SNR and congestion. Even lightweight 
        subflow-aware prioritization significantly improves QoE under 
        adverse conditions.}
        \label{fig:tmobile}
\end{figure}

5G New Radio (NR) has seen rapid adoption in recent years, promising higher bandwidth and lower 
latency than previous generations. Meanwhile, Real-Time Communication (RTC) 
applications, such as video conferencing, VoIP, and emerging LLM-based 
voice interaction, are increasingly deployed over 5G
\cite{voipstats2026, edge_intelligence}. Despite these advances, RTC 
performance over 5G often remains unsatisfactory in compromised 
wireless channels and under contention, with issues like lag, 
frame rate (FPS) degradation, and resolution drops observed both in our 
experiments (\cref{fig:tmobile}), and in 
prior studies~\cite{domino, 10.1145/3696348.3696889, webrtc_experiment_mobile, webrtc_over_5g, low_latency_video_5g}. 

Many 5G benefits, especially the high bandwidth promised by 
Enhanced Mobile Broadband (eMBB)~\cite{itu_m2410,3gpp_tr38913}, 
manifest mostly under ideal conditions (few users and high-signal-to-noise
wireless channel).
In practice, however, signal 
quality is often suboptimal due to factors like blockage, 
fading, and mobility~\cite{understand5g, 5g_divide, 5g_sa_meas, 5G_Metamorphosis, cellular_on_wheel}. 
Moreover, in 
dense areas, cell towers (gNBs) frequently serve many users (UEs) 
simultaneously and 
operate near full resource utilization~\cite{understand5g,urban5g,nrscope}, 
leading to significant congestion.
Similarly, while Ultra-Reliable Low Latency Communication (URLLC)~\cite{urllc_concept} aims
to support low-latency RTC traffic, it is not fully standardized 
nor widely implemented or deployed,
and inefficiently consume spectrum when used naively
\cite{maghsoudnia:urllc-distant-goal} (more in~\S\ref{s:related}). 
Instead of leveraging eMBB and URLLC, 
most carriers treat all Internet traffic, including VoIP, 
as a single best-effort traffic class \cite{qos_not_used1, qos_not_used2, qos_not_used3}, 
without prioritizing QoE-critical RTC data, in practice. As a result, particularly 
under poor wireless 
conditions and congestion, such traffic suffers, leading 
to degraded QoE or even unusable sessions, as \cref{fig:tmobile} shows.

Modern RTC applications such as Zoom~\cite{zoom2026} and WebRTC~\cite{webrtc2026} follow the Application-Level Framing (ALF) principle~\cite{alf-ccr90}. They use RTP headers to define Application Data Units (ADUs), the smallest units of data meaningful to the application. ALF recognizes that not all ADUs need to be delivered perfectly for correct application functionality at the receiver. In RTC scenarios, ALF empowers the network to treat different parts of a flow differently under limited resources, prioritizing critical components so that they are delivered in time and can be continuously rendered at the receiver.

In RTC, a flow consists of multiple ADU types, including video, audio, redundancy, and control messages, and each packet carries a unique ADU type~\cite{zoom_measurement, rfc3550, rfc7656, rfc8834}. Techniques such as Scalable Video Coding (SVC)~\cite{wiki-scalable-video-coding, getstream_svc_glossary} 
further split video into a critical \textit{base layer} and optional \textit{enhancement layers}. Losing base-layer frames disrupts decoding and causes freezes, while losing enhancement layers degrades quality (\eg resolution or smoothness) but preserves continuity.

\begin{figure}
    \centering
        \includegraphics[width=0.99\linewidth]{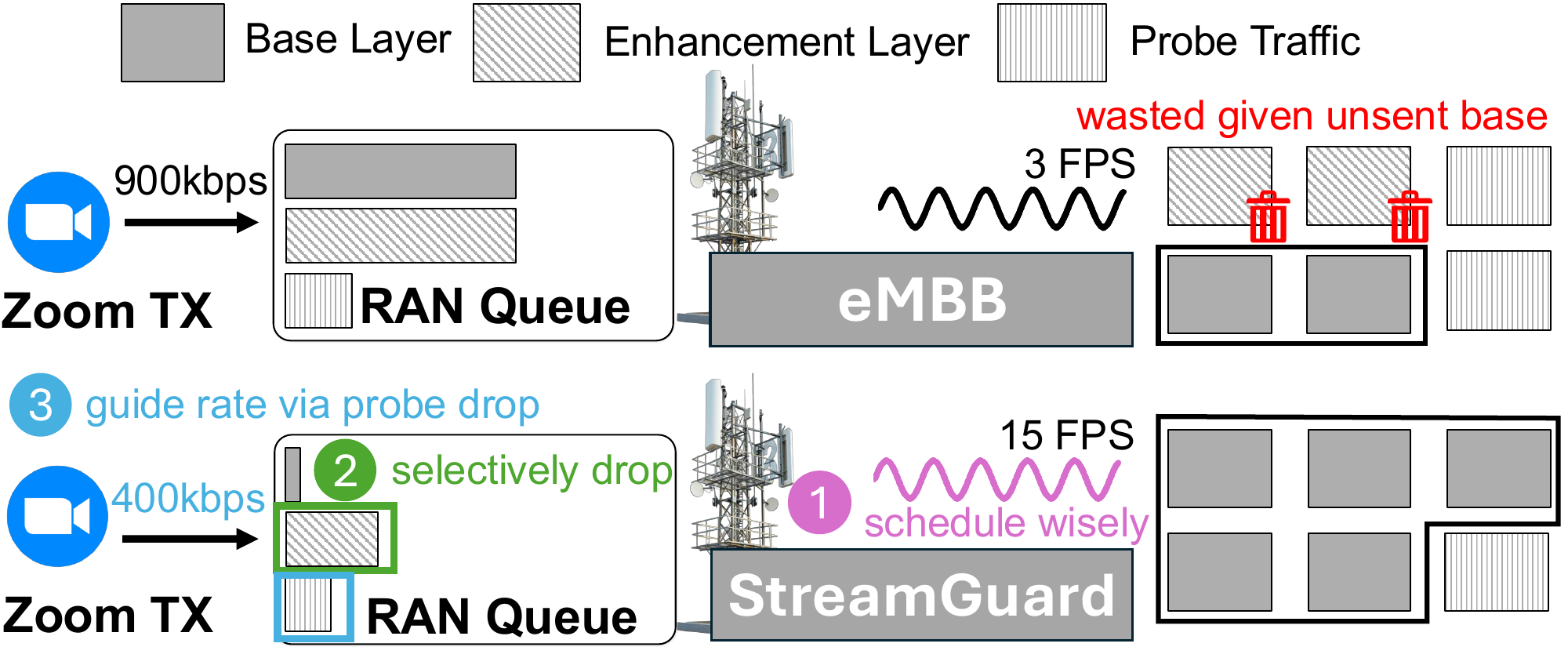}
        \caption{Under limited capacity, 5G eMBB (\emph{upper}) schedules Zoom ADUs indiscriminately, leading to insufficient base-layer delivery, undecodable enhancement layers, and poor QoE. 
        \systemname{} (\emph{lower}) performs base-layer prioritization, enhancement-layer dropping, and probe-based rate control to optimize RAN resources for QoE. 
        }
        \label{fig:intro}
\end{figure}

Importantly, 5G NR QoS filters operate only at flow-level granularity (transport layer) using 3GPP packet detection rules~\cite{etsi_ts_129_244}. Since RTP headers reside in the application layer, the RAN does not distinguish between different RTP stream types. As a result, the RAN cannot treat substreams differentially based on their importance. For example, it will allocate equal spectrum resources to both base-layer and enhancement-layer packets, and expend unneeded retransmissions on enhancement layers, resources that would be better used to protect QoE-critical base-layer frames. 5G QoS therefore leads to suboptimal application performance, as visualized in \cref{fig:intro} and further shown in \S\ref{s:eval}.

By understanding application content, 5G networks can prioritize different ADUs, ensuring that QoE-critical data is delivered first while less important or non-interactive traffic may be delayed or dropped. For example, under poor channel conditions or congestion, transmission of the video base layer should be prioritized (\cref{fig:intro}).
This idea generalizes across applications. In cloud gaming, user inputs and synchronization data should be prioritized; in AR/VR, central views should take precedence over peripheral ones; and even mice traffic (\eg IoT or web) can benefit from prioritization at a granularity smaller than an entire flow~\cite{dchannel} to reduce latency.

\parabreak{}This paper presents the design and implementation of \textbf{\systemname{},} a NextG RAN design for subflow-level granularity prioritization in 5G that leverages application semantics to improve RTC QoE while balancing fairness. \systemname{} is an extensible framework, where application-specific logic (\eg application-layer subflow identification) is modularized via pluggable components, enabling support for diverse interactive applications. This paper presents the design and implementation of our \systemname{} prototype for Zoom, and makes the following contributions:

\parahead{Zoom Subflow Identification.} We analyze Zoom’s ADU structure to identify its constituent subflows: audio, base-layer video, and enhancement-layer video, despite the lack of explicit header information. Using traffic analysis and studies~\cite{zoom_measurement, 10.1145/3696348.3696889}, we derive deep packet inspection (DPI) rules for subflow classification. 

\parahead{QoE Monitoring in the RAN.} We integrate DPI into the gNB to efficiently infer Zoom QoE in real time. The \systemname{} Monitor tracks key QoE metrics (\eg FPS, delay, bitrate) together with RAN signals (\eg MCS, CQI), providing inputs for the controller.

\parahead{Making Prioritization Decisions.} The controller selects subflows to prioritize by evaluating candidate actions by predicting QoE gains and fairness costs, based on current QoE and RAN conditions. It then chooses the scheme that best balances QoE and fairness.

\parahead{Standards-Compliant Prioritization.} \systemname{} realizes subflow prioritization within existing 5G mechanisms: in the downlink via DPI-based \emph{QoS Flow Identifier} (QFI) rewriting, and in the uplink via a lightweight DPI and ToS-marking shim, mapping subflows to different priority queues.

\parabreak{}Beyond prioritization, \systemname{} also shapes application behavior through selective subflow dropping and probe-based rate control, aligning sending rates with constrained radio resources (\cref{fig:intro}). We implement \systemname{} on a real 5G testbed and evaluate it across diverse scenarios (\eg varying traffic directions and numbers of Zoom and background UEs), comparing against the standard 5G QoS framework and prior subflow-aware approaches \cite{tcran}. Across all scenarios, \systemname{} improves Zoom QoE from unusable levels ($\approx$10–40) to smooth performance ($\approx$60–80) at comparable background TCP throughput, while also preserving up to 2× higher background throughput at similar QoE. Overall, \systemname{} consistently achieves Pareto-optimal tradeoffs between QoE and background throughput across all settings.

\section{Primer: Quality-of-Service in 5G}
\label{s:background}

For context, we introduce the QoS architecture used in current 5G systems.
To support applications with diverse latency, reliability, and throughput requirements, 5G provides a user-plane QoS architecture that separates traffic into multiple logical paths with differentiated treatment. The three key abstractions are \emph{QoS flows}, \emph{Data Radio Bearers} (DRBs), and \emph{Logical Channel Groups} (LCGs). A QoS flow is the basic QoS unit in the 5G core and RAN, specifying desired service characteristics via the \emph{5G QoS Identifier} (5QI) and uniquely identified by the \emph{QoS Flow Identifier} (QFI)~\cite{larry5gbook, 3gpp_ts_23_287}. In the RAN, packets from one or more QoS flows are mapped onto DRBs, which realize differentiated treatment of packets over the radio interface. In the UE, LCGs group data to reflect different QoS classes for scheduling and buffer reporting.

\begin{figure}
    \centering
        \includegraphics[width=0.99\linewidth]{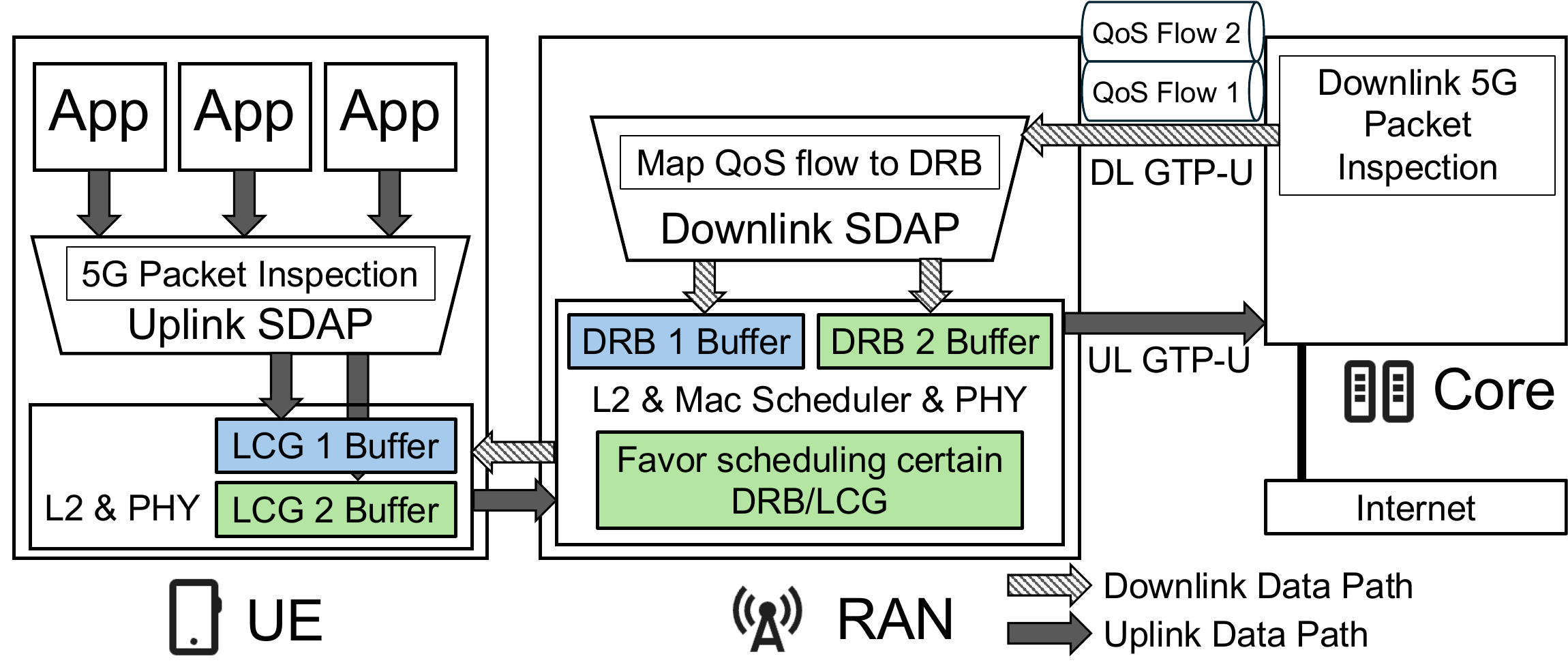}
        \caption{5G QoS mechanisms. In the downlink, core classifies packets to different QoS flows, which are then mapped to DRBs in the RAN for differentiated treatment. In the uplink, modem classifies packets into different LCGs, enabling differentiated treatment as well.}
        \label{fig:qos_drb_5g}
\end{figure}

In the downlink, the standard 5G data path operates as shown in~\fig{fig:qos_drb_5g}, where an application packet first traverses the core network, in which packet detection rules determine the QoS flow to which it should be assigned. The core network then encapsulates the packet using the \emph{GPRS Tunneling Protocol (User Plane)} (GTP-U), the standard tunneling protocol used to carry user-plane packets between the core and the RAN, and inserts the corresponding QFI value in the GTP-U header. When the packet arrives at the gNB, the downlink \emph{Service Data Adaptation Protocol} (SDAP) layer inspects the QFI and maps the packet to the DRB associated with that QoS flow. In this way, 5G provides a built-in mechanism for differentiated treatment of downlink traffic.

Turning to the uplink, when a UE joins the network, the RAN installs a set of packet inspection rules at the uplink SDAP layer inside the UE. These rules, similar to the downlink ones, allow the UE to classify traffic into different LCGs (as shown in \cref{fig:qos_drb_5g}). Upon uplink transmission opportunities, the UE reports its uplink buffer occupancy to the gNB via a \emph{Buffer Status Report} (BSR), which explicitly indicates the amount of queued data per LCG. Based on this report, the gNB scheduler can prioritize uplink grants for higher-priority LCGs, while the UE correspondingly prioritizes transmissions from those LCGs with the granted resources.

\section{Related Work}
\label{s:related}

\paragraph{Existing 5G modes.} The default 5G mode, eMBB, uses a coarse-grained QoS architecture (\S\ref{s:background}) and typically treats all traffic uniformly, leading to suboptimal video conferencing QoE under a given resource budget (\S\ref{s:eval}). URLLC concepts promise strict latency and reliability guarantees but are difficult to actualize, requiring significant modifications to the radio technology (\eg fast RF chain switching for mini-slot scheduling~\cite{sharetechnote_urllc}), and may over-provision resources for reliability (\eg PHY-layer duplication), leading to reduced spectral efficiency~\cite{urllc_concept, maghsoudnia:urllc-distant-goal,toward6g}. 
Network slicing provides service-level isolation but remains coarse-grained and cannot adapt quickly enough for dynamic QoE optimization~\cite{network_slicing_challenges, network_slicing_concept, network_slicing_flexibility}. From first principles, network slicing does not refine QoS beyond flow-level granularity and thus reduces to the behavior of existing QoS mechanisms.
\paragraph{Proposed 5G architectures.} Several prior works explore 5G designs for a diversified set of applications.
EdgeRIC~\cite{edgeric} enables real-time RAN control via xApps with enhanced observability, but does not provide application-specific designs for video conferencing traffic and does not support L7 subflow-level differential handling. RadioSaber~\cite{285076} and RadioNinja~\cite{radioninja} focus on efficient slicing and resource allocation without explicitly optimizing QoE. Overall, these systems operate at flow-level QoS granularity and therefore cannot distinguish QoE-critical ADUs from less critical ones. From first principles, such designs reduce to the existing 5G QoS mechanisms that we evaluate in~\S\ref{s:eval}. 

DChannel~\cite{dchannel} selectively offloads a subset of traffic to URLLC to speed up web page load time, but does not target RTC application performance, lacks application-level subflow prioritization, and evaluates URLLC via Ethernet emulation rather than realizing it over a wireless medium. TC-RAN~\cite{tcran} leverages application subflow differentiation, but does not incorporate QoE feedback on its decisions, nor does it incorporate inter-flow optimization of traffic across all users in the RAN, removing its ability to make adaptive and QoE-aware decisions. We compare \systemname{} against both in~\S\ref{s:eval}. 

Octopus~\cite{10419234} drops less critical subflows (\eg enhancement layers) under congestion but does not offer a mechanism to tradeoff and manage the contention between RTC traffic versus non-RTC traffic. Further, Octopus requires modifications to the endpoint RTC application, making it inapplicable to closed-source systems like Zoom. QCON~\cite{qcon} similarly tracks flow-level application and radio metrics in the RAN to selectively activate a secondary cellular link (\ie dual connectivity~\cite{3gpp_dual_connectivity}) for stabilizing RTC QoE. However, it requires operator-enabled dual connectivity, consumes additional resources by always attempting to deliver all RTC traffic rather than selectively dropping less QoE-critical subflows, and estimates bandwidth usage using simplified equal-allocation assumptions instead of modeling actual scheduler behavior, despite access to scheduler information. Finally, most of these prior works~\cite{10419234,tcran,dchannel,edgeric,qcon} focus on the downlink only, whereas \systemname{} addresses both directions. 

\paragraph{End-host application solutions.} Several approaches improve QoE via endpoint adaptation but require modifying the application. Salsify~\cite{salsify} adapts encoding and transmission to network conditions on a per-frame level, while Tyrus~\cite{Tyrus} and Vidaptive~\cite{Vidaptive} selectively schedule or drop less important frames. Systems such as Mowgli~\cite{Mowgli}, onRL~\cite{onrl}, Loki~\cite{loki}, and Concerto~\cite{Concerto} 
enhance codec and congestion control decisions using learning-based techniques to better predict and adapt to varying network capacity. More recently, works like Gemino~\cite{Gemino, neur_video1, neur_video2} explore neural compression methods for highly efficient video encoding and transmission. All these methods require access to application source code and often coordination between the application and the network, making them inapplicable to closed-source systems such as Zoom, which \systemname{} targets.

\section{Design}

\begin{figure}
    \centering
        \includegraphics[width=0.99\linewidth]{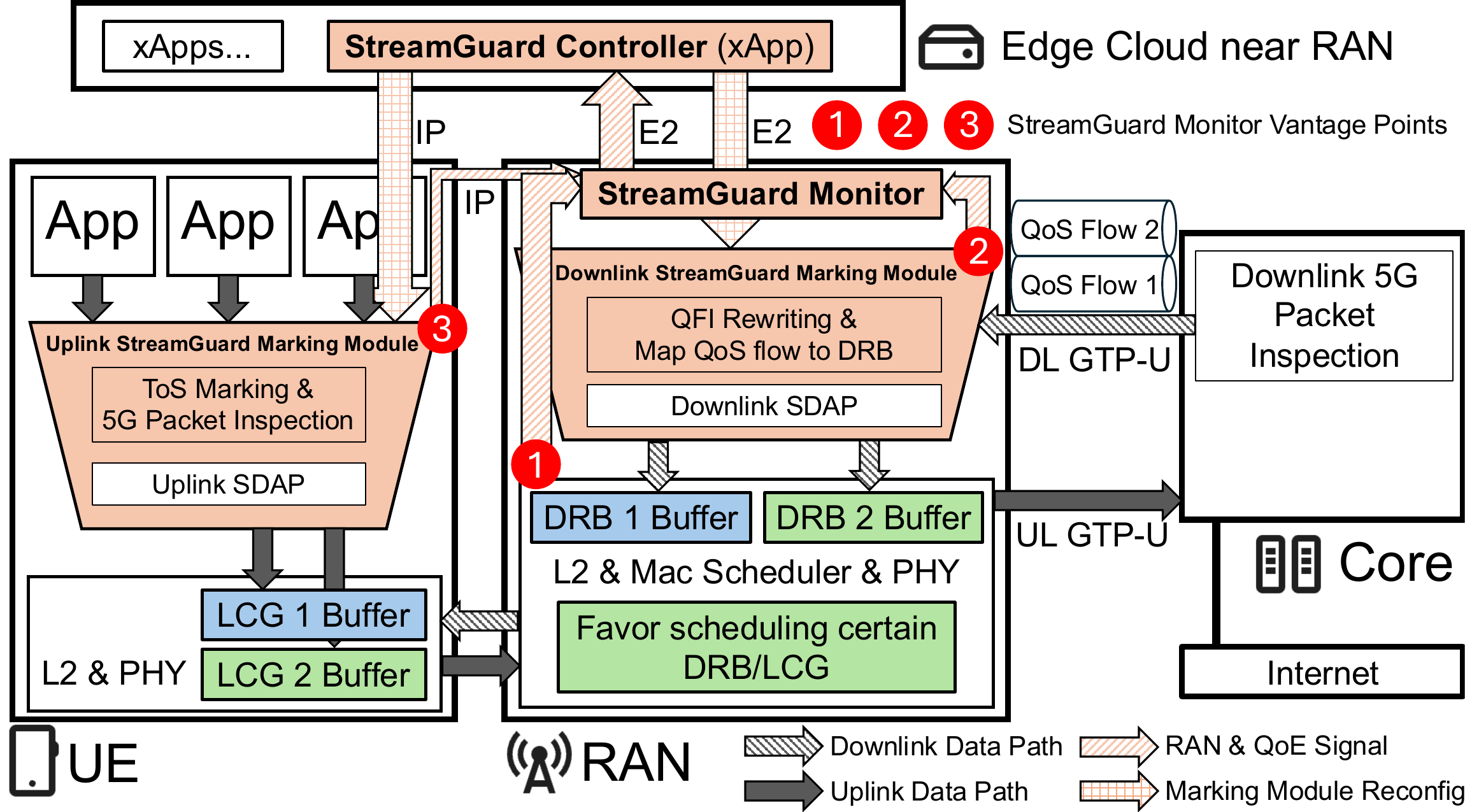}
        \caption{\systemname{} components (orange). The \systemname{} Monitor continuously tracks RAN-level metrics and inferred application QoEs. These measurements are sent to the \systemname{} Controller (xApp), which makes prioritization decisions and enforces them by reconfiguring the \systemname{} Downlink and Uplink Marking Module.}
        \label{fig:control_loop}
\end{figure}

\systemname{} improves QoE while maintaining fairness by selectively 
prioritizing and protecting QoE-critical application data under resource contention.
We realize this application-aware, subflow-level prioritization within the 5G stack~\cite{etsi_5g}. 
As packets traverse the RAN, \systemname{} inspects them, classifies them into subflows, and steers them into appropriate priority queues, while a control loop continuously monitors QoE and RAN conditions, updating these decisions. 
\paragraph{System overview.} \systemname{} consists of three main components that together form a closed control loop among the near-RT RIC, the RAN, and optionally UE.
All three components leverage \systemname{}'s overarching DPI plugin system (\eg Appendix~\ref{s:zoomdetails：zoomdpi} for Zoom), which enables compatibility with and extensibility for different RTC applications (and possibly other application categories in the future).

\noindent\textbf{\systemname{} Monitor} in the RAN continuously estimates QoE metrics for interactive flows and collects relevant RAN-level measurements in real time (§\ref{s:design:metric_tracking}).

\noindent\textbf{\systemname{} Controller}, implemented as an xApp in the Near-RT RIC (deployed in the edge cloud close to the RAN), consumes these measurements. Based on these measurements, the controller determines subflow-level prioritization and traffic-shaping decisions to better match radio constraints, and periodically issues corresponding control actions, closing the control loop shown in \cref{fig:control_loop} (§\ref{s:design:controller}).

\noindent\textbf{Downlink and Uplink Marking Modules} reside in the 5G user plane at both the RAN and inside the UE, and apply these decisions from the controller by marking packets with the appropriate priority on live traffic (§\ref{s:design:data_plane}).

    
    

\subsection{\systemname{} Monitor}
\label{s:design:metric_tracking}
\systemname{} makes prioritization decisions based on a set of continuously monitored metrics spanning three categories: RAN-layer indicators (\S\ref{s:design:metric_tracking:ran_metrics}), RTC quality metrics used for QoE estimation (\S\ref{s:design:metric_tracking:qoe_metrics}), and statistics on background and competing flows (\S\ref{s:design:metric_tracking:offered_load}, \S\ref{s:design:metric_tracking:bw_hungry_flow}).
RAN metrics help evaluate the resulting fairness impact under the default scheduler from a particular prioritization scheme, while QoE metrics help the controller estimate the potential QoE improvement from prioritizing specific subflow(s).
Collecting application-specific metrics requires application awareness.
Also here, the monitor uses \systemname{}'s plugin interface to invoke per-application modules that then instrument the packet stream at the gNB.
This enables computing application-specific metrics, for example, video frame rates in Zoom.
The metrics the \systemname{} monitor computes and passes to the controller are summarized in \cref{tab:tracked_metrics}.
We will now explain the different classes of metrics and how the monitor collects them.

\begin{table}[t]
\centering
\caption{Metrics tracked by \systemname{} Monitor}
\label{tab:tracked_metrics}
\begin{tabular}{>{\centering\arraybackslash}m{3.8cm}
                >{\centering\arraybackslash}m{1.3cm}
                >{\centering\arraybackslash}m{1.3cm}}
\toprule
\textbf{Metric} & \textbf{\shortstack{Update \\ Interval}} & \textbf{\shortstack{Avg \\ Window}}\\ 
\midrule
CQI, SINR, MCS (\S\ref{s:design:metric_tracking:ran_metrics}) & $\approx <$ 1ms & 50ms\\
\midrule
FPS (\S\ref{s:design:metric_tracking:qoe_metrics}) & \multirow{4}{*}{\shortstack{Packet \\ Inter-Arrival \\ Time}} & 1s\\
Media Delays (\S\ref{s:design:metric_tracking:qoe_metrics}) &  & 1s\\
Offered Loads (\S\ref{s:design:metric_tracking:offered_load}) &  & 1s \\
BW-hungry Flow? (\S\ref{s:design:metric_tracking:bw_hungry_flow}) &  & N/A\\
\bottomrule
\end{tabular}
\end{table}

\subsubsection{RAN-level MAC/PHY Metrics}
\label{s:design:metric_tracking:ran_metrics}

The Monitor continuously tracks key PHY- and MAC-layer indicators, including the selected modulation and coding scheme (MCS), Channel Quality Indicator (CQI), uplink SINR, and other scheduler-relevant metrics.
These metrics are collected upon updates from the RAN and averaged over a 50~ms window, matching the \systemname{} Controller’s decision interval $I$ (§\ref{s:design:controller}), after which the aggregated values are passed to the Controller. Such metrics are intrinsic to scheduling and link adaptation in modern RAN systems and are readily available in both vendor and open-source implementations~\cite{3gpp38214,3gpp38321}.


\subsubsection{Application and Network-Layer Metrics}
\label{s:design:metric_tracking:qoe_metrics}

At the application and network layer, the \systemname{} monitor tracks RTC quality indicators in the context of competing and background traffic.
In particular, we monitor video frame rate (FPS) and per-direction delay by inspecting RTP header fields and leveraging 5G-native mechanisms like HARQ acknowledgments. The exact monitoring mechanism here also depends on the direction of traffic.

\paragraph{Downlink direction.} 
Video frames transported over RTP typically span multiple packets where all packets belonging to one frame share the same RTP timestamp value~\cite{rfc3550}.
Additionally, depending on the application, there may be fields that carry the number of packets in the current frame or an indicator for the last packet of a frame~\cite{michel:passive-zoom}.
This allows the \systemname{} monitor to track frame rate by inspecting packets at the gNB’s PHY layer (\circnum{1} in \cref{fig:control_loop}) after UE acknowledgment, and continuously counting the number of complete video frames delivered within a sliding one-second window.

Going further, to measure RAN-to-UE delay, we track the RAN-side frame-level delay, defined as the sojourn time that all packets belonging to a video frame spend inside the gNB. 
Specifically, the \systemname{} monitor records the interval from when the first packet of a frame enters the highest SDAP layer at the gNB (\circnum{2} in \cref{fig:control_loop}) to when the last packet of that frame is delivered to the UE in the PHY layer (\circnum{1} in \cref{fig:control_loop}).
We infer the delivery time using 5G downlink HARQ mechanism: when packets are acknowledged by the UE at time $t_{\text{ACK}}$, the packets are known to have been successfully received $k_1$ slots earlier, where $k_1$ is the gNB-configured delay between a downlink transmission and the corresponding ACK.
Thus, the delivery time is computed as $t_{\text{deliver}} = t_{\text{ACK}} - k_1$. To our knowledge, this is the first technique that leverages purely in-network signals to reconstruct frame-level RAN-to-UE delivery delay without UE-side instrumentation.

For audio, each packet typically encapsulates a complete audio sample; thus, the RAN-to-UE audio delay is measured as the packet sojourn time from the SDAP layer to its inferred delivery time at the UE, analogous to the video case. 
When an audio sample spans multiple packets, we instead measure delay at the frame level using the same approach as for video.

\paragraph{Uplink direction.} Monitoring frame rate on uplink traffic works in the same way as for downlink.
\systemname{} continuously inspects received packets at the gNB’s PHY layer (\circnum{1} in \cref{fig:control_loop}), identifies uplink frames using the same cues as in the downlink, and tracks FPS over time.

\begin{figure}
    \centering
    \includegraphics[width=0.98\linewidth]{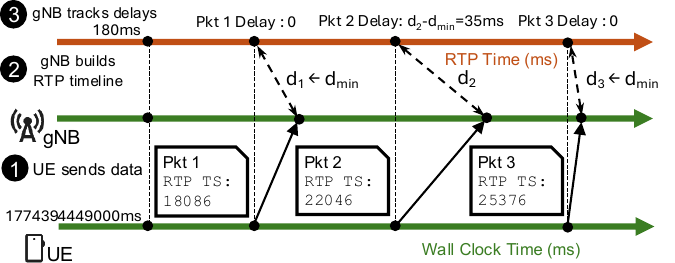}
        \caption{How gNB estimates uplink app-to-gNB delay for RTP packets. Upon receiving RTP packets from UE (\CircledText{\small 1}), gNB converts their RTP timestamps into ms (\CircledText{\small 2}). It then establishes a baseline and estimates lower-bound uplink delays (\CircledText{\small 3}).}
        \label{fig:ul_delay_track}
\end{figure}

Tracking UE-to-RAN delay, however, is more challenging because the gNB cannot directly observe when and what packets are buffered at the UE’s modem. Fortunately, most real-time communication (RTC) applications use RTP, whose RTP timestamp field, while primarily intended for playback pacing, records the generation time of the frame (and its constituent packets) and provides an approximation of when packets are sent. 
Based on this insight, we developed an estimation mechanism for UE-to-gNB delay that compares RTP tiemstamps with gNB arrival times.
This mechanism is depicted in \cref{fig:ul_delay_track} and explained in more detail next.

\textit{RTP Timestamp Meaning.} 
As shown in \cref{fig:ul_delay_track}, each video frame or audio sample is assigned an RTP timestamp, corresponding to its sampling time at the sender application. 
RTP timestamp values, though, are expressed in sampling instants \cite{rfc3550}, rather than wall-clock time. 
Dividing the timestamp by the sampling rate converts unit of time to seconds.
The sampling rate can be readily obtained from application specifications or documentation \cite{rfc3551}, or inferred via simple measurement \cite{uncover_rtp}.
Video always uses 90kHz~\cite{rfc3550}.

\textit{Uplink Delay Tracking Procedure.} Once gNB receives RTP packets (\circnum{1} in \cref{fig:ul_delay_track}) at the PHY layer, the \systemname{} monitor parses their RTP timestamps and converts them into millisecond-scale real time by dividing by the sampling rate (\circnum{2} in \cref{fig:ul_delay_track}).
It then computes the offset between the packet’s arrival time (logged at the PHY layer) and its converted RTP time (\ie the $d$ values in \cref{fig:ul_delay_track}), and tracks the minimum observed offset over time as a baseline delay (\eg initially $d_1$, later $d_3$ in \cref{fig:ul_delay_track}). 
This minimum approximates the smallest UE-to-gNB delay observed so far. Any excess offset beyond this baseline is interpreted as additional uplink latency (\circnum{3} in \cref{fig:ul_delay_track}), arising from events like scheduling delay, buffering, or retransmissions.

The above procedure describes packet-level uplink delay tracking. 
Frame-level delay tracking builds on this by computing the offset between the wall-clock arrival time of the last packet of a frame at the gNB and its converted RTP timestamp, subtracting the baseline offset.

\subsubsection{Offered Loads}
\label{s:design:metric_tracking:offered_load}

In addition to QoE metrics, the \systemname{} controller relies on estimates of offered load (in bits per second), \ie the traffic demand presented to the RAN, to quantify the fairness impact of prioritization decisions. 
We therefore track offered load at two complementary levels in both downlink and uplink: (i) fine-grained layer-7 subflows for video conferencing, and (ii) remaining background traffic.

\paragraph{Downlink direction.} We estimate offered load by monitoring packets entering the gNB’s SDAP layer.
Leveraging application-layer semantics, we track per-subflow offered load for interactive traffic, as well as the remaining per-UE background traffic.
Each estimate is updated upon the arrival of the corresponding packets.

\paragraph{Uplink direction.} To estimate subflow-level offered load for video conferencing in the uplink, we leverage the DPI module (shown in \cref{fig:control_loop}) to identify relevant packets as they traverse the UE kernel and measure their arrival rates at the modem. These measurements are continuously updated upon packet arrival and reported to the \systemname{} Monitor in the RAN via lightweight IP messages over the high-priority uplink data path (\circnum{3} in \cref{fig:control_loop}) every 10 ms. This reporting interval is much shorter than the typical pacing of Zoom audio packets (\eg 20 ms) and video frames (\eg 33 ms) \cite{Interactivity_QoE, 10.1145/3696348.3696889}, ensuring that the \systemname{} Controller operates on fresh offered load estimates.

To ensure scalability, only UEs participating in uplink video-conferencing prioritization under \systemname{} need to deploy the shim layer. However, the \systemname{} Controller requires background uplink offered load from all UEs to maintain a complete view of traffic demand. We therefore design an approach to estimate background uplink offered load using only signals available from existing 5G mechanisms.

\begin{figure}
    \centering
    {\includegraphics[width=0.75\linewidth]{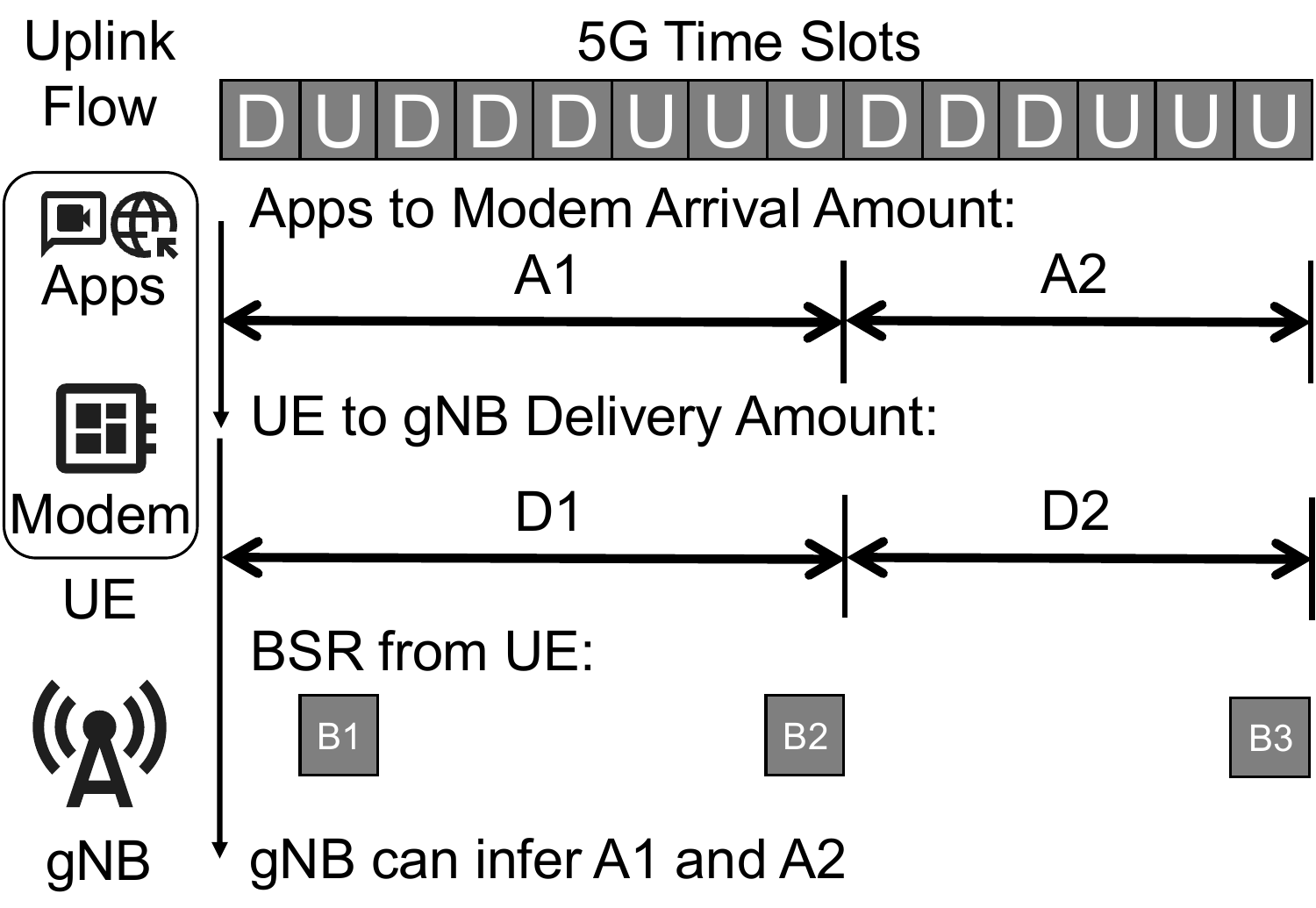}}
        \caption{How the gNB infers uplink offered load. The gNB derives the newly arrived uplink data amount, $A_1$, from changes in reported buffer occupancy, $B_2 - B_1$, and the delivered byte amount $D_1$, using $A_1 = B_2 - B_1 + D_1$, thereby enabling it to track offered load over time. }
        \label{fig:ul_ingress_infer}
\end{figure}

The gNB continuously receives Buffer Status Reports (BSRs) from the UE, each indicating the amount of pending uplink data buffered in the UE’s modem per LCG (with background and video-conferencing traffic assigned to separate LCGs). Consider two consecutive BSRs received by the gNB at time slots \(t_1\) and \(t_2\), reporting pending data amounts \(B_1\) and \(B_2\), respectively. The change in reported buffer occupancy reflects both the amount of data delivered to the gNB, denoted \(D_1\), and the amount of new data arriving from the UE kernel to the modem, denoted \(A_1\), during the interval \([t_1, t_2]\). These quantities are related by:
$B_2 = B_1 - D_1 + A_1$. Since the gNB directly observes \(B_1\), \(B_2\), and \(D_1\), it can readily infer the newly arrived data amount \(A_1\), as shown in \cref{fig:ul_ingress_infer}. Over time, the gNB accumulates such estimates from successive BSR intervals. Summing the inferred \(A\) values over the most recent one-second window yields an estimate of the instantaneous uplink offered load (per second). The rate is updated upon each new BSR received at the gNB’s MAC layer and continuously pulled by the \systemname{} Monitor.

We emphasize that this approach provides a reasonably accurate estimate of the kernel-to-modem arrival rate rather than a precise queueing model. BSR reports are capped, so consecutive maximum reports may hide newly arrived data, yielding a lower-bound estimate in rare cases. In addition, uplink queues may experience drops or resets (\eg radio link failures, PDCP timer discards, or overflows), causing the inferred arrival amount $A$ to be negative; we retain these values when computing the average offered load, as such drops effectively relieve demand on gNB resources.

\subsubsection{Detection of Bandwidth-Hungry Flows}
\label{s:design:metric_tracking:bw_hungry_flow}

Among the background traffic, bandwidth-hungry flows (\eg, bulk transfers or on-demand video fetching) continuously probe available capacity via congestion control. As a result, their instantaneous offered load does not reflect intrinsic demand. To account for this, we incorporate a lightweight detection mechanism in both directions. The \systemname{} Monitor tracks TCP connections by identifying \verb|SYN| or \verb|SYN-ACK| packets. Based on steady-state TCP behavior and prior work \cite{dctcp, nature_data_center_traffic, Hedera, DevoFlow, rivillo_elephant}, we classify a connection as bandwidth-hungry if (i) it persists for at least 2 seconds and (ii) more than 50 large (bytes $\geq 1000$) data packets are observed within one second in a given direction. Flows using other congestion-control protocols such as QUIC~\cite{quic_rfc} can be detected similarly.
Consistent with prior work \cite{rivillo_elephant}, we validate this heuristic using controlled experiments on our testbed, where sustained activities such as file transfers and on-demand video streaming consistently satisfy these criteria, while lightweight traffic such as web browsing does not. Once detected, the presence of a bandwidth-hungry flow is reported to the \systemname{} Controller to guide prioritization decisions more wisely (\S\ref{s:design:controller:fair_model}).

\subsection{\systemname{} Controller}
\label{s:design:controller}

The \systemname{} controller makes subflow-level prioritization decisions based on real-time RTC quality measurements and RAN state.
It decides which subflows to protect and to what extent, while avoiding harm to non-interactive (background) traffic under limited spectrum resources. 
Additionally, the controller has the ability to influence application rate using two mechanisms: (i) selectively dropping subflows that are not required for correct decoding and (ii) shaping application probing traffic~(\S\ref{s:design:controller:behavior_shaping}).
These actions allow \systemname{} to align application demand with radio resources.

\begin{figure}
{\includegraphics[width=0.85\linewidth]{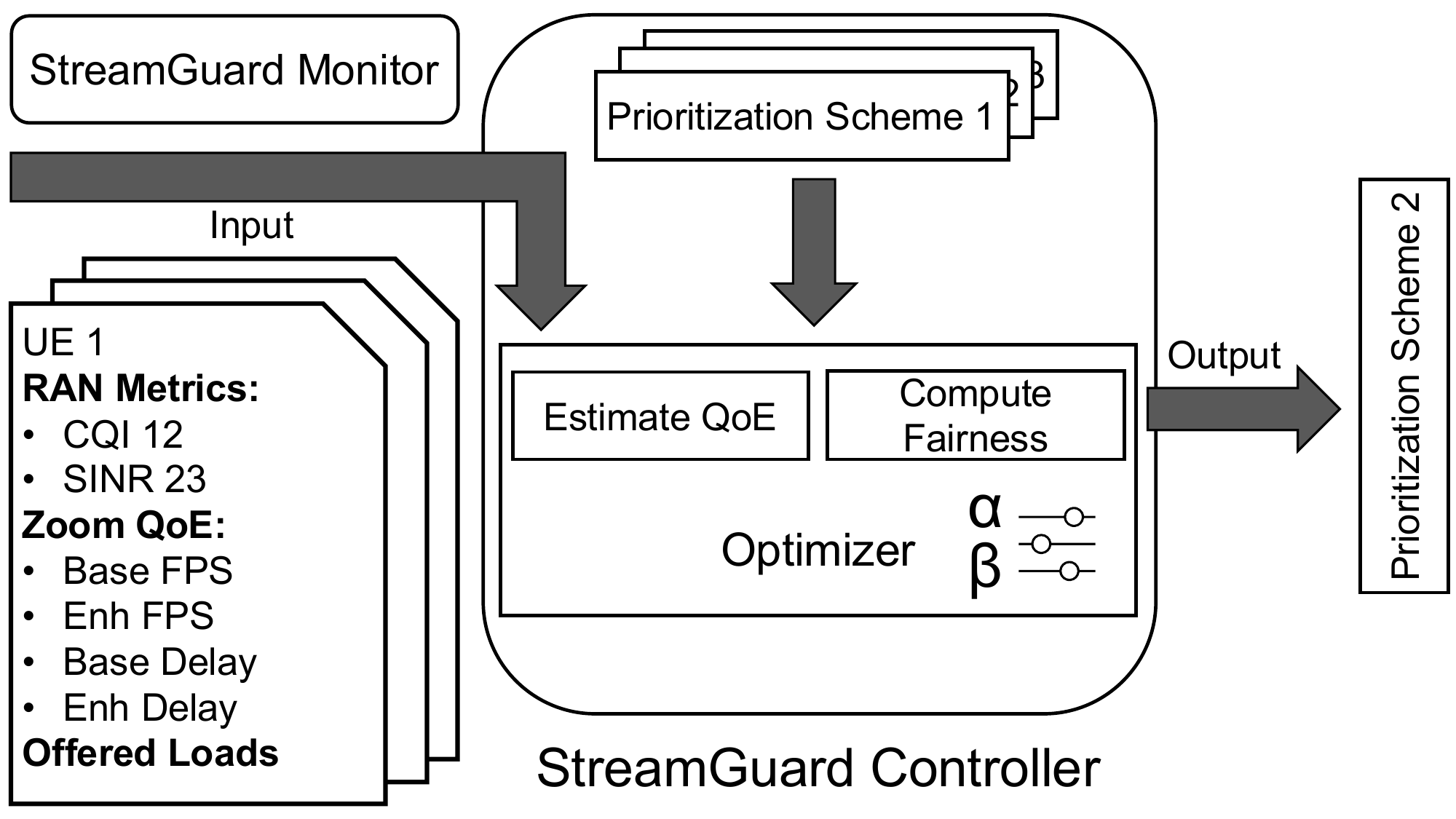}
\caption{\systemname{} controller workflow. It ingests metrics from the monitor, evaluates candidate prioritization schemes, and selects the one that best satisfies the configured QoE–fairness tradeoff via parameters $\alpha$ and $\beta$ (\S\ref{s:design:controller:optimization}).}
\label{fig:controller_workflow}}
\end{figure}

The controller continuously performs ``what-if'' analysis on a small set of subflow-level prioritization schemes for uplink and downlink separately; it selects the scheme that best balances QoE improvement for interactive applications with fairness to other traffic.
Operator-tunable parameters control the ``aggressiveness'' of the prioritization. 
Specifically, this decision process consists of three steps:
QoE gain modeling (\S\ref{s:design:controller:qoe_model}), fairness impact modeling (\S\ref{s:design:controller:fair_model}), and joint optimization (\S\ref{s:design:controller:optimization}), as shown in \cref{fig:controller_workflow}.
The exact prioritization schemes depend on the application and are defined in \systemname{} application plugins.
For illustration, in Zoom, the options are: (1) prioritize all traffic, (2) prioritize all except the enhancement layer, and (3) no prioritization.


\subsubsection{QoE Gain Modeling}
\label{s:design:controller:qoe_model}

In order to perform the before-mentioned ``what-if'' analysis, the controller estimates how an interactive application’s QoE will change under different subflow prioritization choices.
Using real-time QoE indicators from the \systemname{} monitor (\cref{s:design:metric_tracking}), it predicts the marginal benefit of prioritizing specific subflows. 
For each flow $f$, the controller evaluates a set of candidate actions. Let
$
\mathbf{k}_f \in \{0,1\}^{M_f}
$
indicate which of the $M_f$ subflows are prioritized. Given the current QoE state $\mathbf{s}_f(t)$, the controller assigns a normalized gain
\begin{equation}
Q_f(\mathbf{k}_f \mid \mathbf{s}_f(t)) \in [0,1]
\label{eq:qoe_score}
\end{equation}
representing the expected QoE improvement in the next decision interval (\S\ref{s:design:controller:optimization}). At a high level, the QoE gain model captures how prioritizing different subflows improves application performance under the current network and application state. It accounts for both the relative importance of each subflow and the current level of degradation, so that prioritization decisions focus on subflows that are most likely to yield meaningful QoE improvements.

We implement a QoE gain model for each application that assigns every subflow $i$ an intrinsic quality score $q_i \in [0,1]$ based on frame-rate and delay metrics, and computes a weighted baseline QoE.
Prioritization yields additional gain proportional to  $(1 - q_i)$, capturing diminishing returns. 
Enhancement-layer quality is bounded by base-layer quality to reflect the dependencies in the SVC decoding chain.
Appendix~\ref{s:zoomdetails:qoegainmodel} contains details on this, including pseudocode.

\subsubsection{Fairness Impact Modeling}
\label{s:design:controller:fair_model}

While prioritization can substantially improve QoE for interactive flows, it inevitably consumes spectrum resources that would otherwise be available to other users and applications.
Therefore, the controller continuously estimates the fairness cost associated with each prioritization choice.

Using measurements of offered load and RAN metrics (\eg MCS, CQI, Uplink SINR) from \S\ref{s:design:metric_tracking:offered_load}, the controller first models how radio resources would be allocated under the default scheduler, without \systemname{}. 
This establishes a baseline expectation for how much spectrum non-prioritized traffic would receive given current load and channel conditions.
For each candidate prioritization scheme, the controller estimates the total radio resources $R(\mathbf{K}, t)$, measured in \emph{Physical Resource Blocks} (PRBs) per controller interval (\S\ref{s:design:controller:optimization}), reserved for high-priority subflows under decision $\mathbf{K}$, where $\mathbf{K} = (\mathbf{k}_1, \ldots, \mathbf{k}_S)$ denotes the joint prioritization decision across the $S$ active interactive flows.
These resources are effectively ``reserved'', ensuring timely delivery even under congestion.
Let $U(t)$ denote the total PRBs that non-prioritized traffic would receive under the default scheduler over a controller interval, and define the spare capacity as $S(t) = C - U(t)$, where $C$ is the total PRB budget per controller interval.
The fairness impact is captured by the normalized fairness loss score
\begin{equation}
F(\mathbf{K} \mid t) = \frac{\max\big(0, R(\mathbf{K}, t) - S(t)\big)}{U(t)} \in [0,1].
\label{eq:fairness_model}
\end{equation}
By construction, $F(\mathbf{K} \mid t)=0$ when prioritization fits entirely within the available spare capacity. Otherwise, $F(\mathbf{K} \mid t)$ increases as protected subflows begin to consume resources that would otherwise serve ordinary traffic.
The detailed pseudocode is provided in Appendix~\ref{s:fairness_impact_model_impl}.

\paragraph{Allocation modeling for bandwidth-hungry flows.}
The controller is informed whether a UE carries a bandwidth-hungry flow (\eg bulk transfer) by the \systemname{} monitor (\S\ref{s:design:metric_tracking:offered_load}). 
Unlike RTC traffic, such flows continuously probe available capacity and converge to a fair-share throughput under the default scheduler.
To capture this behavior, we model their demand as unbounded by setting their demands to $\infty$. 
This abstraction enables stable estimation of their steady-state resource usage, avoiding misleading short-term fluctuations from congestion control and ensuring that prioritization does not unfairly starve background traffic.

\subsubsection{Joint QoE--Fairness Optimization}
\label{s:design:controller:optimization}

With both QoE gains and fairness costs estimated, the controller selects a subflow-level prioritization strategy that jointly accounts for user experience and resource fairness.
This decision is formulated as a lightweight optimization over a discrete set of candidate prioritization actions and is executed periodically,  every \emph{decision interval} $I$.

The decision interval $I$ is configurable and reflects a tradeoff between responsiveness and stability as it controls how often the controller re-evaluates conditions and updates the prioritization scheme.
A shorter interval allows the controller to react quickly to changes in application demand and wireless conditions, while a longer interval reduces control overhead and avoids unnecessary reconfiguration.
As an example, for Zoom traffic, we perform a sensitivity analysis and set $I = 50\text{ms}$ (\S\ref{s:eval:benchmark:sensitivity_analysis} and \cref{fig:sensitivity_analysis}).
For other applications, $I$ can be tuned per workload.


The controller’s objective is governed by two tunable parameters, $\alpha$ and $\beta$:
\begin{itemize}
    \setlength{\itemsep}{0pt}
    \item $\alpha$ controls how QoE improvements are distributed across RTC flows:
    a large $\alpha$ improves the worst-performing flow; a smaller $\alpha$ optimizes average QoE.
    \item $\beta$ controls the tradeoff between QoE gains and fairness: a smaller $\beta$ enables more aggressive RTC traffic prioritization; a larger $\beta$ makes \systemname{} more conservative.
\end{itemize}

\noindent
For each interactive flow $f$, the controller computes a QoE score $Q_f(\mathbf{k}_f \mid \mathbf{s}_f(t))$ (Eqn.~\ref{eq:qoe_score}) for each prioritization decision. We aggregate per-flow QoE into a system-level objective as
\[
Q_{\text{agg}} =
\alpha \cdot \min_{f \in \mathcal{F}} Q_f
+
(1-\alpha) \cdot \frac{1}{|\mathcal{F}|} \sum_{f \in \mathcal{F}} Q_f.
\]
At each decision interval,  the controller selects
\[
\mathbf{K}^*(t)
=
\arg\max_{\mathbf{K}}
\left[
(1-\beta)\, Q_{\text{agg}}
\;+\;
\beta \, F(\mathbf{K} \mid t)
\right],
\]
where $F(\mathbf{K} \mid t)$ captures the fairness impact (Eqn.~\ref{eq:fairness_model}). 
The resulting decision determines which subflows are prioritized for each RTC flow and is applied by reconfiguring the marking module (\S\ref{s:design:data_plane}), closing the control loop.

\subsubsection{Increase Control-Loop Stability}
To mitigate oscillations in the control loop, \systemname{} applies three stabilizing mechanisms.

First, \systemname{} introduces hysteresis for prioritization decisions. After a transition to a new state is proposed, a timer of duration $5I$ is started. The transition is applied only if the same decision is consistently proposed throughout this period; otherwise, the timer is reset with the new proposal. With $I=50$\,ms for Zoom, this corresponds to 250\,ms of confirmation.
This choice of hysteresis over five control intervals filters out short-term fluctuations and avoids oscillations~\cite{2010feedback}, while remaining responsive to sustained changes, safely exceeding the timescales of RTC sender adaptation dynamics~\cite{realtime_webrtc_control,analysis_of_gcc,investigation_gcc}.

Second, the controller smooths received offered load measurements using an exponentially weighted moving average (EWMA). We set the smoothing factor to $\alpha=0.2$. With decision interval $I=50$\,ms, this $\alpha$ corresponds to a strong memory of approximately 250\,ms from the model in \cite{2018forecasting}, capturing recent trends while filtering out short-term noise from sender dynamics~\cite{analysis_of_gcc,investigation_gcc}. This operates on a timescale similar to that of the hysteresis, ensuring a consistently stable control loop. 

Third, when a new RTC flow is first observed, it is temporarily granted full prioritization for two seconds (in the case of Zoom). This warm-up period protects an application's initial probing traffic for bandwidth estimation, which lasts two seconds in Zoom (\S\ref{s:eval:benchmark}). Without such protection, probe packets may be suppressed by existing congestion, causing the sender to adapt to an unnecessarily low rate.

\subsubsection{RAN-Assisted Application Behavior Shaping}
\label{s:design:controller:behavior_shaping}

While subflow-level prioritization improves the delivery of QoE-critical traffic, it does not by itself prevent excess non-essential traffic from occupying scarce radio resources. We further introduce two RAN-assisted mechanisms: selective subflow dropping and probe-based rate control. Both shape application behavior to work in concert with the RAN control loop under constrained resources.

\paragraph{Selective subflow dropping.}
\label{s:design:controller:app_shape:selective_drop}
When only a subset of an application’s subflows is prioritized under resource constraints, low-priority subflows are often severely delayed and provide little value to the receiver. 
In some cases, these delayed packets can even be harmful.
In addition to consuming scarce radio resources, their delayed arrival affects the receiver's jitter-buffer dimensioning algorithm, ultimately further delaying playback.
We validate this behavior for Zoom in our system.
As shown in~\S\ref{s:eval:benchmark}, dropping poorly delivered enhancement-layer packets in the 5G uplink, when only base-layer video and audio are prioritized, significantly improves playback frame rate and visual quality at the receiver.

Based on this observation, \systemname{} drops low-priority subflows under specific prioritization conditions, thereby avoiding spectrum waste and improving end-to-end QoE.
For Zoom, we drop the enhancement layer in the uplink when only the base layer and audio are prioritized.

\paragraph{Probe-based rate control.}
An application’s offered load may exceed its allocated capacity, causing excess traffic to build up in queues and resulting in long delays.
For RTC traffic, such delayed packets are often no longer useful at the receiver, thereby wasting radio resources.
Many video-conferencing applications adapt their transmission rate using probing traffic to infer available bandwidth~\cite{webrtc_probing1}. 
We again validate this mechanism through controlled experiments on Zoom. 
As shown in~\S\ref{s:eval:benchmark:rate_control_via_probe_drop}, the sender’s transmit bitrate strongly correlates with probe packet delivery: increasing probe drop rates consistently leads to lower transmission rates.

Based on this, the \systemname{} controller instructs the marking module (\S\ref{s:design:data_plane}) to proportionally drop probe packets (when supported by the application), thereby fine-tuning the sender’s rate to better align with the allocated goodput.
This ensures that the flow adapts to the available resource budget, especially when only a subset of subflows is prioritized.

\subsection{\systemname{} Marking Module}
\label{s:design:data_plane}

The \systemname{} Marking Module analyzes the deep structure of traffic, identifies relevant subflows, and maps them to priority levels specified by the controller (\S\ref{s:design:controller}) for differentiated handling. For extensibility, modules leverage packet subflow identification rules that they read from application-specific \systemname{} plugins. Based on these rules, the marking module then rewrites the relevant uplink or downlink QoS field, overcoming the limitations of the coarse-grained 5G QoS architecture (see~\S\ref{s:background}). This inspection and tagging logic differs between uplink and downlink as outlined next.

\begin{figure}
    \centering
        \subfigure[Downlink direction.]{\includegraphics[width=0.48\linewidth]{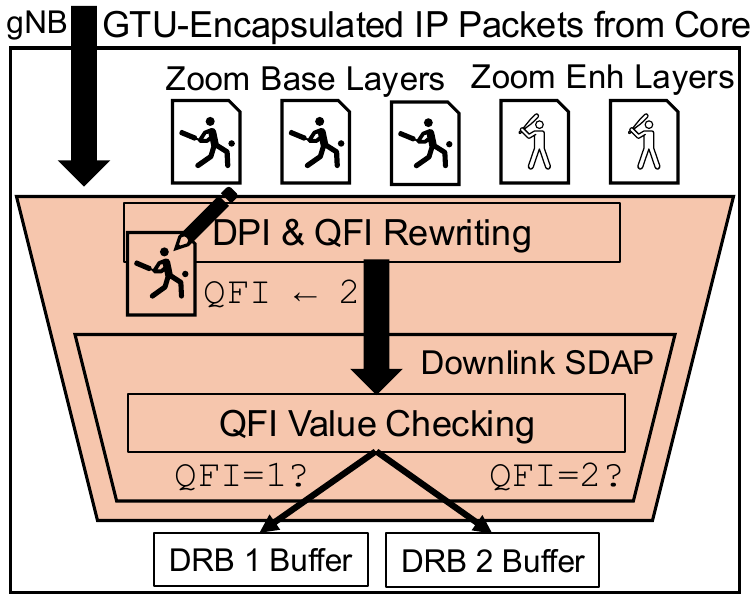}\label{fig:dl_arch}}
        \subfigure[Uplink direction.]{\includegraphics[width=0.48\linewidth]{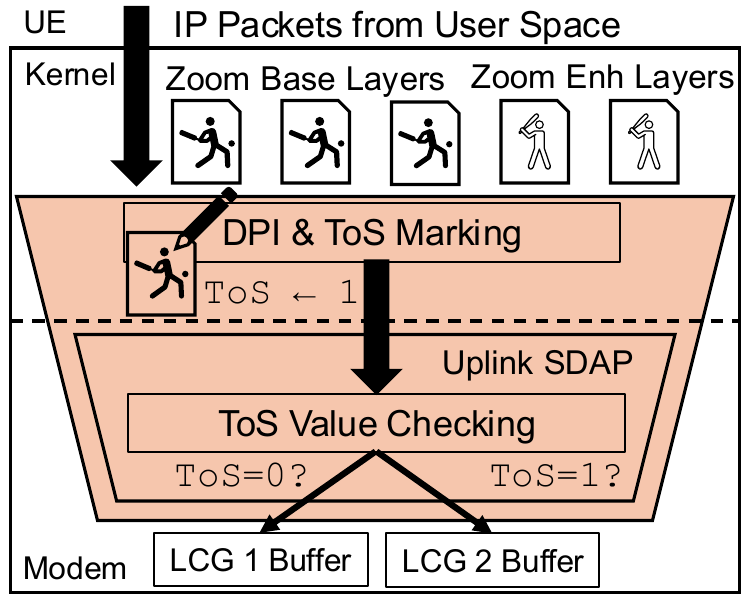}\label{fig:ul_arch}}
        \caption{\systemname{} Marking Module details. For downlink, we perform DPI and rewrite QFI to asssign L7 subflows to different DRBs. For uplink, we perform DPI and ToS marking in a kernel shim layer. These ToS-marked packets are then mapped to their corresponding LCGs by the modem.}
        \label{fig:arch}
\end{figure}

\paragraph{Downlink packet marking.}
Downlink marking operates immediately before the SDAP layer, as shown in \cref{fig:dl_arch}.
The DPI rules and QFI rewriting policies can be dynamically configured by the \systemname{} controller (\S\ref{s:design:controller}) via the E2 interface between the Near-RT RIC and RAN, as shown in \cref{fig:control_loop}. Based on the rewritten QFI values, the SDAP layer in the downstream steers packets associated with different application subflows into distinct DRBs, enabling the underlying scheduler to serve them with different priorities. The scheduler first serves pending data in high-priority DRBs, following its vendor-specific downlink UE scheduling policy, and then proceeds to lower-priority DRBs in the same manner.

\paragraph{Uplink traffic marking.}
In the uplink direction, we propose a lightweight extension to the operating system kernel performing \systemname{} marking.
This is positioned 
immediately before packets enter the cellular modem, as shown in \cref{fig:ul_arch}. 
This module can be readily updated with modern kernel's programmability and extensibility (\eg via eBPF hooks at the traffic control layer) without modifying application code.
This module performs DPI and marks uplink packets with distinct ToS values accordingly. The uplink SDAP layer in the modem can be configured, using existing 5G mechanisms, to map packets with different ToS values into different LCGs with corresponding priorities. The scheduler in the RAN then prioritizes giving uplink grants for high-priority LCGs, following its vendor-specific uplink UE scheduling policy, and subsequently serves lower-priority LCGs in the same manner. The DPI rules and ToS-marking policies are dynamically configured by the \systemname{} Controller through a lightweight IP control channel, which is itself transmitted over a high-priority RAN bearer to ensure timely and reliable delivery.

\section{Implementation}
\label{s:impl}


We implement \systemname{} across multiple software components. We build the \systemname{} Monitor (\S\ref{s:design:metric_tracking}) and the modified MAC scheduler (\ie priority-aware scheduling for marked packets) on top of srsRAN \cite{srsran}. We collect relevant Zoom- and RAN-level metrics directly within the RAN stack, log them efficiently, and pipe them to an external PyPy~\cite{pypy} (compiled Python) program for fast post-processing and forwarding to the \systemname{} Controller (\S\ref{s:design:controller}). We implement the controller in Python and leverage Google OR-Tools~\cite{google_ortools} to efficiently solve the prioritization decision as a constraint-based optimization problem. The detailed formulation is provided in Appendix~\ref{app:controller_optimization}. For the data plane, we implement the downlink \systemname{} Marking Module (\S\ref{s:design:data_plane}) in srsRAN’s SDAP layer, and deploy the uplink \systemname{} Marking Module (\S\ref{s:design:data_plane}) as an eBPF program on the UE. We use Open5GS \cite{open5gs} as the 5G core and extend it to support QoS flow filter installation based on IP ToS values at the UE for uplink flow classification (\S\ref{s:design:data_plane}), addressing a missing capability in the current Open5GS implementation. We also implement the DPI and ToS-marking shim layer on the UE as an eBPF program. The lines of code are summarized in 
Appendix~\ref{s:line_of_codes}.

\begin{figure}
    \centering
        \subfigure[Evaluation testbed setup.]{\includegraphics[width=0.6\linewidth]{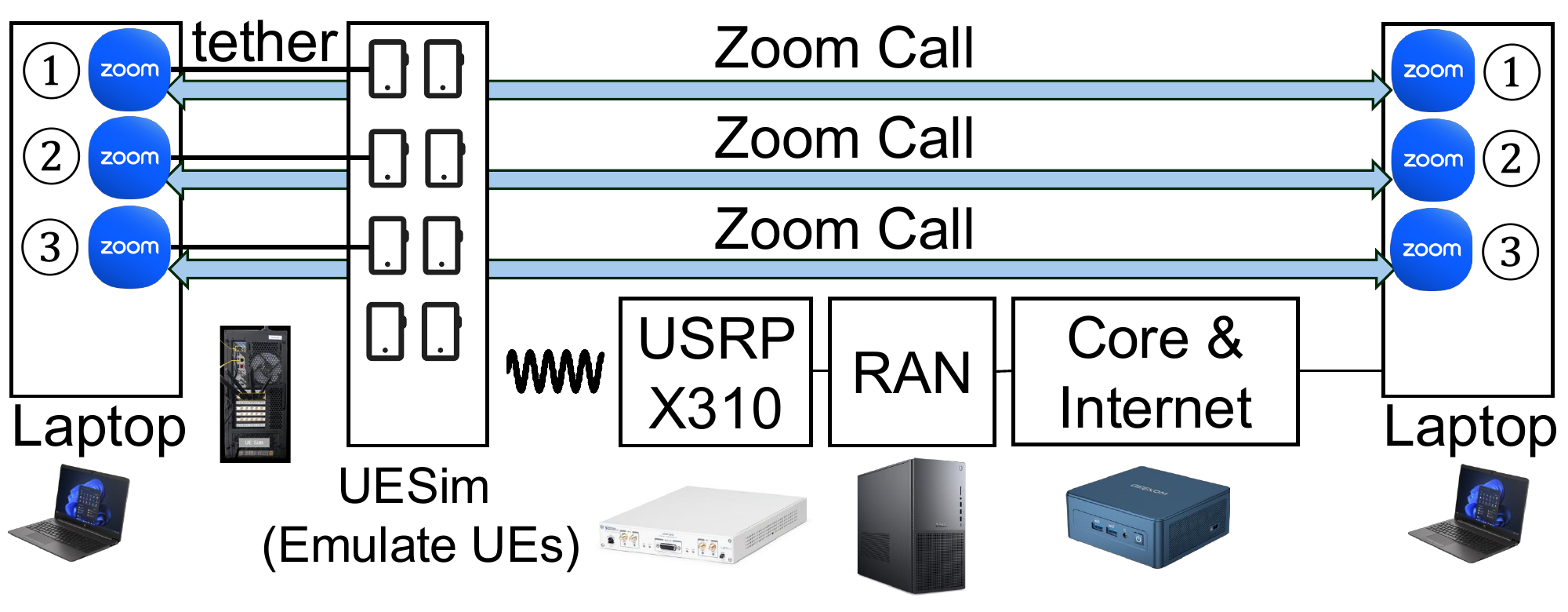}\label{fig:testbed_setup}}
        \subfigure[Prioritization solve time.]{\includegraphics[width=0.39\linewidth]{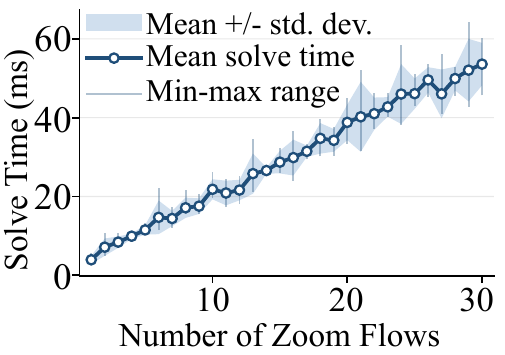}\label{fig:profile_solve_time}}
        \caption{The testbed and the prioritization decision solve time as a function of the Zoom flow number with the setup.}
        \label{fig:setup}
\end{figure}



\section{Evaluation}
\label{s:eval}

In this section, we evaluate \systemname{} against the vanilla 5G QoS framework and state-of-the-art subflow-aware approaches inspired by TC-RAN and DChannel across diverse scenarios. We improve these baselines to make them applicable to our setting (\eg by adding missing uplink support) while preserving their core design principles. We also present micro-benchmarks to justify key \systemname{} design choices.

\subsection{Evaluation Setup}
\label{s:eval:setup}

\paragraph{Experiment testbed.}
As shown in~\cref{fig:testbed_setup}, we use Amarisoft UESim~\cite{amarisoft_uesim}, a high-performance server capable of emulating up to 64 full-stack 5G UEs, to act as the UEs connecting to the RAN. A laptop with an Intel Core i5 CPU and 16\,GB RAM, tethered to UESim, runs Zoom sessions over these emulated UEs. We deploy a USRP X310 as the gNB RF front end, connected to a desktop with an Intel Core i7 CPU and 32\,GB RAM running the srsRAN gNB with our \systemname{} implementation. The gNB operates as a 15\,MHz cell, as supported by srsRAN. The setup is connected to an Open5GS core on another machine (Intel Core i7 CPU, 32\,GB RAM) that provides Internet connectivity. A separate laptop with an Intel Core i5 CPU and 16\,GB RAM, connected via a stable Internet link, hosts the remote Zoom endpoints. Across this setup, Zoom, 5G data processing, and measurement pipelines all run smoothly. The primary potential bottleneck is solving the prioritization decision problem in the \systemname{} controller, which is NP-hard. To evaluate this, we profile the solve time as a function of the number of Zoom calls. As shown in~\cref{fig:profile_solve_time}, with a controller decision interval of 50ms for Zoom (\S\ref{s:design:controller:optimization}), the controller computes the optimal decision within this deadline for up to $\approx$25 Zoom calls, representing a typical high load in a 15\,MHz cell. For larger 100\,MHz cells with more concurrent interactive sessions, the problem search space can be efficiently pruned by the Google OR-Tools solver~\cite{google_ortools} and partitioned into independent subproblems. This results in a parallel structure that can be solved using 6–10 worker threads on a commodity CPU or GPU.

\begin{figure*}
    \centering
    {\includegraphics[width=0.85\linewidth]{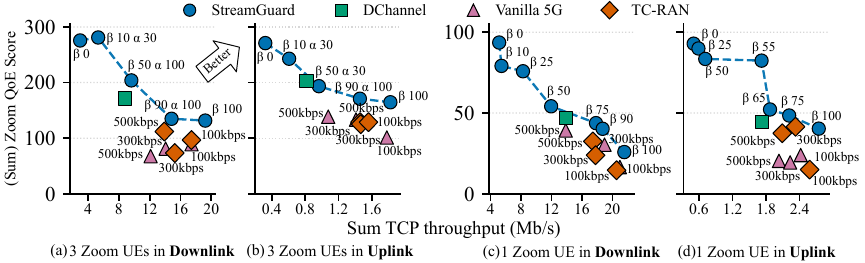}}
         \caption{Sum Zoom QoE and TCP goodput for downlink (receive-only) and uplink (upload-only), in both multi- and single-Zoom-session settings (left to right: \CircledText{2}, \CircledText{4}, \CircledText{1}, \CircledText{3} in \S\ref{s:eval:setup}). \systemname{} achieves a Pareto-optimal frontier over the baselines.}
         \label{fig:qoe_tcp_summary}
\end{figure*}

\begin{figure*}[t]
    \centering
        \subfigure[Scanning $\alpha$ under different $\beta$s in the downlink multi-Zoom-session experiment (\CircledText{2} in \S\cref{s:eval:setup}).\label{fig:multi_dl_qoe_alpha_line}]{\includegraphics[width=0.24\linewidth]{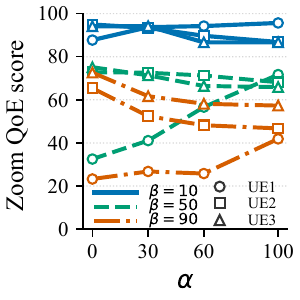}}
         \hspace{0.02\linewidth}
         \subfigure[Per-UE Zoom QoE (3 UEs) vs. aggregate TCP goodput (7 UEs) in the downlink multi-Zoom-session experiment (\CircledText{2} in \S\cref{s:eval:setup}).\label{fig:multi_dl_qoe_tcp_separate}]{\includegraphics[width=0.72\linewidth]
         {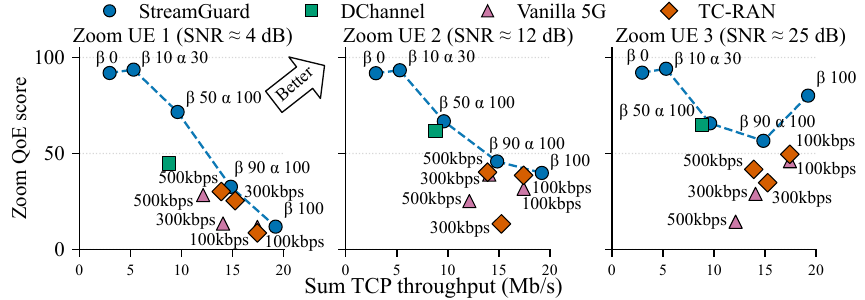}}

         \vspace{-10pt}
         
         \subfigure[Scanning $\alpha$ under different $\beta$s in the uplink multi-Zoom-session experiment (\CircledText{4} in \S\cref{s:eval:setup}).\label{fig:multi_ul_qoe_alpha_line}]
        {\includegraphics[width=0.24\linewidth]{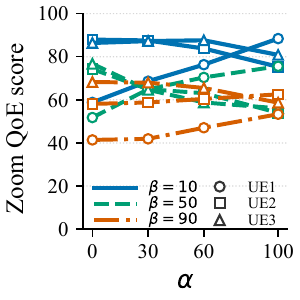}}
         \hspace{0.02\linewidth}
         \subfigure[Per-UE Zoom QoE (3 UEs) vs. aggregate TCP goodput (3 UEs) in the uplink multi-Zoom-session experiment (\CircledText{4} in \S\cref{s:eval:setup}).\label{fig:multi_ul_qoe_tcp_separate}]{\includegraphics[width=0.72\linewidth]{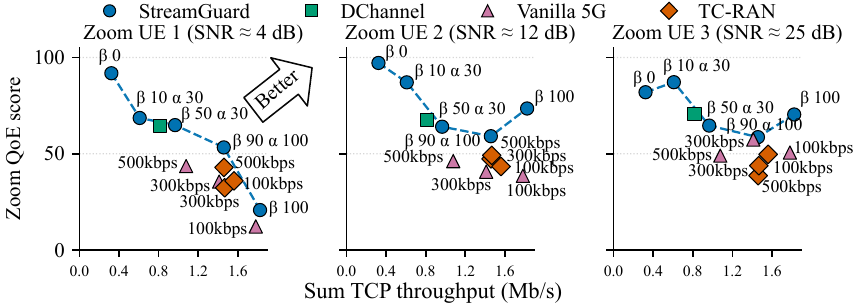}}
        \caption{Downlink-only (first row) and Uplink-only (second row) multi-Zoom-session results. We sweep $\alpha$ under different $\beta$ values (subplots (a) and (c)) and select a representative $\alpha$ for each $\beta$. Subplots (b) and (d) show per-UE Zoom QoE and aggregate TCP goodput for each run. \systemname{} achieves a Pareto-optimal frontier over the baselines.}
        \label{fig:multi_ul_main}
\end{figure*}

\paragraph{Evaluation scenarios.} 
We evaluate in five scenarios: 

\noindent\circnum{1} a single downlink (receive-only) Zoom call on a UE with poor SNR ($\approx$4 dB), with 7 competing TCP flows. 

\noindent\circnum{2} three downlink Zoom calls on UEs with heterogeneous SNRs ($\approx$4, 12, 25 dB), with 7 competing TCP flows.

\noindent\circnum{3} a single uplink (send-only) Zoom call on a UE with poor SNR ($\approx$4 dB), with 3 competing TCP flows.

\noindent\circnum{4} three uplink Zoom calls on UEs with heterogeneous SNRs ($\approx$4, 12, 25 dB), with 3 competing TCP flows.

\noindent\circnum{5} three bidirectional Zoom calls with heterogeneous SNRs, with both downlink (7 flows) and uplink (3 flows) TCP competition.

Across all scenarios, we scale TCP flows to ensure persistent congestion. In multi-Zoom scenarios (\circnum{2}, \circnum{4}, \circnum{5}), we use as many concurrent sessions as supported by the testbed.

We vary $\beta$ and $\alpha$ to study their impact. Single-Zoom scenarios (\circnum{1}, \circnum{3}) focus on $\beta$, while multi-Zoom scenarios (\circnum{2}, \circnum{4}) evaluate $\alpha$ under different $\beta$ values. We then select representative $\alpha$–$\beta$ pairs and apply them to the bidirectional case (\circnum{5}), emulating operator tuning. We compare against three baselines. 

\noindent\textbf{DChannel-style:} Always prioritizes Zoom base-layer video and audio; the enhancement layer is assigned lower priority, following and enhancing DChannel’s design of always prioritizing a certain subset of webpage loading traffic.  

\noindent\textbf{TC-RAN–style:} Prioritizes base-layer video and audio up to a bitrate quota (100, 300, 500 kbps), with excess traffic demoted; the enhancement layer is always low priority, following TC-RAN’s use of SLA-based priority provisioning. 

\noindent\textbf{Vanilla 5G:} Prioritizes the entire Zoom flow up to a quota (100, 300, 500 kbps), with excess traffic demoted, consistent with the standard flow-level 5G QoS mechanism.

\begin{figure}
    \centering  {\includegraphics[width=1\linewidth]{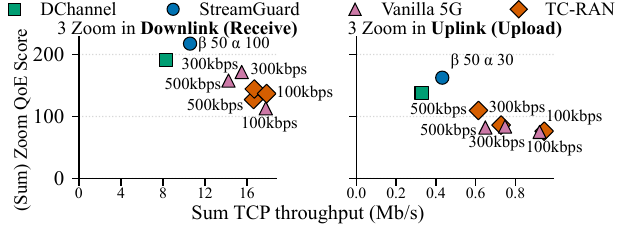}}
         \caption{Sum Zoom QoE and TCP goodput for downlink and uplink, in the bidirectional scenario (\CircledText{5} in \S\ref{s:eval:setup}).}
        \label{fig:bi_qoe_tcp}
\end{figure}

\begin{figure}[t]
    \centering
    {\includegraphics[width=0.99\linewidth]{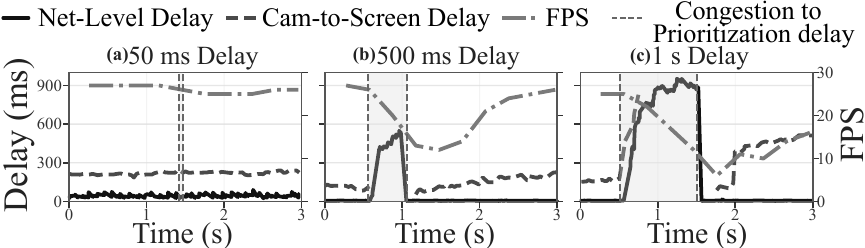}}
         \caption{Sensitivity analysis: varying the delay before competing traffic arrives to induce congestion and trigger prioritization. Gaps in the cam-to-screen delay indicate freezes.}
         \label{fig:sensitivity_analysis}
\end{figure}

\begin{figure*}[t]
    \centering
         \subfigure[Zoom transmit bitrates (average and time-series) under varying probe drop rates.\label{fig:zoom_tx_under_different_probe}]
         {\includegraphics[width=0.71\linewidth]{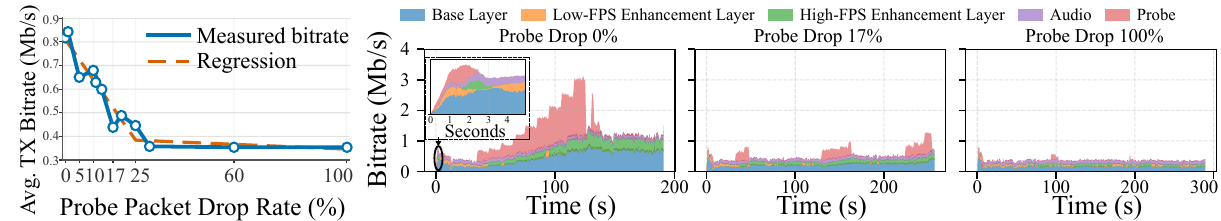}}
         \hspace{0.02\linewidth}
         \subfigure[Zoom enhancement layer drop result.\label{fig:selective_enh_drop}]{\includegraphics[width=0.26\linewidth]{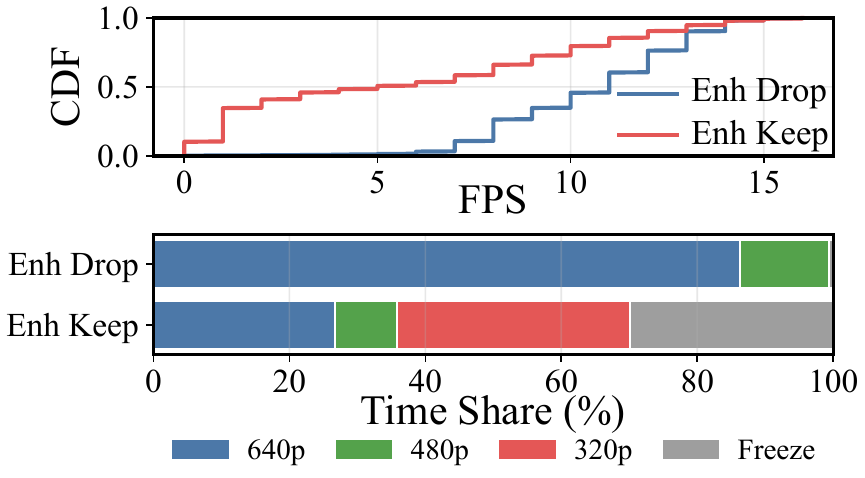}}
        \caption{Experiments on RAN-assisted shaping of Zoom application behavior.}
        \label{fig:probe_drop}
\end{figure*}



\paragraph{Measurement setup.} On the Zoom TX host (Linux), we use a virtual camera (\verb|v4l2loopback|~\cite{v4l2loopback}) and FFmpeg to stream a pre-recorded video where each frame contains a QR code for identification. We log presentation timestamps (PTS) and frame numbers 
at FFmpeg and capture Zoom traffic using tcpdump. On the RX host, we capture Zoom traffic with tcpdump and record the screen, including the video playback and a real-time GUI clock. We use high screen refresh rates and frame-rate recording to ensure precise timing, while keeping overhead minimal. All logging starts before the Zoom session. Both hosts are NTP-synchronized. Each experiment runs a 4–5 minute Zoom session with concurrent TCP traffic. From the data, we extract metrics such as camera-to-screen delay, audio delay, video resolution (from Zoom console), and FPS.

\paragraph{QoE score model in evaluation.} We evaluate user experience using a composite QoE score in the range $[0,100]$, tailored for Zoom and constructed following established practice in prior QoE modeling studies \cite{CFA, can_you_see, demystify_video_conference, domino, salsify, dashlet, pensieve}. The overall score is defined as the sum of 4 equally weighted terms (25 points each):
$
\text{QoE} = Q_{\text{audio}} + Q_{\text{video}} + Q_{\text{fps}} + Q_{\text{res}}.
$

\noindent\textbf{Audio and video delay.} The audio ($Q_{\text{audio}}$) and video ($Q_{\text{video}}$) components capture timeliness and stability of media delivery. Each is computed based on the mean one-way delay and mean RTP jitter, where lower delay and jitter yield higher scores, and values beyond application-specific thresholds are clipped. 

\noindent\textbf{Frame rate.} The frame rate component ($Q_{\text{fps}}$) reflects visual smoothness. It is computed from the average rendered FPS over time, with penalties for temporal variations and frame drops. The score increases with FPS and saturates once a target frame rate is reached. 

\noindent\textbf{Resolution.} The resolution component ($Q_{\text{res}}$) captures visual quality. It is computed based on the fraction of time the video is rendered at each resolution, excluding frozen periods. Higher resolutions contribute more to the score. 

All components are normalized to $[0,25]$. Detailed formulations, parameter values, and freeze detection rules are provided in Appendix~\ref{s:eval_qoe_details:qoe_model_details}.

\subsection{Evaluation Results}

For the downlink single-Zoom scenario (\CircledText{1}), increasing $\beta$ allocates more resources to Zoom, producing a clear Pareto frontier between QoE and TCP throughput (\cref{fig:qoe_tcp_summary}(c)). This is enabled by \systemname{}’s feedback design, which prioritizes critical subflows when they matter most. In contrast, TC-RAN and Vanilla 5G perform similarly and poorly under congestion, as their coarse quotas fail to protect key subflows, leading to high delay and freezes. \systemname{} improves QoE from unusable ($\approx$20–40) to smooth levels ($\approx$60–80) with comparable or higher TCP throughput. DChannel achieves reasonable QoE but offers only a single operating point.

The uplink case (\CircledText{3}, \cref{fig:qoe_tcp_summary}(d)) shows similar trends. \systemname{} again traces a Pareto frontier and improves QoE by 15–30 points over baselines at comparable throughput. TC-RAN and Vanilla 5G often fail to sustain even smooth audio transmission under tight quotas, causing large audio delays and self-inflicted video pauses. DChannel remains competitive but inflexible. We observe a steep QoE drop when $\beta$ increases from 55 to 65 because resource reduction at this point triggers Zoom low-FPS modes. 

In the multi-UE downlink scenario (\CircledText{2}), increasing $\alpha$ reduces QoE disparity across UEs, consistent with its role in shifting toward max-min fairness  (\cref{fig:multi_dl_qoe_alpha_line}). For $\beta$=10, the gap shrinks but slightly re-expands in the other way due to inevitable in-network QoE estimation error, which can lead to mild over-protection of the worst-condition UE. Following an operator-style parameter sweep, we select $\alpha = 30$ for $\beta= 10$, $\alpha = 100$ for $\beta = 50$, and $\alpha = 100$ for
$\beta = 90$ for comparison against baselines. Note that $\alpha$ has little impact at the extremes $\beta = 0$ and $\beta = 100$, where the system either does not prioritize Zoom at all or fully prioritizes it
(with sufficient cell capacity), respectively. Across $\beta$ values, \systemname{} consistently outperforms TC-RAN and Vanilla 5G, improving QoE by 20–40 points at similar TCP throughput, or achieving higher throughput at similar QoE (\cref{fig:multi_dl_qoe_tcp_separate}). UEs with better channel conditions achieve higher QoE. Interestingly, UE 3 shows a QoE rebound as $\beta$ increases from 90 to 100. At $\beta=100$, its strong channel allows sufficient goodput for high-quality Zoom delivery. When $\beta<100$, prioritization toward weaker UEs reduces its resources and degrades performance. An outlier occurs for UE 2 under TC-RAN (300 kbps), where Zoom pauses video for an extended period instead of adapting its bitrate when delivery is impaired.

The uplink multi-UE scenario (\CircledText{4}) shows similar behavior (\cref{fig:multi_ul_main}). Following the same
procedure as in the downlink, we select $\alpha = 30$ for $\beta = 10$, $\alpha = 30$ for $\beta = 50$, and $\alpha = 100$ for $\beta = 90$ for comparison against baselines. \systemname{} improves QoE by 15–35 points and achieves up to $\approx$2$\times$ higher throughput than TC-RAN and Vanilla 5G at similar QoE. DChannel achieves comparable performance to \systemname{} but provides only a single operating point. Increasing $\alpha$ again reduces QoE disparity, and with \systemname{} all UEs trace Pareto frontiers between QoE and throughput.

The bidirectional case, as shown in~\cref{fig:bi_qoe_tcp}, follows similar trends. A representative configuration ($\beta=50,\alpha=100$) achieves high QoE with reasonable TCP throughput. DChannel underperforms, while TC-RAN and Vanilla 5G achieve higher throughput but much lower QoE, placing them below the Pareto trend observed in the single-direction scenarios.

The detailed QoE breakdown for all experiments is provided in Appendix~\ref{s:eval_qoe_details:eval_details}.

\subsection{Micro-Benchmarks}
\label{s:eval:benchmark}
We present the following micro-benchmarks to justify key design choices and mechanisms in \systemname{}.

\noindent
\textbf{Sensitivity analysis of the controller interval.}
\label{s:eval:benchmark:sensitivity_analysis}
As shown in \cref{fig:sensitivity_analysis}, we study the impact of the controller interval $I$ during a congested downlink Zoom session. 50\,ms intervals maintain low delay and stable FPS, while longer intervals (\ie 500\,ms or 1\,s) allow congestion to persist, causing delay spikes, enlarged jitter buffers, and visible freezes before recovery. This motivates our choice of a short control interval $I=50$ for Zoom (\S\ref{s:design:controller}). To better understand how network dynamics affect video playback, we develop \emph{Flow2Frame}, which maps packets to their corresponding video frames (Appendix~\ref{s:flow2frame}). This shows that delayed prioritization prolongs packet delays, causing frames to miss playback deadlines and not be rendered at the receiver.

\noindent
\textbf{Zoom TX rate control via probe dropping.}
\label{s:eval:benchmark:rate_control_via_probe_drop}
We conduct controlled experiments by selectively dropping Zoom probe traffic at varying drop rates and observing the resulting sender behavior. As shown in \cref{fig:zoom_tx_under_different_probe}, there is a clear relationship between probe drop ratio and the reduction in sending rate. \systemname{} exploits this by proportionally dropping probes to steer the sending rate toward available capacity (\S\ref{s:design:controller:behavior_shaping}). \cref{fig:zoom_tx_under_different_probe} also shows representative time series under different drop rates.

\noindent
\textbf{Selective dropping of enhancement layer.}
\label{s:eval:benchmark:selective_drop}
We conduct a controller experiment comparing dropping versus not dropping the video enhancement layer when only the base layer and audio are prioritized under heavy uplink
congestion in 5G. As shown in \cref{fig:selective_enh_drop}, enabling dropping significantly improves FPS and resolution at the receiver. This indicates that delayed enhancement-layer packets can negatively affect the receiver’s playback logic, and removing
them avoids such degradation and saves radio resources. \systemname{} incorporates this mechanism (\cref{s:design:controller:behavior_shaping}).
\section{Conclusion}

We have presented \systemname{}, a practical 5G architecture for subflow-level, QoE-aware prioritization of real-time applications. By leveraging application semantics, \systemname{} dynamically prioritizes QoE-critical subflows through a closed control loop that ingests in-network QoE estimates and RAN signals, models QoE gains and fairness impact under candidate actions, and enforces the most appropriate decisions in a standards-compliant manner. Our implementation, tailored for Zoom and deployed on a real 5G testbed, shows that \systemname{} consistently improves video-conferencing QoE under congestion and poor wireless conditions while preserving strong fairness for competing traffic, achieving a Pareto frontier between QoE and background TCP throughput across scenarios compared to prior approaches. More broadly, \systemname{} provides a 5G platform that naturally respects Application-Level Framing (ALF), enabling efficient support for emerging interactive applications with diverse QoE requirements. As future work, we plan to extend \systemname{} to support WebRTC-based applications.

This work does not raise any ethical issues.
\section{Acknowledgements}

This material is based upon work supported by the National Science Foundation under grants AST-2232457, CNS-2223556, and OAC-2429485.
\clearpage
\let\oldbibliography\thebibliography
\renewcommand{\thebibliography}[1]{%
  \oldbibliography{#1}%
  \setlength{\parskip}{0pt}%
  \setlength{\itemsep}{0pt}%
}
\bibliographystyle{concise2}
\bibliography{reference}
\clearpage
\begin{appendices}

\section{Flow2Frame: A Methodology of Mapping Network Packets to Video Frames}
\label{s:flow2frame}
To enable the analysis of whether network dynamics truly impact the rendering of corresponding video frames, we introduce a methodology called \textit{\textbf{Flow2Frame}}, which maps RTP packets to the video frames they carry in interactive video services using the RTP protocol. Crucially, this approach does not require access to application source code, making it applicable to commercial, closed-source platforms like Zoom. As mentioned in~\S\ref{s:eval:benchmark:sensitivity_analysis} for the , Flow2Frame reveals how network-induced packet issues affect video playback, providing critical diagnostic insight to guide system design improvements. Without loss of generality, we present Flow2Frame in the context of Zoom as a representative example.


\subsection{Experiment Setup}

We adopt the same measurement methodology as described in \S\ref{s:eval:setup}. Also, we use strace on the Zoom TX host to trace \verb|ioctl(camera_fd, VIDIOC_DQBUF, ...)| and \verb|sendmsg()| system calls from the Zoom process. These correspond to the moments when a frame is dequeued from the virtual camera for encoding and when the encoded video packets are transmitted over the network, respectively.

\subsection{Methodology}

\begin{figure}
    \centering
        {\includegraphics[width=0.99\linewidth]{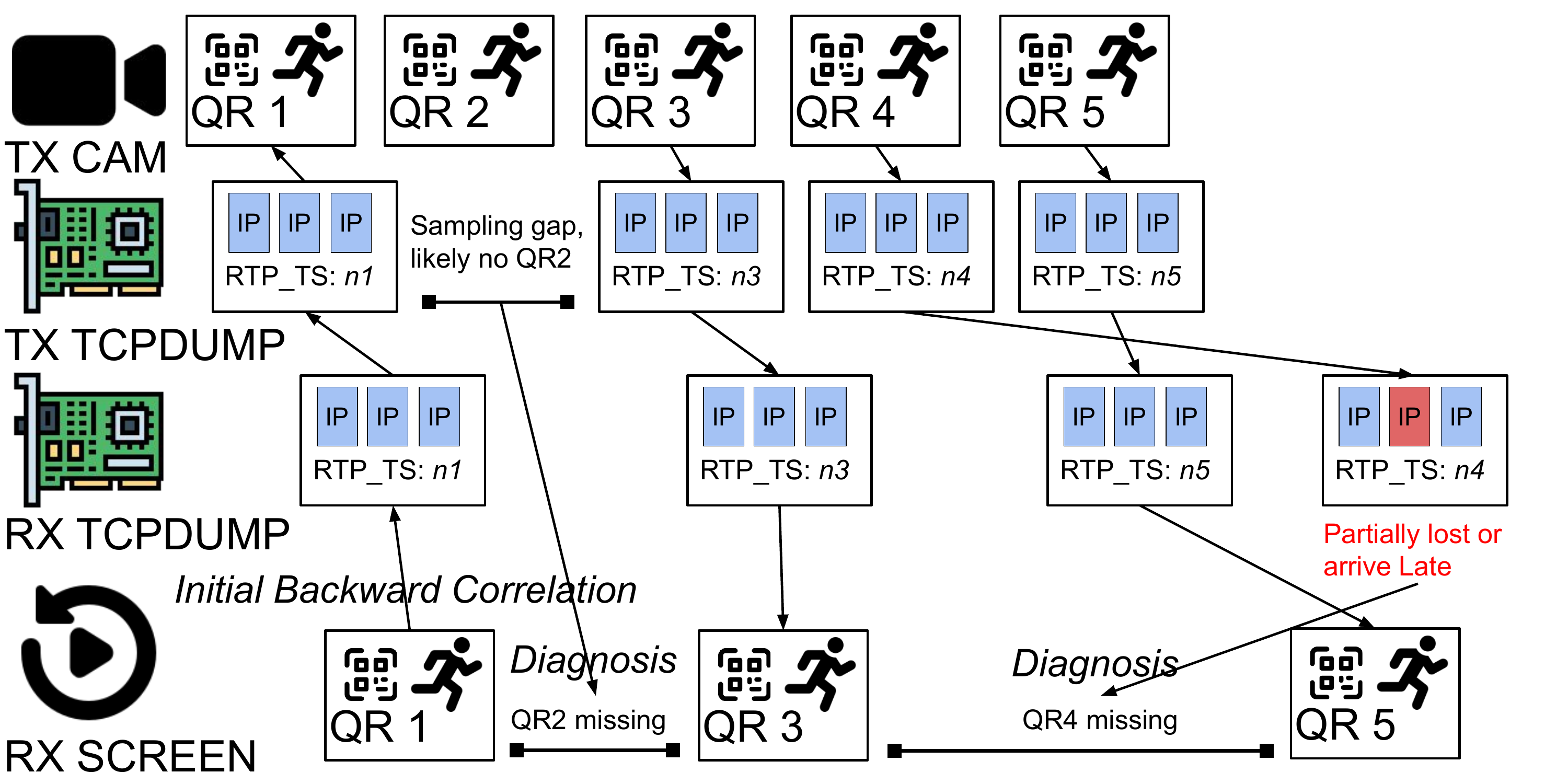}\label{fig:screen_measure}}
        \caption{How to map video frames to their corresponding RTP packets.}
        \label{fig:screen_measure1}
\end{figure}

With the setup described above, we obtain multiple layers of measurement data, as illustrated in \fig{fig:screen_measure1}. Since Zoom uses RTP for video transmission, we begin by identifying the first RTP timestamp and its associated batch of RTP packets, and correlating them with the first video frame rendered on the Zoom RX screen. We refer to this step as the initial backward correlation, shown in the same figure.

By specification, an RTP timestamp marks the sampling instant of the first octet of a frame, but this timestamp is derived from a media clock that differs from the system clock (albeit running at the same rate). This discrepancy presents a key challenge: determining when, on the system clock axis, the first RTP-sampled frame was actually captured. Solving this allows us to detect if any camera frames were skipped by the Zoom encoder, as visualized in \fig{fig:screen_measure1}.

To address this, we employ a sliding window inference technique to align the RTP timeline with the camera's system PTS timeline. Given the shared linearity of the media clock and system clock, the first RTP timestamp should fall between the first and second camera frame PTSs. We discretize this range into 1-ms bins and iteratively test candidate placements. For each, we evaluate the resulting mapping and select the one that minimizes false positives—cases where a supposedly skipped frame unexpectedly appears on the RX side. Details of this algorithm and illustrative examples are provided in Appendix A. Across many measurement sessions lasting several minutes, our method typically finds an alignment with zero false positives (or one in rare cases), while suboptimal alignments exhibit significantly more. This sharp contrast reinforces the correctness of the inferred alignment and provides strong guidance on how to match RTP timestamps to frame PTS values. Further validation is presented in \S\cref{s:flow2frame:rtp_point_placement}.

\begin{figure*}
    \centering
         \subfigure[Prioritization is applied 50ms after competing traffic joins.]{\includegraphics[width=0.45\linewidth]{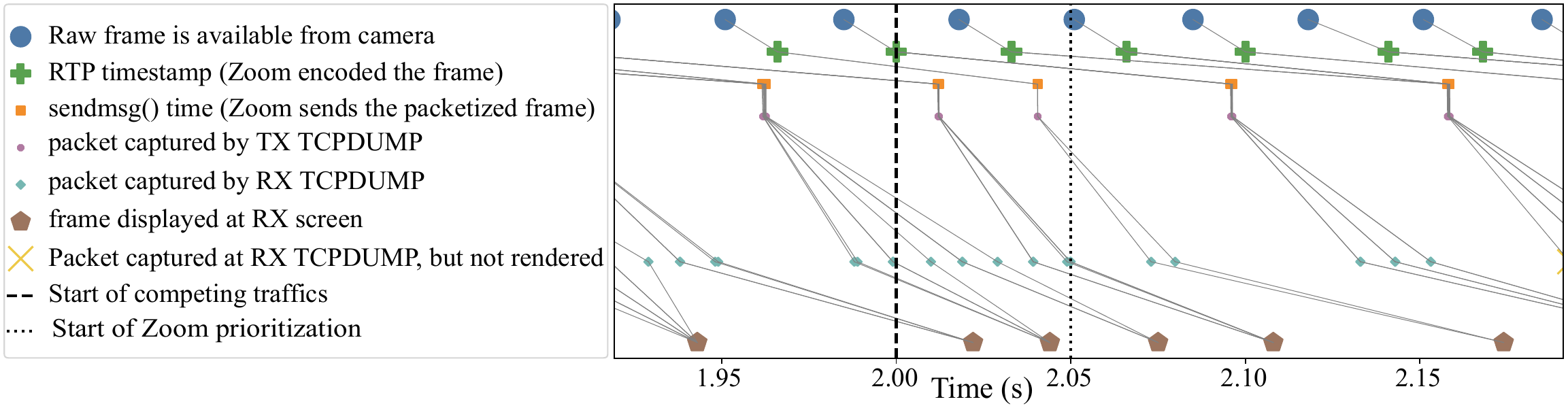}\label{fig:0p05_flow2frame}}
        \subfigure[Prioritization is applied 1s after competing traffic joins.]{\includegraphics[width=0.50\linewidth]{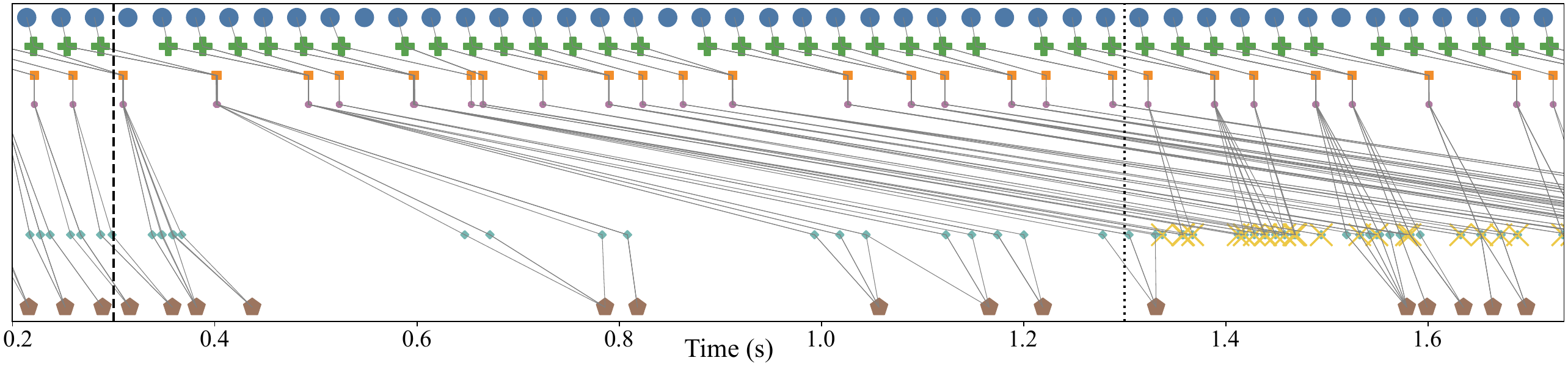}\label{fig:1_flow2frame}}
        \caption{Zoom QoE degradation under different prioritization delays, measured with Flow2Frame.}
        \label{fig:migration_effect}
\end{figure*}

Once the initial RTP timestamp is correctly aligned, we can trace which RTP timestamp corresponds to which camera frame, and detect gaps (\eg when frames are skipped due to encoder decisions). This allows us to associate each video frame with its RTP timestamp and packet batch. Consequently, if a frame is missing on the Zoom RX screen, we can determine whether it was never encoded at the TX side or dropped due to other causes such as network loss or RX-side decisions, as shown again in \fig{fig:screen_measure1}.

As demonstrated in \fig{fig:migration_effect}, Flow2Frame serves as a powerful diagnostic framework, enabling us to pinpoint whether display artifacts are indeed caused by network issues and to inspect behavior at each stage of the video delivery pipeline.

\subsection{RTP Point Placement in Flow2Frame and Further Validation}
\label{s:flow2frame:rtp_point_placement}

\begin{figure}[!ht] 
\centering
\begin{lstlisting}
# cam1_t: presentation timestamp for 1st frame
# cam2_t: presentation timestamp for 2nd frame
# relative_rtp_points: 
# RTP timestamps, and the first has value 0
def put_rtp_points(cam1_t, cam2_t, relative_rtp_points):
    max_delta, max_accuracy = None, 0
    # we iterate every millisecond bin
    for every ms in range(cam2_t - cam1_t):
        t = cam1_t + ms
        abs_rtp_points = guess(t, relative_rtp_points)
        tx_skipped = link(abs_rtp_points, all_cam_t)
        # rx_skipped: frames not displayed by RX
        accuracy = verify(tx_skipped, rx_skipped)
        if accuracy >= max_accuracy:
            best_ms, max_accuracy = ms, accuracy
    # cam1_t + best_ms is the optimal base time
    # which should be the first RTP ts
    return best_ms

def guess(t, relative_rtp_points):
    # add the guessed base time t to every point
    return [t + point for point in relative_rtp_points]

# all_cam_t: presentation timestamps for frames
def link(abs_rtp_points, all_cam_t):
    sampled = set()
    for p in abs_rtp_points:
        find cam_t closest on the left of p in all_cam_t
        sampled.add(cam_t)
    # the complement are those camera frames
    # not sampled by Zoom tx
    return all_cam_t - sampled

def verify(tx_skipped, rx_skipped):
    false_positive = 0
    for qr in tx_skipped:
        if qr not in rx_skipped:
            false_positive = false_positive + 1
    return (tx_skipped - false_positive)/tx_skipped
\end{lstlisting}
\caption{Pseudocode for placing the RTP points in the system clock axis.}
\label{listing:placement_code}
\end{figure}

\begin{figure*}
    \centering
         \subfigure[Good RTP point placement]{\includegraphics[width=0.33\linewidth]{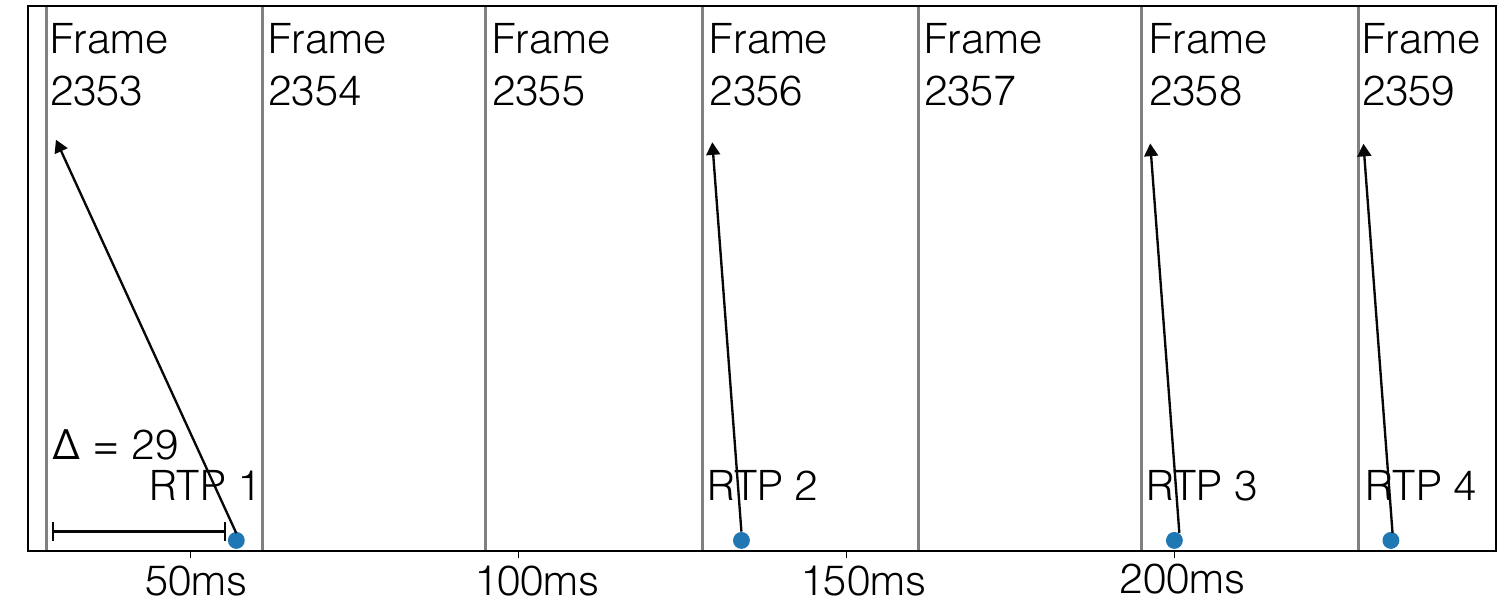}\label{fig:good_placement}}
        \subfigure[Bad RTP point placement]{\includegraphics[width=0.33\linewidth]{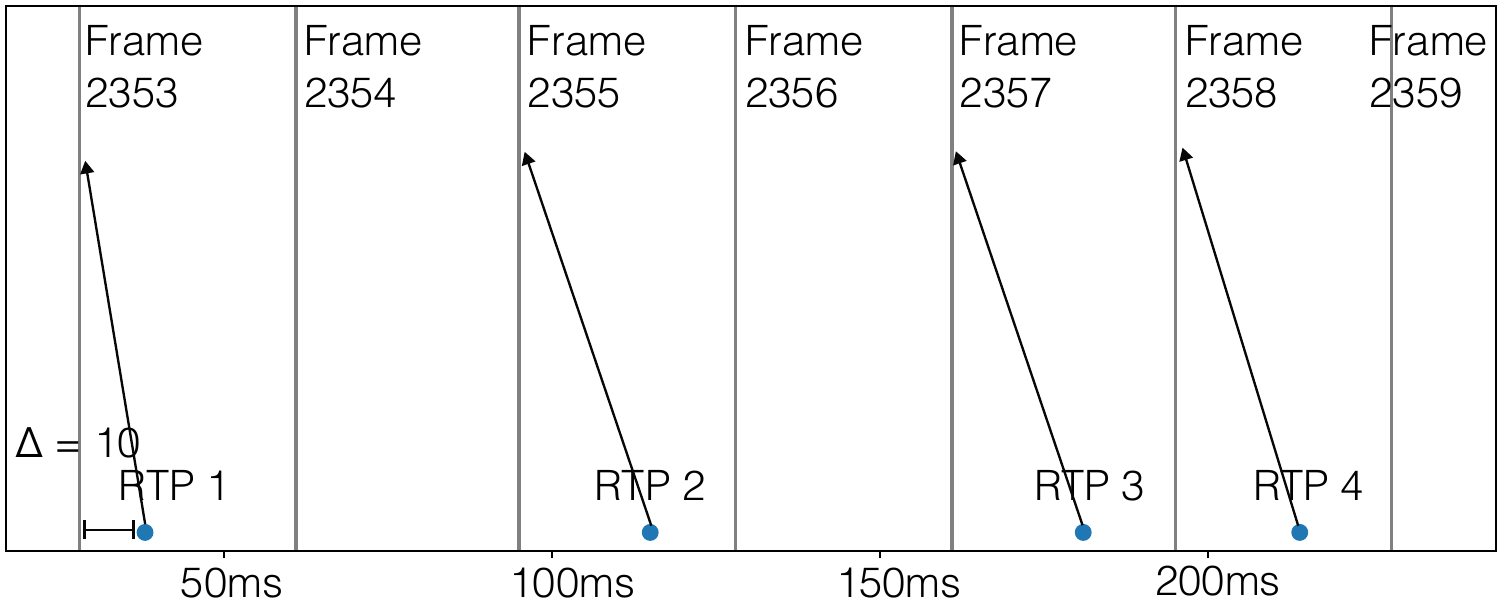}\label{fig:bad_placement}}
        \subfigure[False positive count for different $\Delta$]{\includegraphics[width=0.33\linewidth]{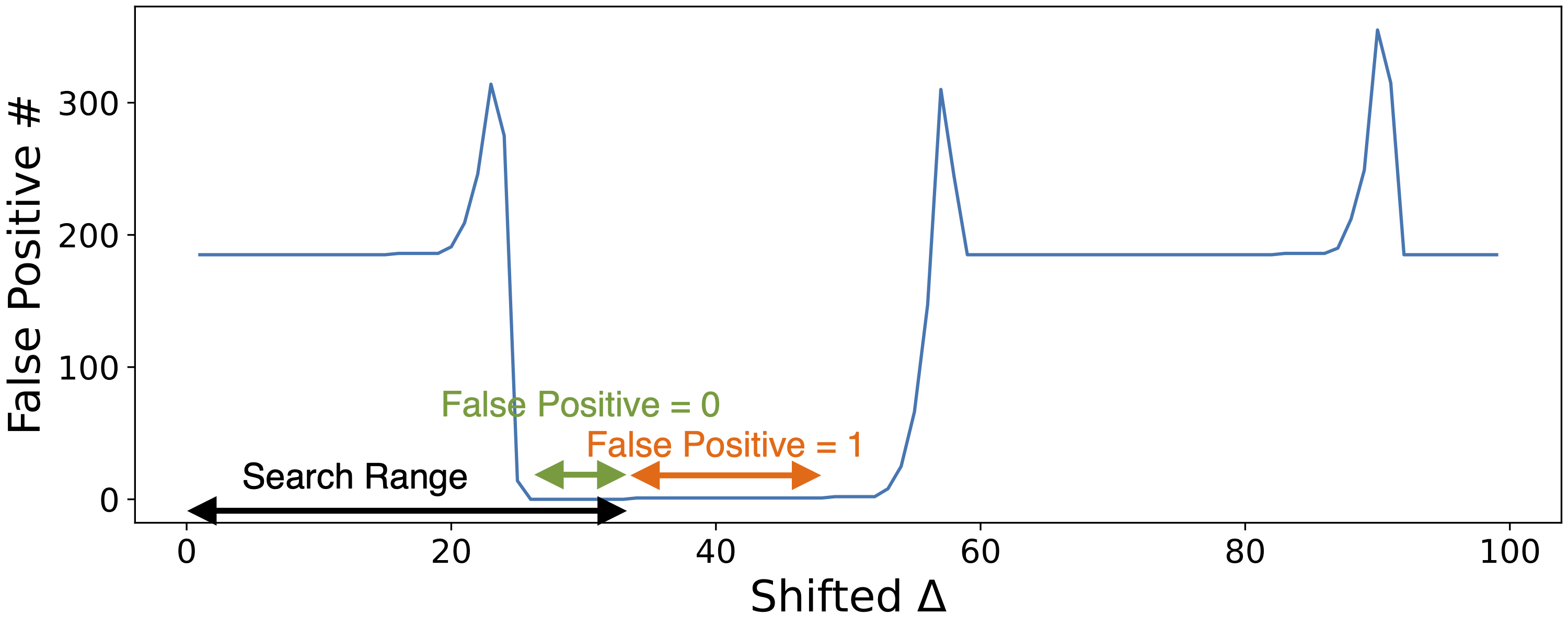}\label{fig:false_positive_count}}
        \caption{Illustration of the RTP timestamp placement procedure. In this frame snippet, frames 2354, 2355, and 2357 are not rendered by the Zoom receiver. With a placement offset of $\Delta = 29$, all frames presumed to be skipped by the sender are indeed absent on the receiver side. In contrast, with $\Delta = 10$, there are 185 instances where frames expected to be skipped by the sender appear at the receiver—indicating mismatches. Thus, we can confidently conclude that (a) represents a valid placement, whereas (b) does not.}
        \label{fig:placement_example}
\end{figure*}

The pseudocode in Listing~\ref{listing:placement_code} outlines the procedure for mechanically determining the optimal placement of RTP timestamp points. We use $\Delta$ to refer to the tick index. Specifically, the method attempts to align the first RTP timestamp at each millisecond tick between the PTS of the first and second camera frames (as implemented in \verb|put_rtp_points|). Once this base timestamp is chosen, the full sequence of RTP points on the system clock timeline can be extrapolated (via \verb|guess|). Each RTP timestamp is then associated with the camera frame it samples by locating the nearest frame on its left (in \verb|link|). This mapping reveals which frames the Zoom TX encoder is expected to skip (visualization can be seen in \fig{fig:placement_example}). We then assess the number of cases where such skipped frames nonetheless appear on the RX screen—termed \textit{false positives} (checked in \verb|verify|). The optimal placement is defined as the one that minimizes the false positive count. As shown in \fig{fig:false_positive_count}, the optimal placement typically results in a false positive count of 0 throughout the entire measurement period, which spans several minutes.

~\fig{fig:good_placement} and ~\fig{fig:bad_placement} illustrate examples of good and bad RTP timestamp placements, respectively. 

\begin{figure*}
    \centering
         \subfigure[Frame 5399]{\includegraphics[width=0.33\linewidth]{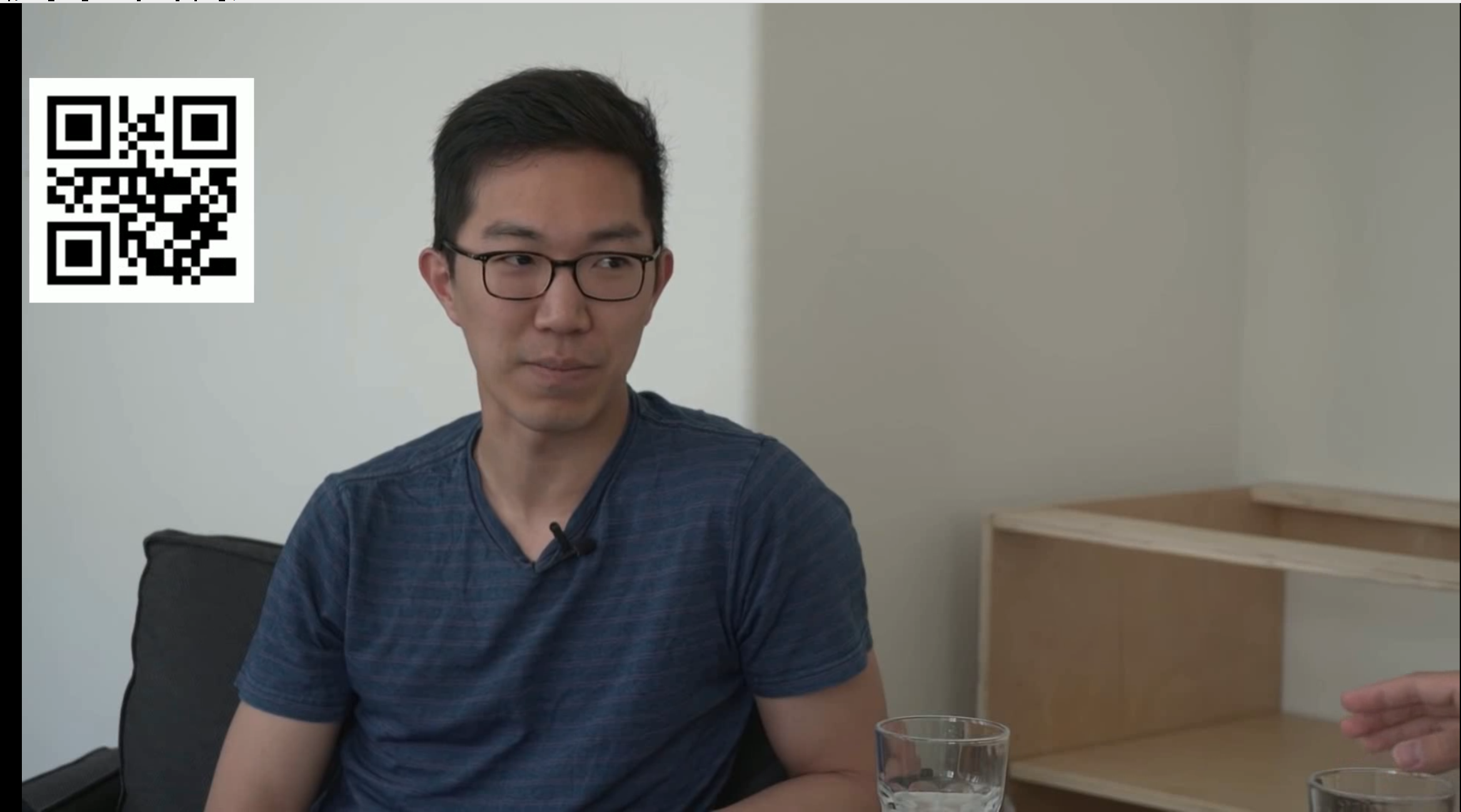}\label{fig:frame_5399}}
        \subfigure[Frame 5400]{\includegraphics[width=0.33\linewidth]{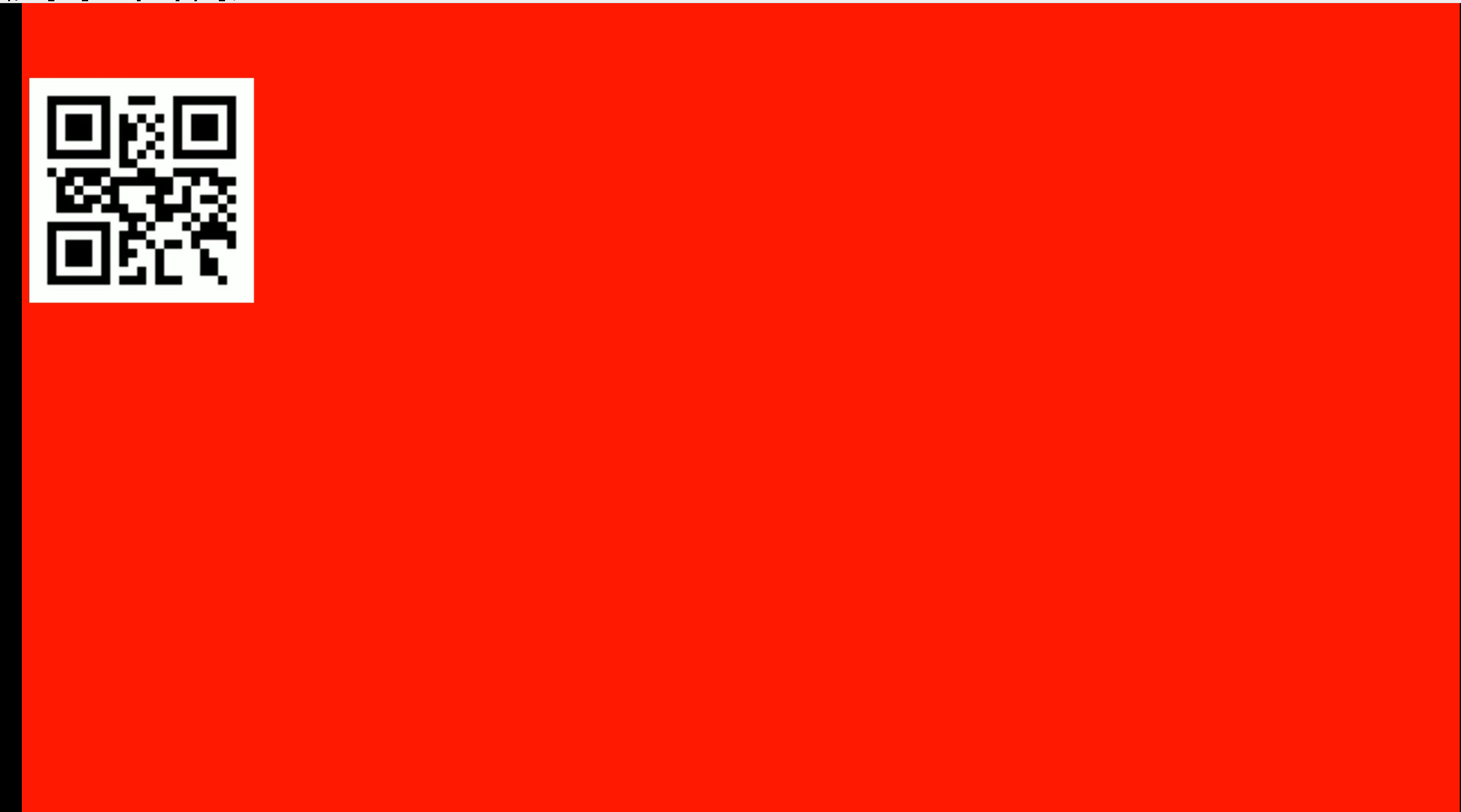}\label{fig:frame_5400}}
        \subfigure[Frame 5401]{\includegraphics[width=0.33\linewidth]{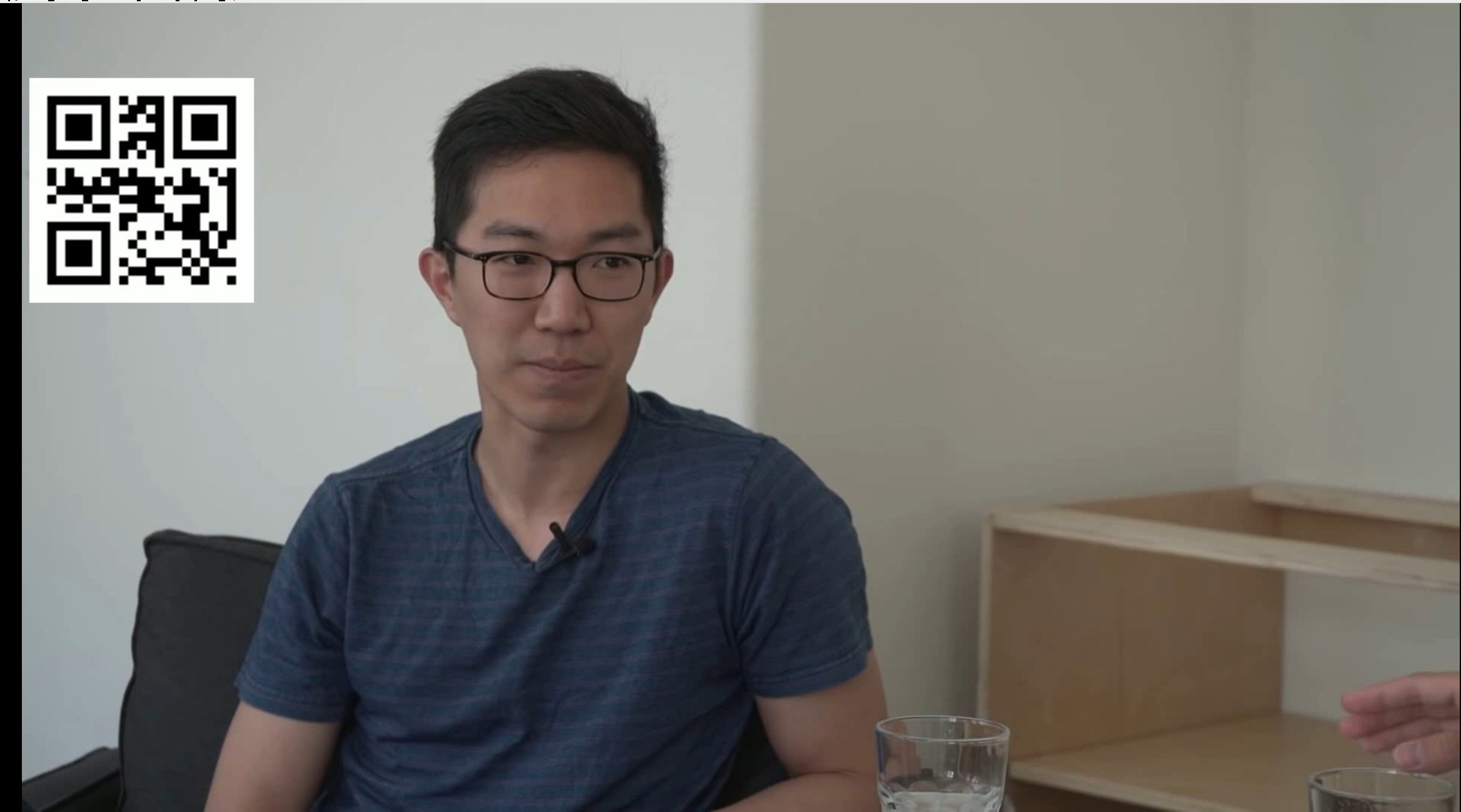}\label{fig:frame_5401}}
        \subfigure[Wireshark display of IP packets for the single-color frame, as well as its preceding and following frames. In "Extension Data" field, "50000" means the packet is for a base-layer frame, while "5ff777" means the packet is for an enhancement-layer frame.]{\includegraphics[width=0.99\linewidth]{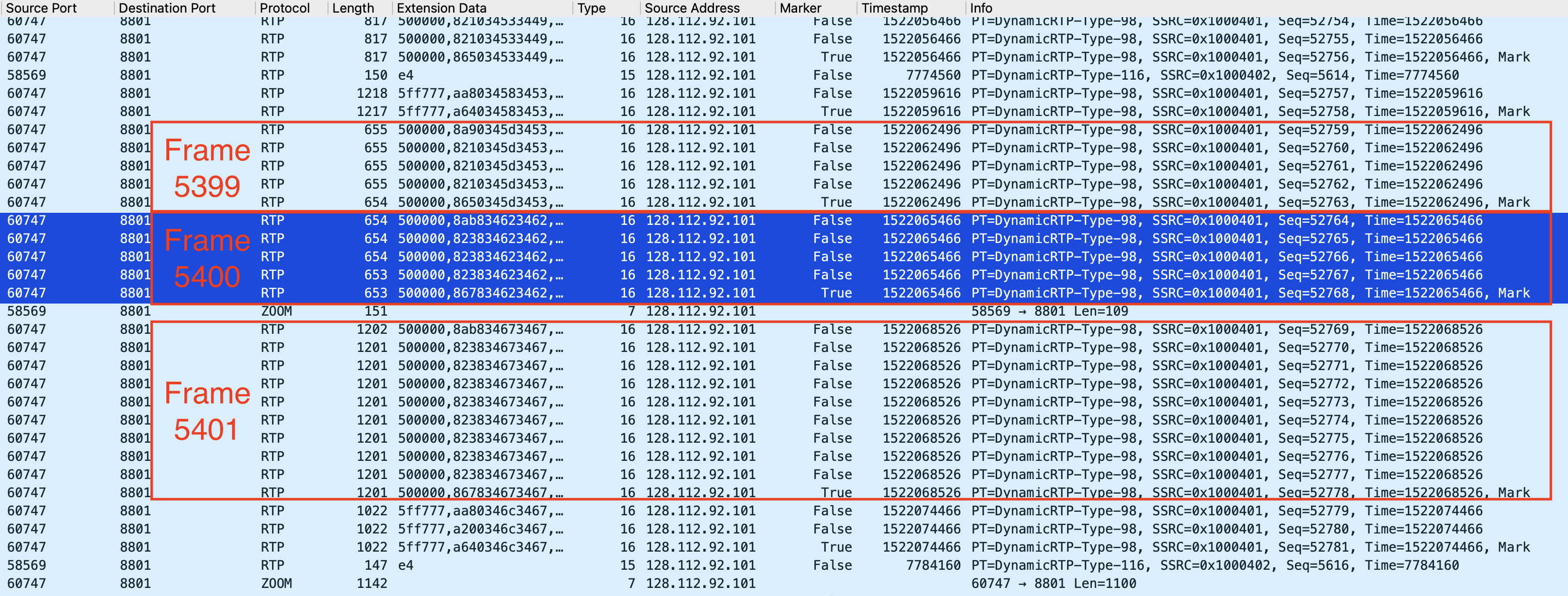}\label{fig:wireshark_injection}}
        \caption{We inject a single-color frame into the middle of the test video to examine whether the IP packets mapped to this frame using our technique exhibit any distinguishing characteristics.}
        \label{fig:single_color_test}
\end{figure*}

To further demonstrate the robustness of the sliding guess technique for RTP timestamp placement, we inject a special single-color frame (embedded with a QR code) into the middle of the test video, as shown in~\fig{fig:single_color_test}. We then examine whether the corresponding mapped IP packets exhibit certain characteristics—such as relatively small payload sizes due to the frame’s low information entropy. 

\fig{fig:wireshark_injection} presents the Wireshark analysis of the mapped IP packet batches for the single-color frame, along with those of the preceding and following frames. The single-color frame is characterized by relatively small packet sizes (654/653 bytes). Typically, base-layer frames (marked as "500000" in the "Extension Data" field) and enhancement-layer frames ("5ff777") alternate. However, for frames 5399, 5400, and 5401, all corresponding IP packets are base-layer, suggesting that the Zoom encoder detected a content discontinuity and designated frame 5400 as a base-layer, likely due to the absence of a usable delta. Additionally, the packet batch for frame 5401 is unusually large, comprising ten 1202/1201-byte packets, hinting at an I frame or a substantial delta region. These observations further reinforce the validity of our packet-to-frame mapping approach.

\end{appendices}
\begin{appendices}

\section{Zoom-Specific Details}
\label{s:zoomdetails}

\subsection{Zoom QoE Gain Model}
\label{s:zoomdetails:qoegainmodel}

The Zoom QoE gain model's pseudocode, used by the \systemname{} Controller, is specified in \cref{listing:qoe_model}. 

We model Zoom's current QoE as a weighted combination of per-subflow quality for base-layer video, enhancement-layer video, and audio, where base-layer video and audio are assigned higher importance than enhancement. Each component’s quality is computed from normalized frame rate and delay, and video enhancement-layer's quality is capped by the base-layer quality to reflect decoding dependency. Prioritization provides additional gain proportional to the remaining quality gap for each subflow, capturing diminishing returns when quality is already high.

\begin{figure}[!ht]
\centering
\begin{lstlisting}
# action = (p_base, p_enh, p_audio)
# three booleans, whether to prioritize 
# base layer, enh layer, and audio
# output is clipped to [0,1]

# clip(x) = max(min(x, 1), 0)

# constants used in the model:
# TARGET_BASE_FPS = 8
# TARGET_ENH_FPS  = 16
# MAX_VIDEO_DELAY = 80   # ms
# MAX_AUDIO_DELAY = 50   # ms
# W_BASE = 0.4
# W_ENH  = 0.2
# W_AUD  = 0.4
# BOOST  = 0.5

# input: a prioritization scheme, QoE metrics
# output: the predicted enhanced QoE score 
def compute_zoom_qoe_gain(action, base_fps, enh_fps,
                base_delay, enh_delay, audio_delay):

    # 1. Assess current FPS
    q_b_fps = clip(base_fps / TARGET_BASE_FPS)
    q_e_fps = clip(enh_fps / TARGET_ENH_FPS)
    # enhancement depends on base layer
    q_e_fps = min(q_e_fps, q_b_fps)

    # 2. Assess current delays
    q_b_delay = clip(1 - base_delay / MAX_VIDEO_DELAY)
    q_e_delay = clip(1 - enh_delay  / MAX_VIDEO_DELAY)
    q_audio   = clip(1 - audio_delay / MAX_AUDIO_DELAY)

    # 3. Intrinsic quality
    q_base = 0.7 * q_b_fps + 0.3 * q_b_delay
    q_enh  = 0.5 * q_e_fps + 0.5 * q_e_delay

    # 4. Baseline QoE
    Q0 = W_BASE * q_base + W_ENH * q_enh + W_AUD * q_audio

    # 5. Prioritization gain
    p_base, p_enh, p_audio = action
    gain = BOOST * (
        W_BASE  * p_base  * (1 - q_base) +
        W_ENH   * p_enh   * (1 - q_enh)  +
        W_AUD * p_audio * (1 - q_audio)
    )

    return clip(Q0 + gain)
\end{lstlisting}
\caption{Pseudocode for Zoom QoE gain computation model in \systemname{} Controller.}
\label{listing:qoe_model}
\end{figure}

\subsection{Zoom DPI Plugin}
\label{s:zoomdetails：zoomdpi}

\cref{listing:qoe_model} presents the DPI plugin used for Zoom subflow identification in both the \systemname{} Marking Module (\cref{s:design:data_plane}) and the \systemname{} Monitor (\cref{s:design:metric_tracking}). By parsing L7 fields, the plugin identifies packets as Zoom audio, video base layer, high-FPS enhancement layer, low-FPS enhancement layer, small-screen video, probe traffic, and other control data.

\begin{figure}[!ht]
\centering
\begin{lstlisting}
Base layer values:
    {50 00 00, 5f f0 00, 5f 0f ff, 5f 7f ff}

High-FPS enhancement layer values:
    {5f f7 77, 5f fa aa, 5f f5 55}

Low-FPS enhancement layer values:
    {57 77 77}

Algorithm: ZoomSubflowClassify(packet)

if packet is not IPv4 or not long enough:
    return NOT_ZOOM_MEDIA
if neither src/dst port is 8801 (not related to Zoom):
    return NOT_ZOOM_MEDIA

l7_payload = packet.payload
zoom_pkt_type = l7_payload[8]

if zoom_pkt_type == 0x0f:
    return AUDIO
if zoom_pkt_type == 0x10:
    bytes = RTP's first extension data
    if bytes match base layer values:
        return BASE
    if bytes match high-FPS enhancement layer values:
        return HIGH_FPS_ENHANCEMENT
    if bytes match low-FPS enhancement layer values:
        return LOW_FPS_ENHANCEMENT
    return SMALL_WINDOW_VIDEO
if zoom_pkt_type == 0x15 and l7_payload size > 1000:
        return PROBE

return CONTROL
\end{lstlisting}
\caption{Zoom subflow identification rules.}
\label{listing:zoomdpi}
\end{figure}

\end{appendices}

\begin{appendices}

\section{Fairness Impact Model}
\label{s:fairness_impact_model_impl}

\cref{listing:fairness_model} gives the fairness impact model details in our implementation tailored for Zoom. The model estimates how much a candidate prioritization decision $\mathbf{K}$ reduces the resources available to ordinary (non-interactive) traffic.

It first approximates the baseline resource allocation under the default scheduler by performing a simulated allocation based on current offered loads and channel quality. Bandwidth-hungry flows are modeled with unbounded demand to capture their steady-state behavior under congestion. From this allocation, the controller estimates $U(t)$, the PRBs that would serve ordinary traffic, and derives the spare capacity as $S(t) = C - U(t)$, where $C$ is the total PRB budget over the interval.

Given a prioritization decision $\mathbf{K}$, the model then computes $R(\mathbf{K}, t)$, the PRBs reserved for protected Zoom subflows. The fairness impact is quantified as the excess reservation beyond spare capacity, \ie $\max(0, R(\mathbf{K}, t) - S(t))$.

Finally, this excess is normalized by $U(t)$ to produce a fairness loss score $F(\mathbf{K}\mid t) \in [0,1]$, representing the fraction of ordinary traffic that would be displaced by prioritization.

\begin{figure}[!ht]
\centering
\begin{lstlisting}
# Input: decision K, UEs
# Output: F(K|t) in [0,1]

# TOTAL_PRB: PRBs per interval
# clip(x) = max(min(x,1),0)

def fairness_loss(K, ues):

    # 1. Default allocation simulation
    offered_loads = ues.offered_loads
    ran_states = ues.ran_states
    for ue in ues:
        if ue.bw_hungry:
            offered_loads[ue.id] = INF

    # simulate the default scheduler
    PRBs = alloc_sim(offered_loads, ran_states, TOTAL_PRB)

    # 2. Compute slack PRB capacity
    U = 0
    for ue in ues:
        # compute the non-zoom loads' PRBs
        U += PRBs[ue.id] * (ue.nonzoom_loads / ue.loads)
    slack = TOTAL_PRB - U

    # 3. Reserved PRBs under K
    R = 0
    for ue in ues:
        # whether to prioritize base, enh, audio
        p_base, p_enh, p_audio = K[ue.id]
        load = 0
        if p_base:
            load += ue.zoom_base_load
        if p_enh:
            load += ue.zoom_enh_load
        if p_audio:
            load += ue.zoom_audio_load
        # simulate the high-priority allocation
        R += high_priority_alloc_sim(load, ue.ran_state)
        
    # 4. Excess and normalize
    excess = max(0, R - slack)
    F = excess / max(1, U)
    return clip(F)
\end{lstlisting}
\caption{Pseudocode for fairness impact modeling in \systemname{} Controller, tailored for Zoom.}
\label{listing:fairness_model}
\end{figure}

\end{appendices}
\begin{appendices}

\section{Optimization Formulation}
\label{app:controller_optimization}

The prioritization decision in \systemname{} is formulated as a multiple-choice binary integer optimization problem. The “multiple-choice” structure arises because the solver selects exactly one prioritization action from a candidate set for each Zoom UE, while “binary integer” refers to representing each action selection using binary variables. The QoE and resource usage associated with each action are precomputed from the current application and RAN states and incorporated into the optimization as parameters and constraints. We solve this problem efficiently using Google OR-Tools CP-SAT, a constraint-based integer solver~\cite{google_ortools}. At each decision interval, the controller applies the prioritization actions returned by the solver.

\paragraph{Decision Variables.}
Let $\mathcal{Z}$ denote the set of Zoom UEs (\ie Zoom calls). Let $\mathcal{A}$ denote the set of candidate prioritization actions. We define binary decision variables
\[
x_{i,j} \in \{0,1\}, \quad \forall i \in \mathcal{Z},\; j \in \mathcal{A},
\]
where $x_{i,j} = 1$ indicates that action $j$ is selected for Zoom call $i$.

\paragraph{Precomputed Quantities.}
For each Zoom UE $i$ and action $j$, we precompute:
\begin{itemize}
    \item $q_{i,j} \in [0,100]$: the QoE score achieved under action $j$ (explained in \S\ref{s:design:controller:qoe_model}).
    \item $r_{i,j} \in \mathbb{Z}_{\ge 0}$: the PRB demand for prioritization (resource usage) under action $j$ (via straightforward simulation).
\end{itemize}

We also define (per decision interval):
\begin{itemize}
    \item $C$: total PRB capacity.
    \item $S$: estimated spare PRB capacity after serving non-Zoom traffic.
    \item $U$: estimated PRB share attributable to non-Zoom traffic (via simulation, explained in \S\ref{s:design:controller:fair_model}).
    \item $L_{\max} = \max(1, U)$: normalization factor for fairness loss.
    \item $\alpha \in [0,1]$: tradeoff between worst-user QoE and average QoE.
    \item $\beta \in [0,1]$: tradeoff between QoE and fairness.
\end{itemize}

\paragraph{Constraints.}

Each Zoom UE selects exactly one action:
\begin{equation}
\sum_{j \in \mathcal{A}} x_{i,j} = 1, \quad \forall i \in \mathcal{Z}.
\end{equation}

The QoE of each Zoom UE is:
\begin{equation}
s_i = \sum_{j \in \mathcal{A}} q_{i,j} x_{i,j}, \quad \forall i \in \mathcal{Z}.
\end{equation}

Relationship between $C$, $S$, and $U$:
\begin{equation}
S = C - U.
\end{equation}

Define the minimum QoE among interactive UEs:
\begin{equation}
z \le s_i, \quad \forall i \in \mathcal{Z}.
\end{equation}

Total reserved PRBs:
\begin{equation}
R = \sum_{i \in \mathcal{Z}} \sum_{j \in \mathcal{A}} r_{i,j} x_{i,j}.
\end{equation}

Excess PRBs beyond available slack:
\begin{equation}
e = \max(0, R - S).
\end{equation}

We enforce a hard safety constraint on excess resource usage:
\begin{equation}
e \le (1 - \beta)\, U.
\end{equation}

Normalized fairness loss:
\begin{equation}
\ell = \frac{e}{L_{\max}}, \quad f = 1 - \ell.
\end{equation}

Average QoE over interactive UEs:
\begin{equation}
\bar{q} = \frac{1}{|\mathcal{Z}|} \sum_{i \in \mathcal{Z}} s_i.
\end{equation}

\paragraph{Objective.}

We define a composite QoE term:
\begin{equation}
Q_{\text{comp}} = \alpha \cdot z + (1 - \alpha)\cdot \bar{q}.
\end{equation}

The \systemname{} controller solves:
\begin{equation}
\max_{x_{i,j}} \quad (1 - \beta)\, Q_{\text{comp}} + \beta\, f.
\end{equation}

\end{appendices}
\begin{appendices}

\section{Implementation Line Count}
\label{s:line_of_codes}

\cref{tab:loc} shows the effective lines of code for each \systemname{} component in our implementation.

\begin{table}
\centering
\caption{\systemname{} Implementation Line Count}
\label{tab:loc}
\begin{tabular}{l c}
\toprule
\textbf{Component} & \textbf{Effective Lines of Code} \\
\midrule
Monitor &  $\approx 4100$\\
Modified MAC Scheduler & $\approx 800$\\
Controller &  $\approx 1500$\\
Marking Module & $\approx 600$\\
Open5GS Extension & $\approx 100$\\
\midrule
Total &  $\approx 7100$\\
\bottomrule
\end{tabular}
\end{table}

\end{appendices}
\begin{appendices}

\section{Evaluation Details}
\label{s:eval_qoe_details}

\subsection{QoE Score Model Details in Evaluation}
\label{s:eval_qoe_details:qoe_model_details}

We construct a composite Zoom QoE score in the range $[0,100]$ to capture latency, temporal stability, frame-rate smoothness, and spatial fidelity. 
The total score is the sum of four equally weighted components:
\begin{equation}
\mathrm{QoE} = S_{\mathrm{audio}} + S_{\mathrm{video}} + S_{\mathrm{fps}} + S_{\mathrm{res}},
\end{equation}
where each component contributes up to 25 points.

\textbf{Audio and Video Delay.}
For both audio and video streams, we incorporate mean one-way delay and mean RTP interarrival jitter to reflect both latency and delay variability. 
For stream $x \in \{\mathrm{audio}, \mathrm{video}\}$, we define the effective delay as
\begin{equation}
D^{(x)}_{\mathrm{eff}} = \mu^{(x)}_{D} + c \, \mu^{(x)}_{J},
\end{equation}
where $\mu^{(x)}_{D}$ is the session-mean one-way delay, $\mu^{(x)}_{J}$ is the session-mean RTP jitter computed following RFC~3550 \cite{rfc3550}, and $c$ is a dimensionless scaling factor. 
In our evaluation, we set $c=1$, assigning jitter the same perceptual weight as additional delay, reflecting that delay instability degrades playout quality similarly to increased latency.

The delay component score is computed as
\begin{equation}
S_x = 25 \cdot \min\!\left(1,\left(\frac{D^{(x)}_{\mathrm{th}}}{D^{(x)}_{\mathrm{eff}}}\right)^k\right),
\end{equation}
where $k=0.5$ controls the smoothness of degradation beyond the acceptable threshold. 
We set the audio delay threshold to $D^{(\mathrm{audio})}_{\mathrm{th}} = 150$\,ms, consistent with ITU-T G.114 \cite{itu_g114}, which considers one-way delays below 150\,ms good for interactive voice communication. 
For video, we set $D^{(\mathrm{video})}_{\mathrm{th}} = 400$\,ms, aligning with the upper bound of ideal interactive latency for real-time conferencing systems recommended by WebRTC \cite{antmedia_webrtc_benefits_2025, nanocosmos_webrtc_latency_2025}.

\textbf{Frame Rate.}
The frame-rate component captures temporal smoothness and stability of playback. 
We define FPS as the number of distinct rendered frames displayed on screen per second, computed using a sliding 1-second window with a 0.3-second step across the session. Over a session, let $\mu_{\mathrm{fps}}$ and $\sigma_{\mathrm{fps}}$ denote the mean and standard deviation of FPS, respectively. 
The frame-rate score is computed as
\begin{equation}
S_{\mathrm{fps}} = 25 \cdot 
\min\!\left(1,\left(\frac{\mu_{\mathrm{fps}}}{F_{\mathrm{th}}}\right)^\gamma \right)
\cdot 
\max\!\left(0, 1 - \kappa \frac{\sigma_{\mathrm{fps}}}{\mu_{\mathrm{fps}}} \right),
\end{equation}
where $F_{\mathrm{th}} = 28$\,FPS is the target threshold approximating near real-time conferencing performance (close to the standard 30\,FPS target, while accounting for practical encoding and adaptation effects), $\gamma=1$ provides linear scaling, and $\kappa=0.5$ moderately penalizes temporal fluctuation via the coefficient of variation.

\textbf{Resolution.}
To reflect perceived spatial quality while properly penalizing playback stalls, 
we first explicitly detect frozen video frames and exclude their durations from 
resolution credit accumulation.

Let $T_i$ denote the rendering timestamp of frame $i$, and define the frame 
duration as
\begin{equation}
\Delta_i = T_{i + 1} - T_i.
\end{equation}
A rendered frame $i$ is classified as \emph{frozen} if
\begin{equation}
\Delta_i \ge 
\max\!\left(3 \cdot \overline{\Delta}_{30}, \; 
\overline{\Delta}_{30} + 150\,\mathrm{ms}\right),
\end{equation}
where $\overline{\Delta}_{30}$ is the linear average of the frame durations 
over the preceding 30 rendered frames. This definition follows the WebRTC 
statistics specification \cite{w3c_webrtc_stats} and captures abnormal rendering gaps relative to 
recent playback dynamics.

Let $T_{\mathrm{total}}$ denote the total playback duration of the session, 
and let $T_r$ denote the cumulative duration of \emph{non-frozen} frames 
rendered at resolution $r$. Frozen intervals do not contribute to any 
resolution layer. 

We consider three observed spatial layers in Zoom: 640p, 480p, and 320p, 
with assigned perceptual weights $w_{640}=1.0$, $w_{480}=0.6$, and $w_{320}=0.3$, 
reflecting their relative visual fidelity. 
The normalized resolution quality factor is computed as
\begin{equation}
Q_{\mathrm{res}} =
\sum_{r \in \{640,480,320\}}
w_r \cdot \frac{T_r}{T_{\mathrm{total}}},
\end{equation}
and the final resolution score is
\begin{equation}
S_{\mathrm{res}} = 25 \cdot Q_{\mathrm{res}}.
\end{equation}

Under this formulation, frozen durations naturally reduce the resolution 
score since they consume session time without contributing to any spatial 
quality layer, thereby accurately reflecting degraded user experience.

\subsection{Evaluation Result Details}
\label{s:eval_qoe_details:eval_details}

Here, we present detailed evaluation results for the experiments described in~\S\ref{s:eval:setup}. Specifically, we report the relationship between camera-to-screen video frame delay and screen-measured FPS, the distribution of video resolution, and the QoE score breakdown for each UE in each experiment.

\begin{figure*}
    \centering
         \subfigure[Zoom FPS vs screen-to-camera delay.\label{fig:single_dl_fps_delay}]
         {\includegraphics[width=0.33\linewidth]{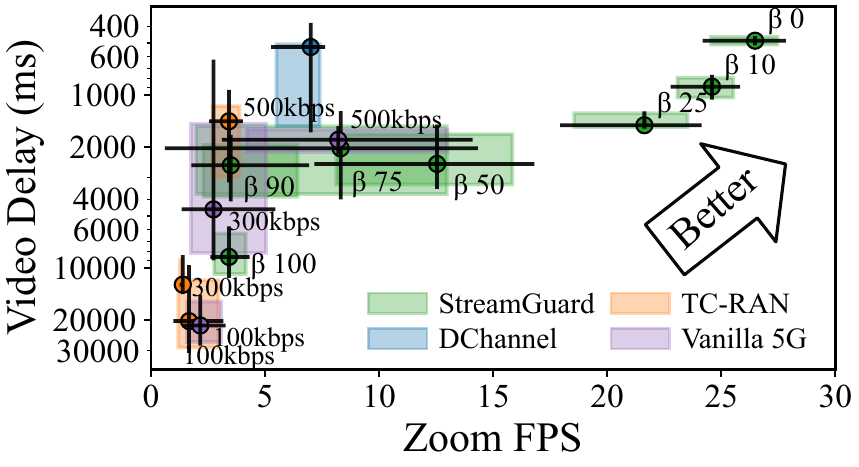}}
         \subfigure[Zoom resolution distributions.\label{fig:single_dl_res}]
         {\includegraphics[width=0.33\linewidth]{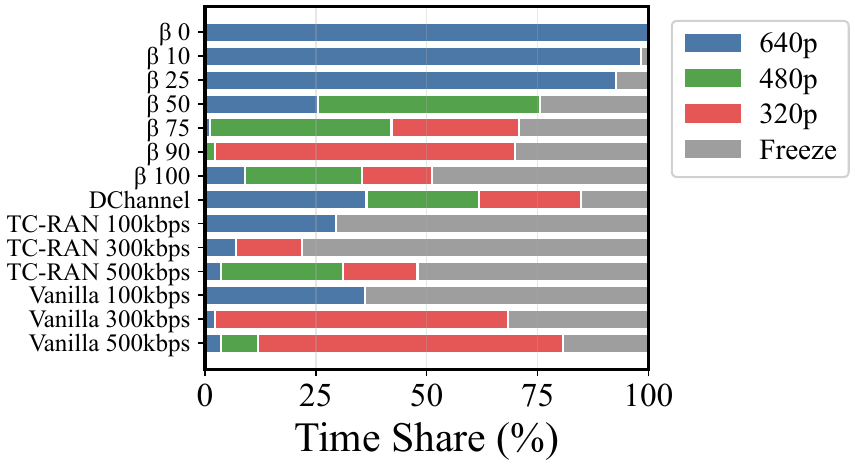}}
         \subfigure[Zoom QoE score breakdowns.\label{fig:single_dl_qoe_breakdown}]
         {\includegraphics[width=0.33\linewidth]{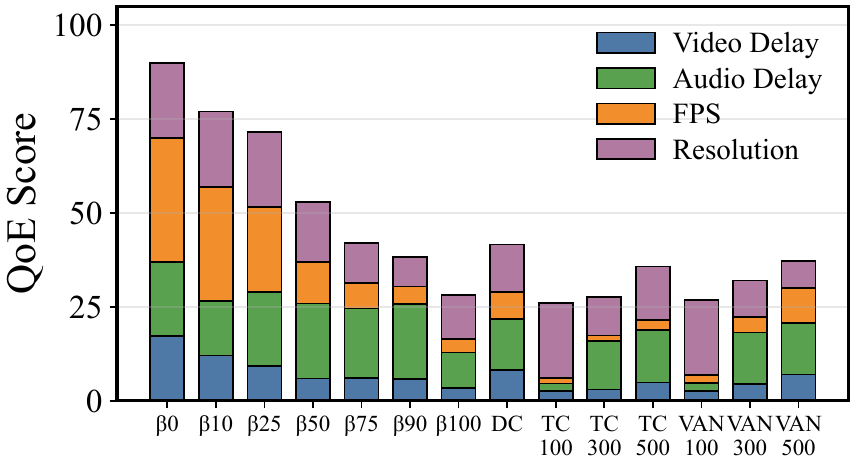}}
        \caption{Detailed evaluation results of downlink single Zoom call experiment.}
        \label{fig:single_dl}
\end{figure*}

\noindent\circnum{1} \textbf{Single Downlink (Receive-Only) Zoom Call.}
We show the video delay–FPS relationship in~\cref{fig:single_dl_fps_delay}, the resolution distribution in~\cref{fig:single_dl_res}, and the QoE score breakdown in~\cref{fig:single_dl_qoe_breakdown}.

\begin{figure*}
    \centering
         \subfigure[$\beta 10$, UE1]
         {\includegraphics[width=0.33\linewidth]{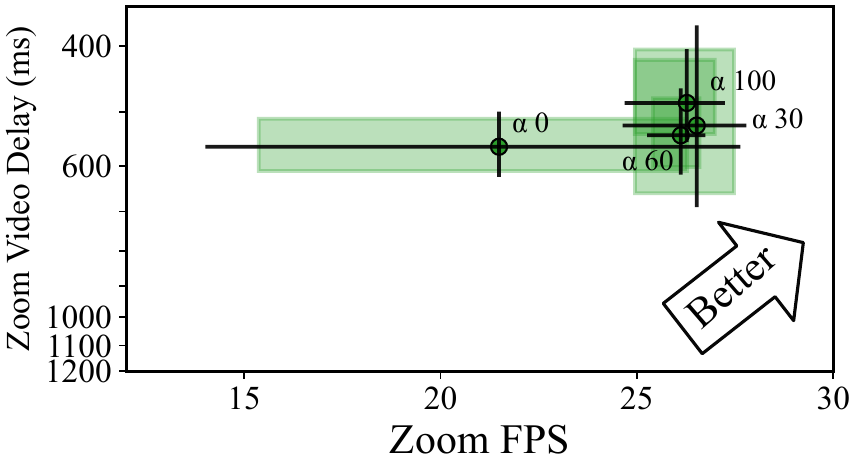}}\label{fig:multi_dl_fps_delay_b10_ue1}
         \subfigure[$\beta 10$, UE2]
         {\includegraphics[width=0.33\linewidth]{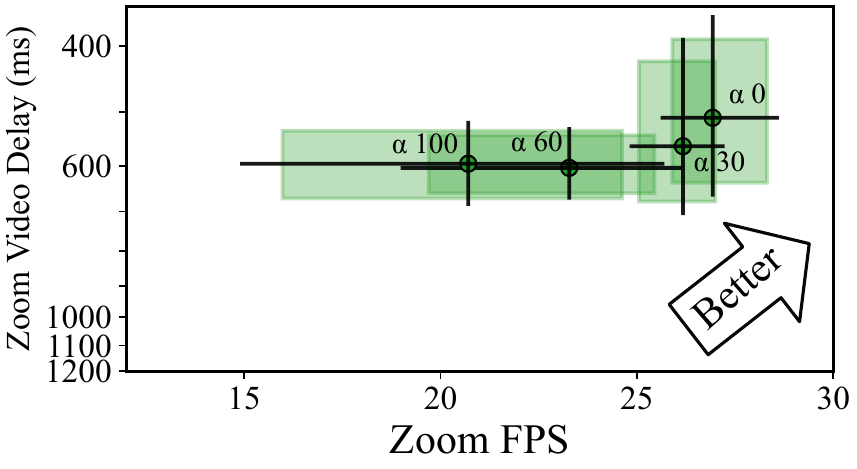}}\label{fig:multi_dl_fps_delay_b10_ue2}
         \subfigure[$\beta 10$, UE3]
         {\includegraphics[width=0.33\linewidth]{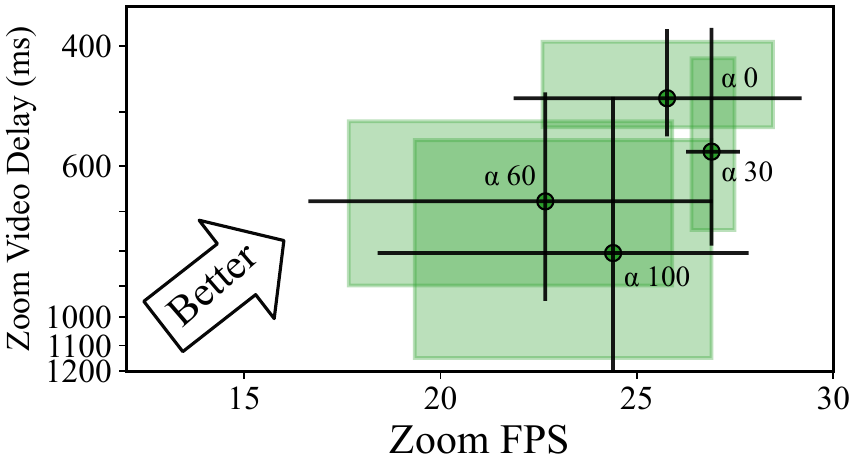}}\label{fig:multi_dl_fps_delay_b10_ue3}
         \subfigure[$\beta 50$, UE1]
         {\includegraphics[width=0.33\linewidth]{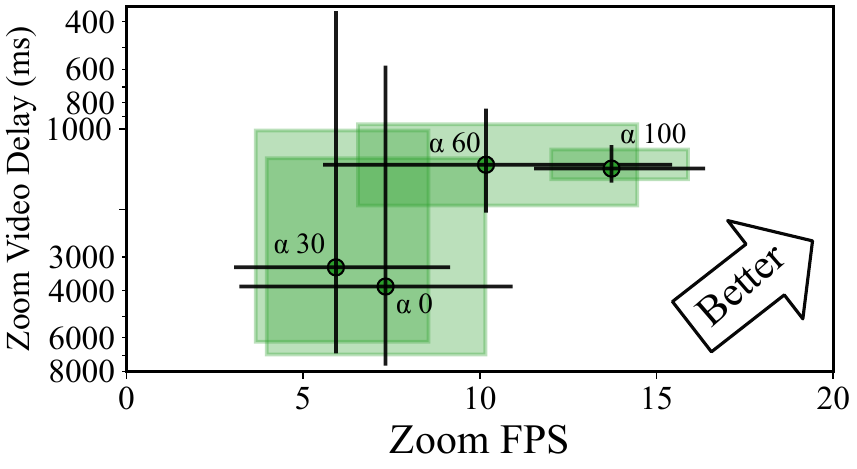}}\label{fig:multi_dl_fps_delay_b50_ue1}
         \subfigure[$\beta 50$, UE2]
         {\includegraphics[width=0.33\linewidth]{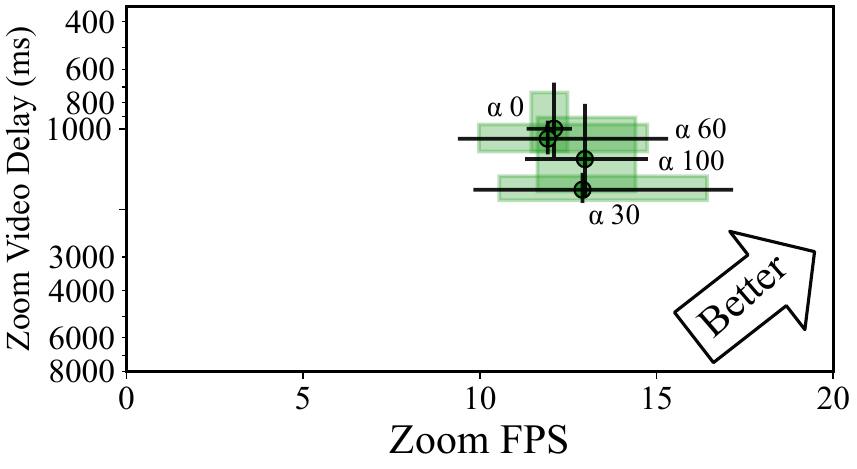}}\label{fig:multi_dl_fps_delay_b50_ue2}
         \subfigure[$\beta 50$, UE3]
         {\includegraphics[width=0.33\linewidth]{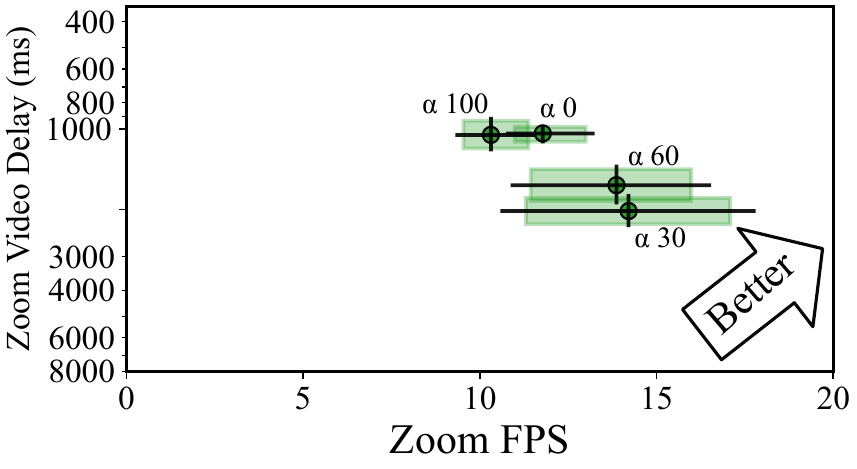}}\label{fig:multi_dl_fps_delay_b50_ue3}
         \subfigure[$\beta 90$, UE1]
         {\includegraphics[width=0.33\linewidth]{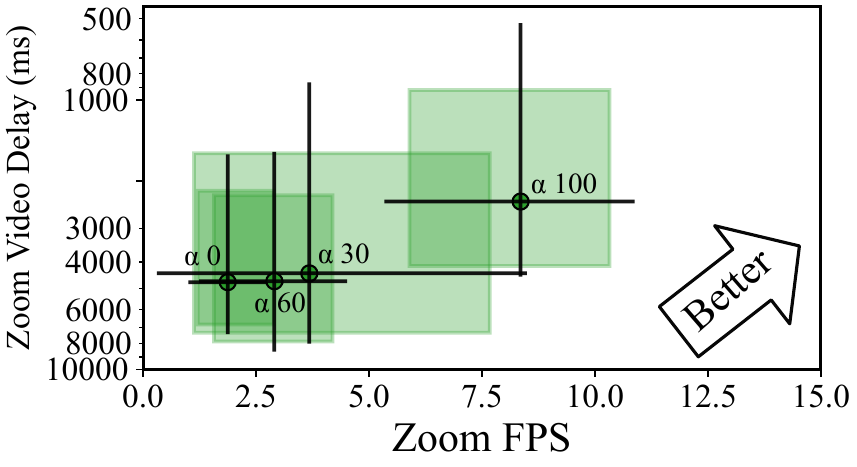}}\label{fig:multi_dl_fps_delay_b90_ue1}
         \subfigure[$\beta 90$, UE2]
         {\includegraphics[width=0.33\linewidth]{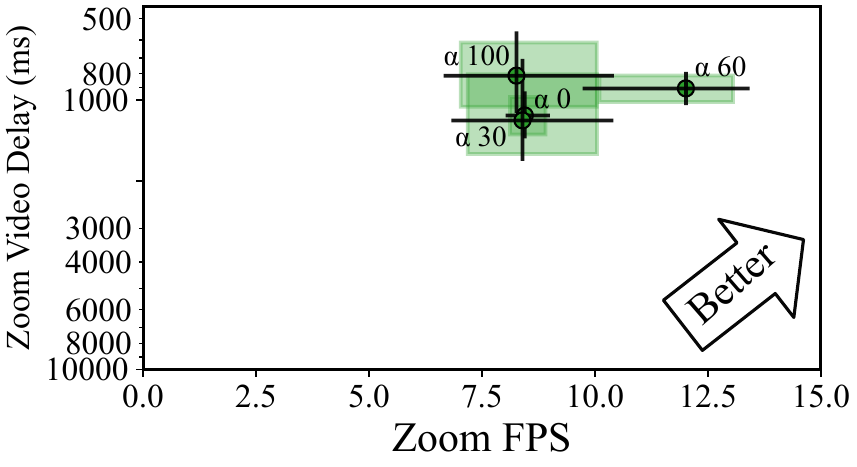}}\label{fig:multi_dl_fps_delay_b90_ue2}
         \subfigure[$\beta 90$, UE3]
         {\includegraphics[width=0.33\linewidth]{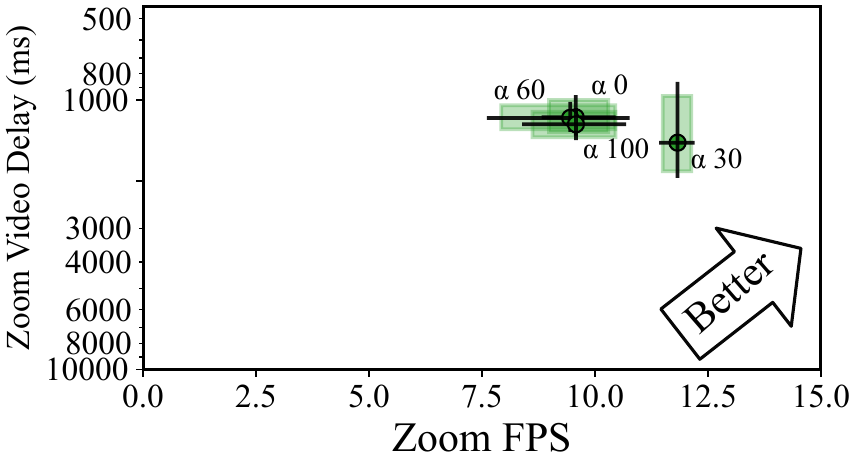}}\label{fig:multi_dl_fps_delay_b90_ue3}
         \subfigure[Optimal $\beta,\alpha$ and baseline, UE1]
         {\includegraphics[width=0.33\linewidth]{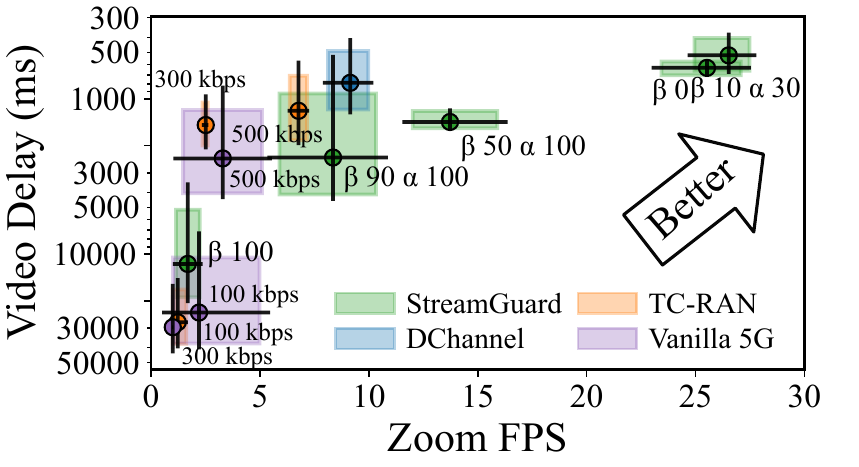}}\label{fig:multi_dl_fps_delay_all_ue1}
         \subfigure[Optimal $\beta,\alpha$ and baseline, UE2]
         {\includegraphics[width=0.33\linewidth]{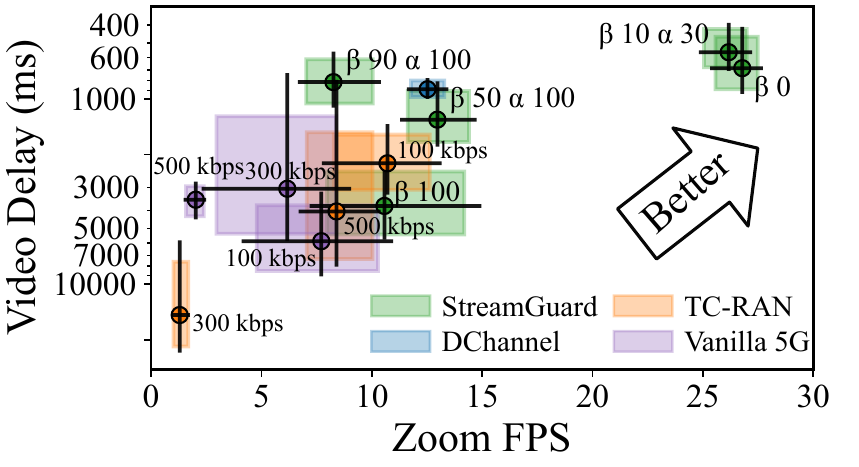}}\label{fig:multi_dl_fps_delay_all_ue2}
         \subfigure[Optimal $\beta,\alpha$ and baseline, UE3]
         {\includegraphics[width=0.33\linewidth]{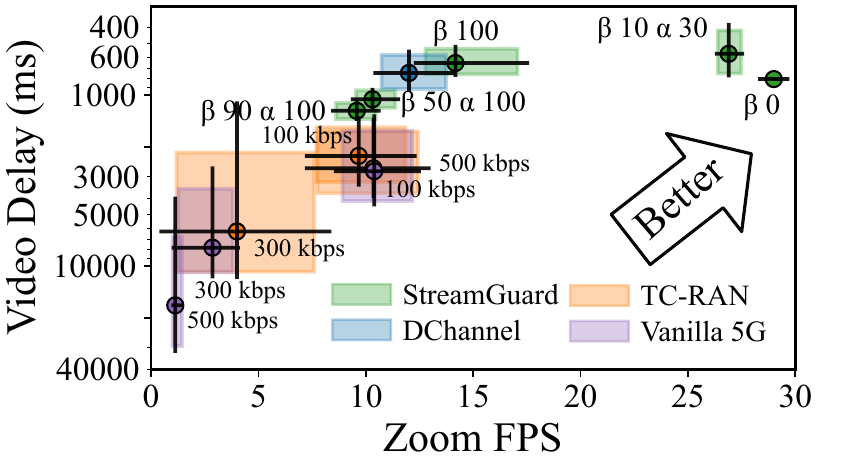}}\label{fig:multi_dl_fps_delay_all_ue3}
        \caption{FPS vs screen-to-camera delay for downlink multi Zoom call experiment.}
        \label{fig:multi_dl_fps_delay}
\end{figure*}

\begin{figure*}
    \centering
         \subfigure[$\beta 10$, UE1]
         {\includegraphics[width=0.33\linewidth]{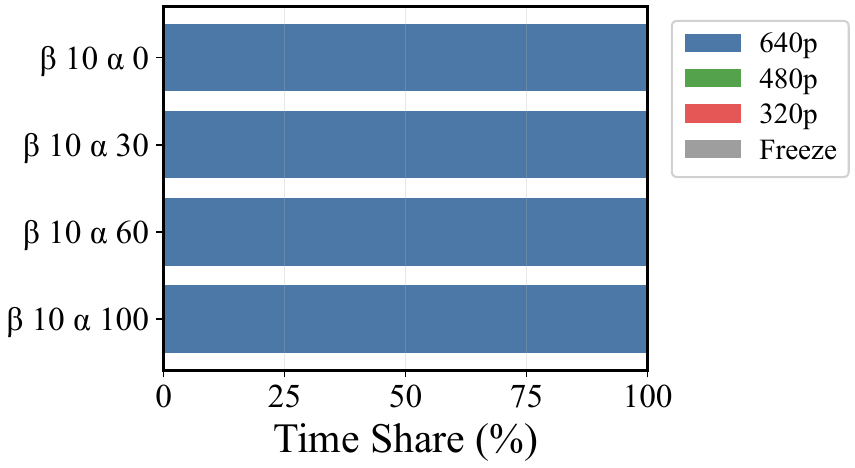}}\label{fig:res_b10_ue1}
         \subfigure[$\beta 10$, UE2]
         {\includegraphics[width=0.33\linewidth]{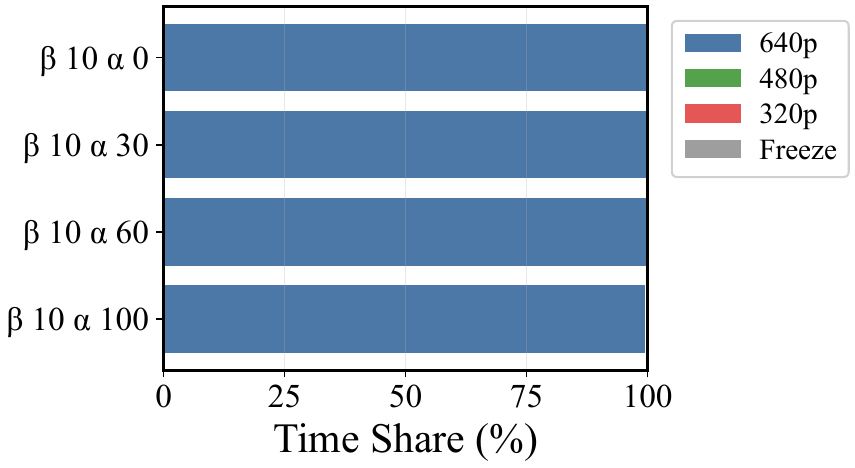}}\label{fig:res_b10_ue2}
         \subfigure[$\beta 10$, UE3]
         {\includegraphics[width=0.33\linewidth]{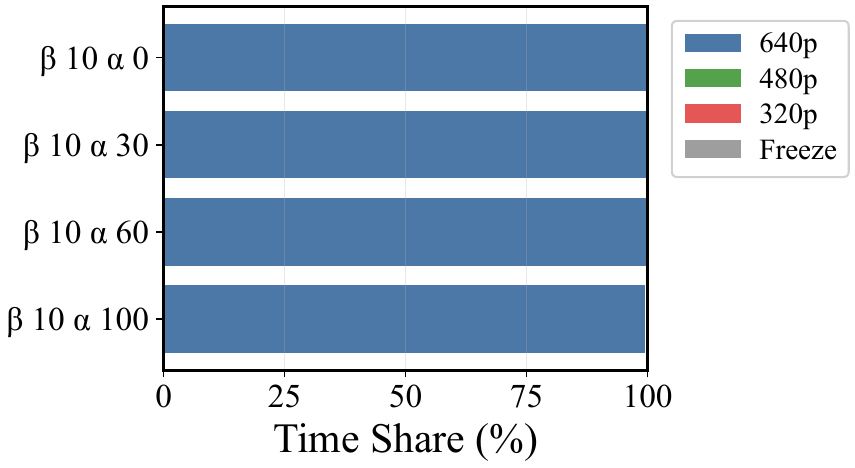}}\label{fig:res_b10_ue3}
         \subfigure[$\beta 50$, UE1]
         {\includegraphics[width=0.33\linewidth]{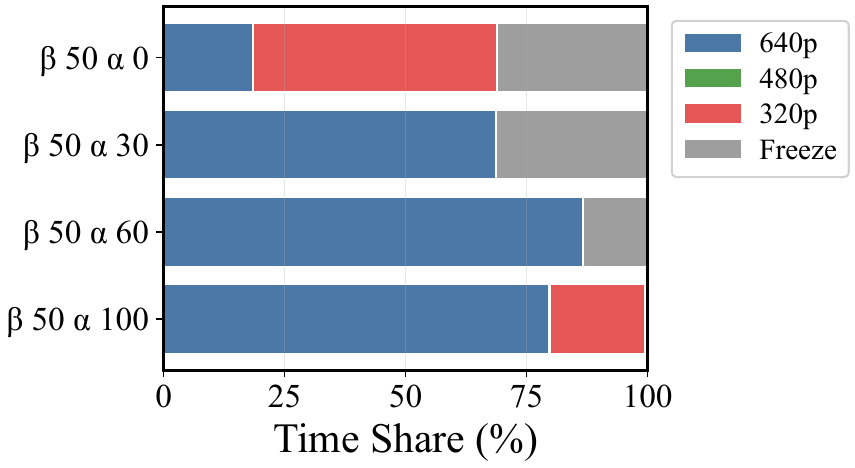}}\label{fig:res_b50_ue1}
         \subfigure[$\beta 50$, UE2]
         {\includegraphics[width=0.33\linewidth]{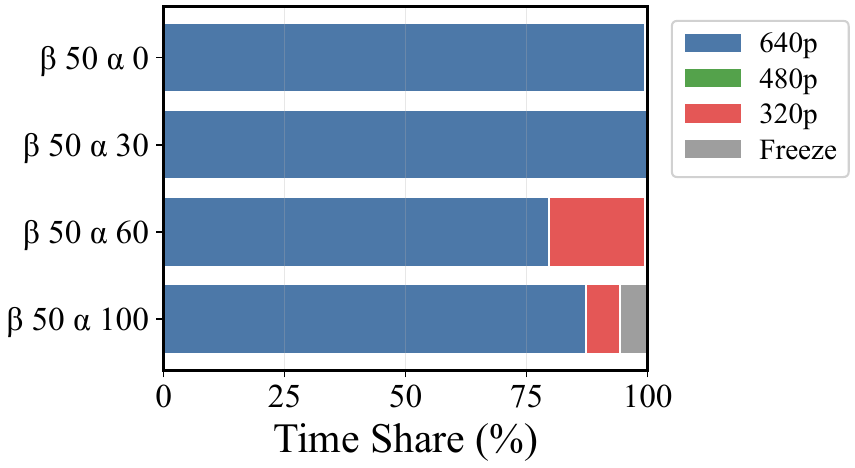}}\label{fig:res_b50_ue2}
         \subfigure[$\beta 50$, UE3]
         {\includegraphics[width=0.33\linewidth]{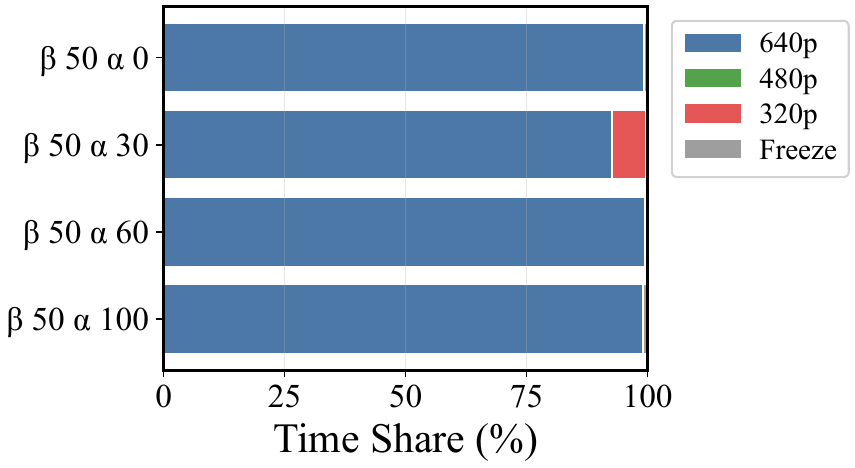}}\label{fig:res_b50_ue3}
         \subfigure[$\beta 90$, UE1]
         {\includegraphics[width=0.33\linewidth]{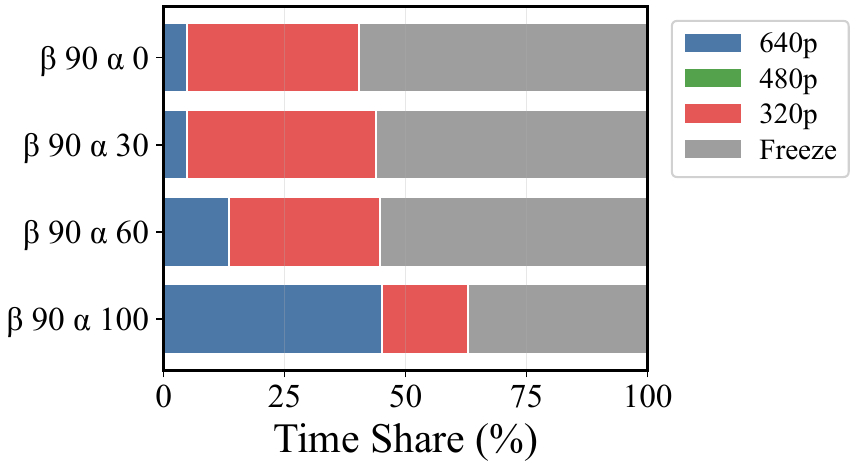}}\label{fig:res_b90_ue1}
         \subfigure[$\beta 90$, UE2]
         {\includegraphics[width=0.33\linewidth]{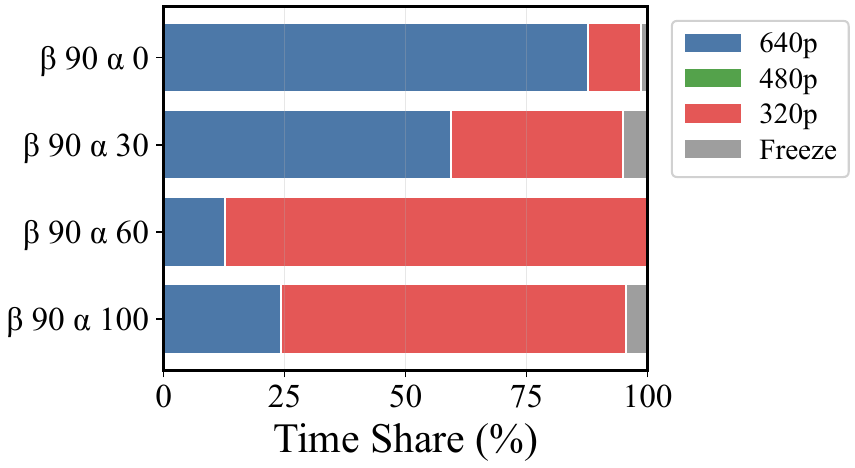}}\label{fig:res_b90_ue2}
         \subfigure[$\beta 90$, UE3]
         {\includegraphics[width=0.33\linewidth]{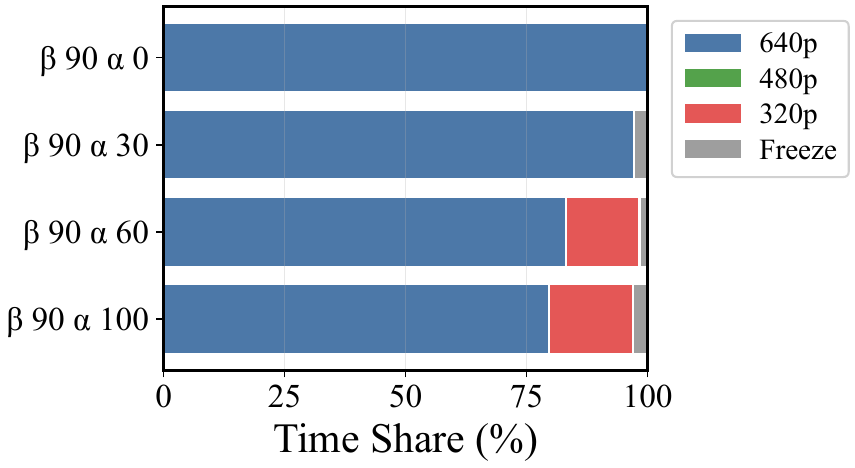}}\label{fig:res_b90_ue3}
         \subfigure[Optimal $\beta,\alpha$ and baseline, UE1]
         {\includegraphics[width=0.33\linewidth]{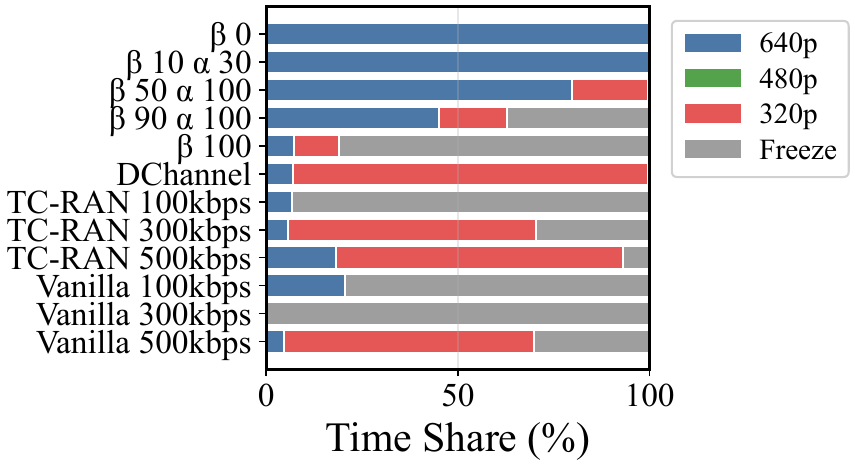}}\label{fig:res_all_ue1}
         \subfigure[Optimal $\beta,\alpha$ and baseline, UE2]
         {\includegraphics[width=0.33\linewidth]{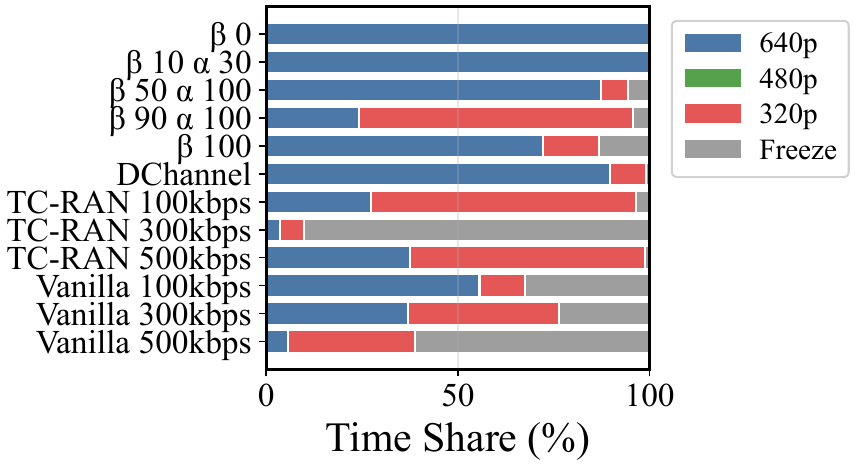}}\label{fig:res_all_ue2}
         \subfigure[Optimal $\beta,\alpha$ and baseline, UE3]
         {\includegraphics[width=0.33\linewidth]{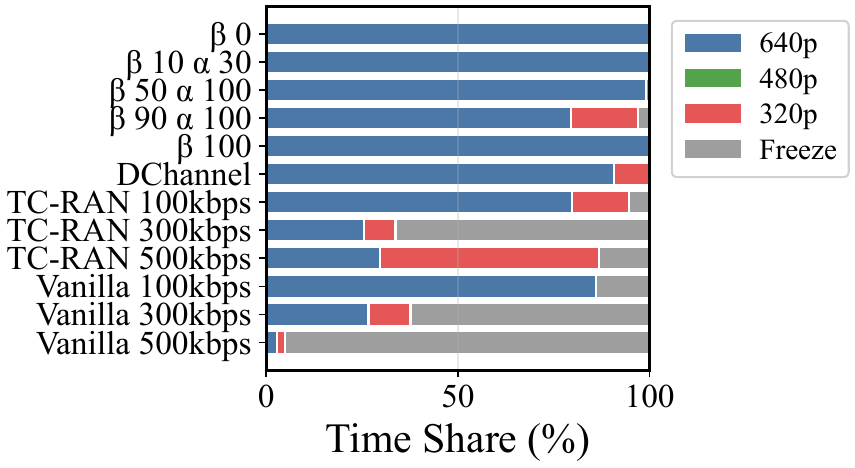}}\label{fig:res_all_ue3}
        \caption{Resolution distributions for downlink multi Zoom call experiment.}
        \label{fig:multi_dl_res}
\end{figure*}

\begin{figure*}
    \centering
         \subfigure[$\beta 10$]
         {\includegraphics[width=0.49\linewidth]{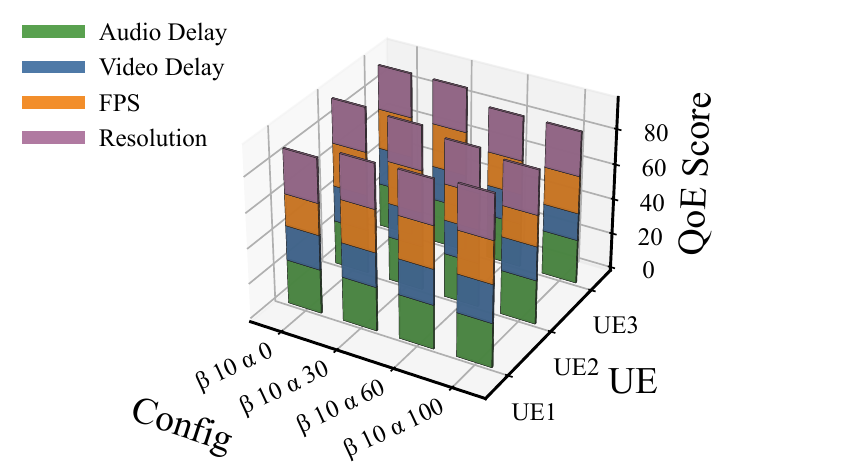}}\label{fig:qoe_breakdown_b10}
         \subfigure[$\beta 50$]
         {\includegraphics[width=0.49\linewidth]{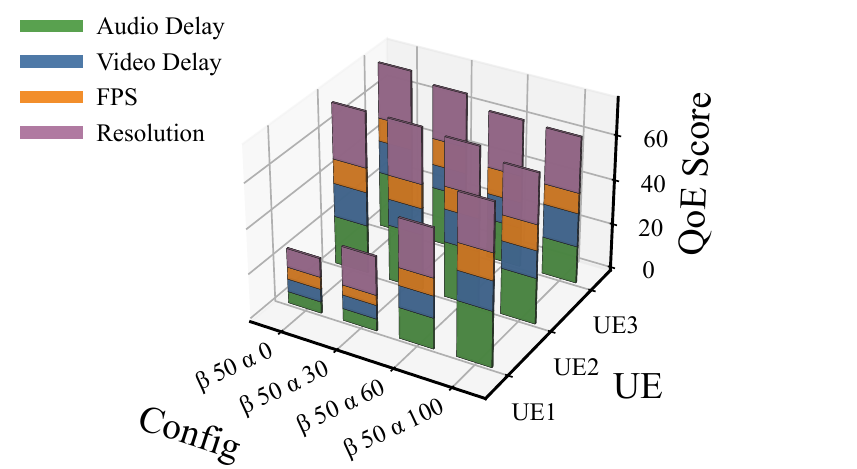}}\label{fig:qoe_breakdown_b50}
         \subfigure[$\beta 90$]
         {\includegraphics[width=0.49\linewidth]{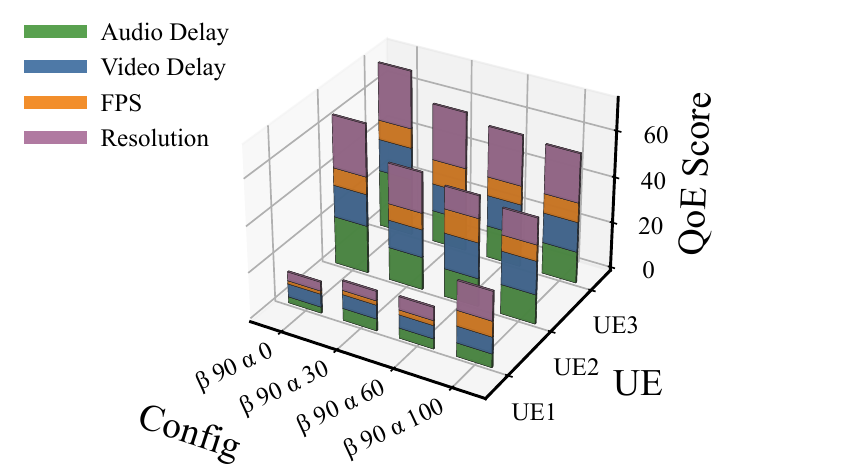}}\label{fig:qoe_breakdown_b90}
         \subfigure[Optimal $\beta, \alpha$ and baseline]
         {\includegraphics[width=0.49\linewidth]{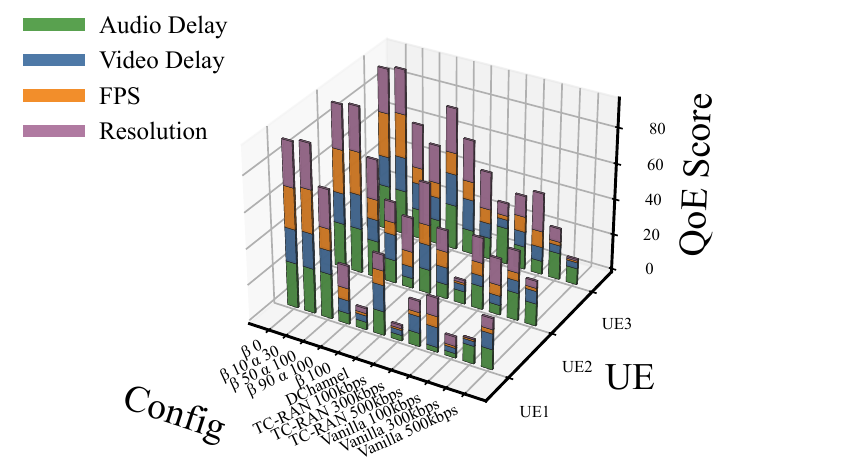}}\label{fig:qoe_breakdown_all}    
        \caption{QoE score breakdowns for downlink multi Zoom call experiment.}
        \label{fig:multi_dl_qoe_breakdown}
\end{figure*}

\noindent\circnum{2} \textbf{Three Downlink Zoom Calls.}
We show the video delay–FPS relationship in~\cref{fig:multi_dl_fps_delay}, the resolution distribution in~\cref{fig:multi_dl_res}, and the QoE score breakdown in~\cref{fig:multi_dl_qoe_breakdown}.

\begin{figure*}
    \centering
         \subfigure[Zoom FPS vs screen-to-camera delay.\label{fig:single_ul_fps_delay}]
         {\includegraphics[width=0.33\linewidth]{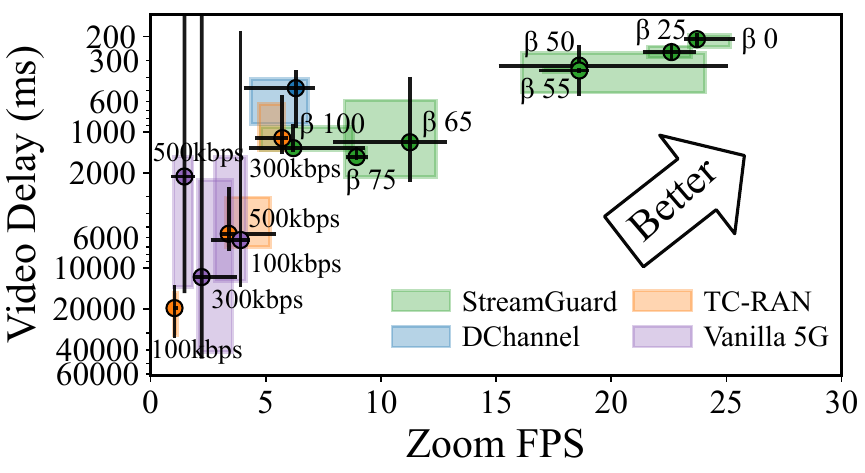}}
         \subfigure[Zoom resolution distributions.\label{fig:single_ul_res}]
         {\includegraphics[width=0.33\linewidth]{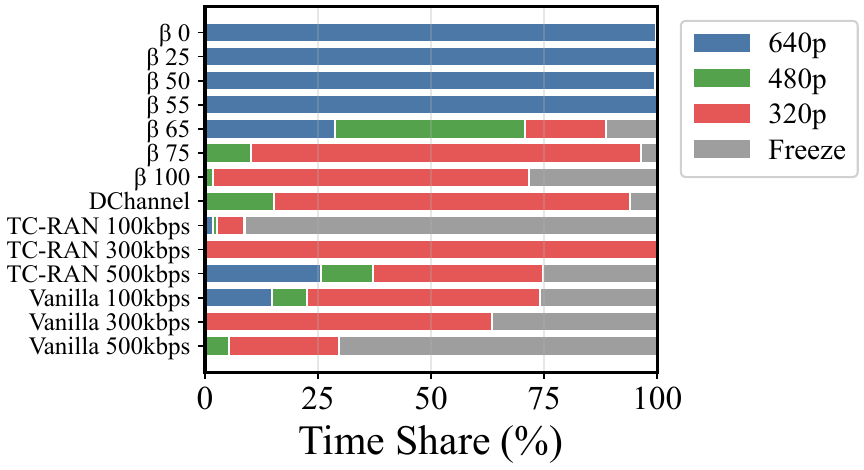}}
         \subfigure[Zoom QoE score breakdowns.\label{fig:single_ul_qoe_breakdown}]
         {\includegraphics[width=0.33\linewidth]{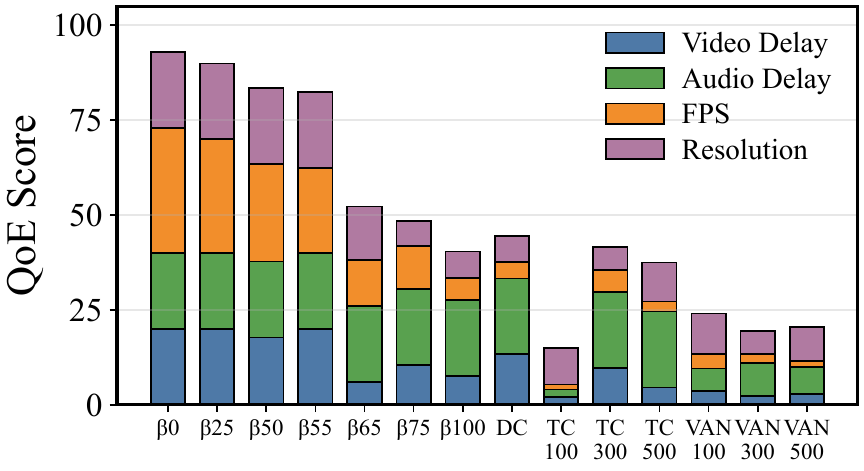}}
        \caption{Detailed evaluation results of uplink single Zoom call experiment.}
        \label{fig:single_ul}
\end{figure*}

\noindent\circnum{3} \textbf{Single Uplink (Send-Only) Zoom Call.}
We show the video delay–FPS relationship in~\cref{fig:single_ul_fps_delay}, the resolution distribution in~\cref{fig:single_ul_res}, and the QoE score breakdown in~\cref{fig:single_ul_qoe_breakdown}.

\begin{figure*}
    \centering
         \subfigure[$\beta 10$, UE1]
         {\includegraphics[width=0.33\linewidth]{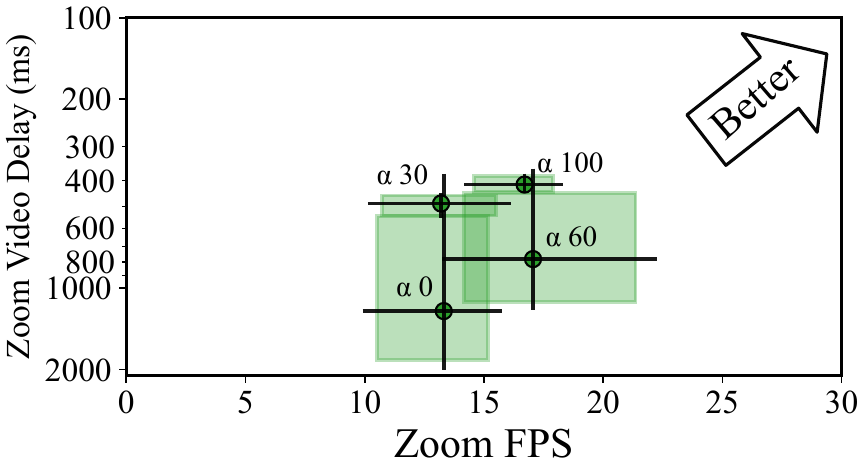}}\label{fig:multi_ul_fps_delay_b10_ue1}
         \subfigure[$\beta 10$, UE2]
         {\includegraphics[width=0.33\linewidth]{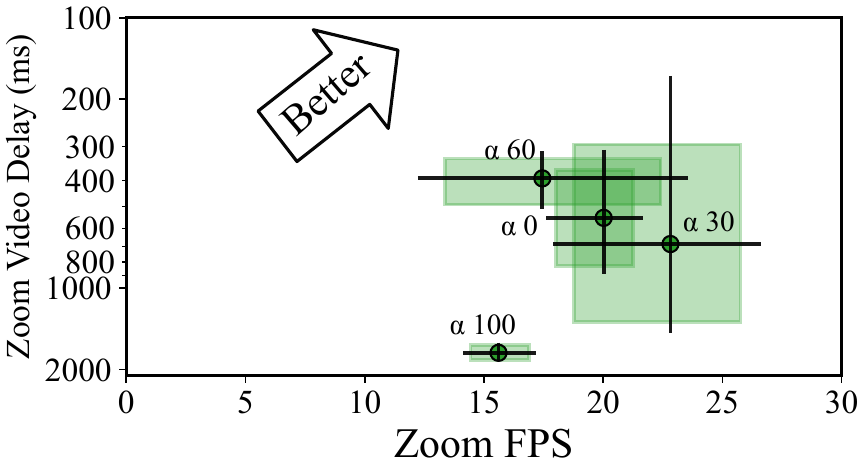}}\label{fig:multi_ul_fps_delay_b10_ue2}
         \subfigure[$\beta 10$, UE3]
         {\includegraphics[width=0.33\linewidth]{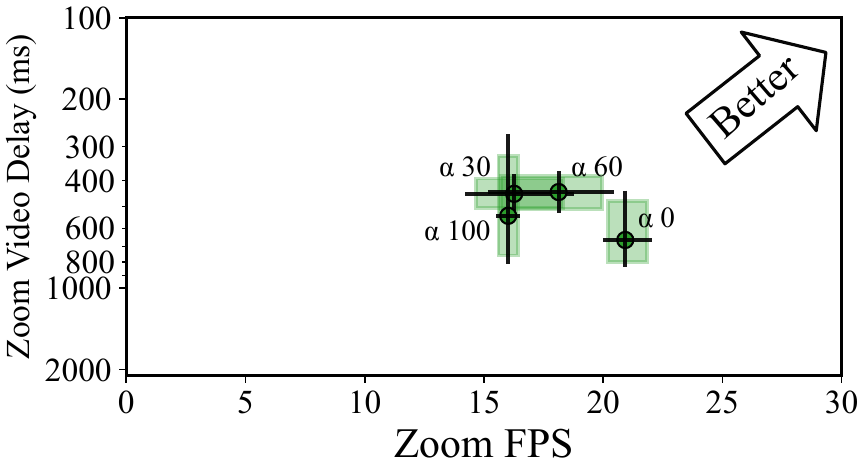}}\label{fig:multi_ul_fps_delay_b10_ue3}
         \subfigure[$\beta 50$, UE1]
         {\includegraphics[width=0.33\linewidth]{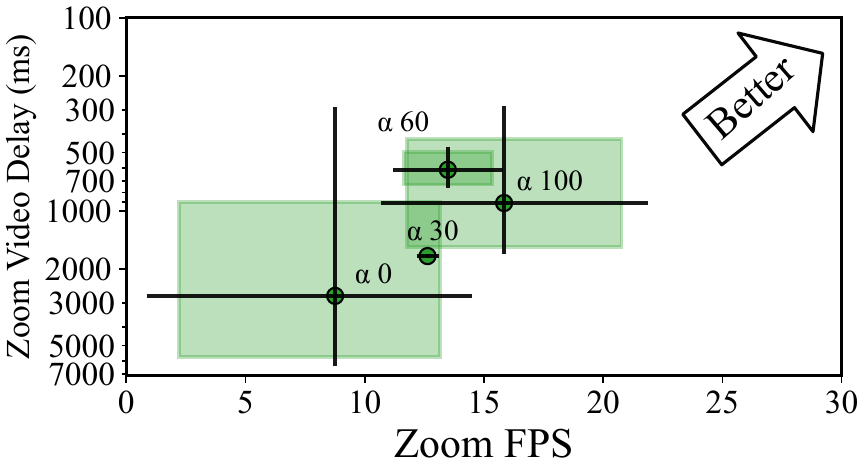}}\label{fig:multi_ul_fps_delay_b50_ue1}
         \subfigure[$\beta 50$, UE2]
         {\includegraphics[width=0.33\linewidth]{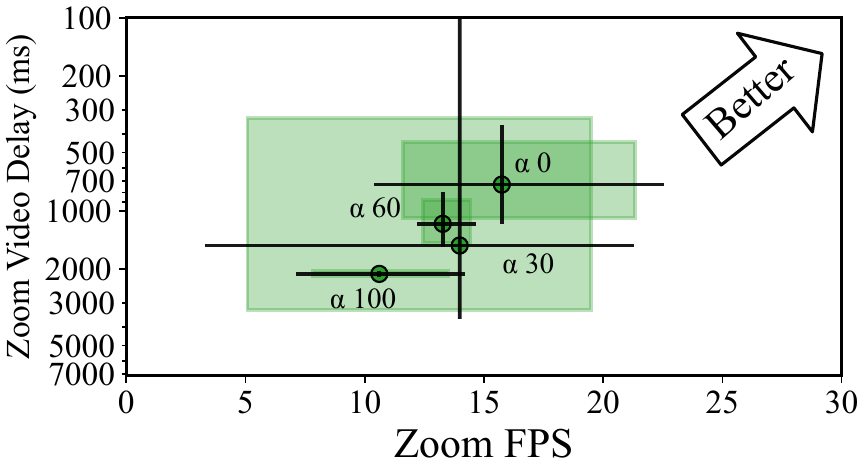}}\label{fig:multi_ul_fps_delay_b50_ue2}
         \subfigure[$\beta 50$, UE3]
         {\includegraphics[width=0.33\linewidth]{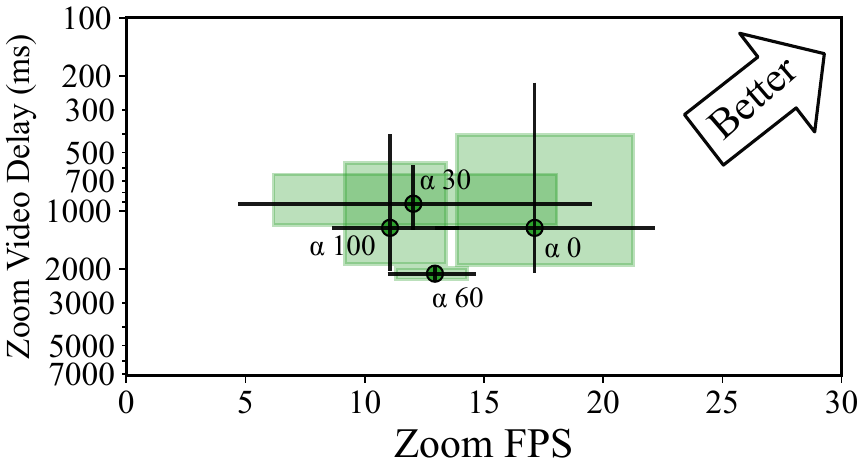}}\label{fig:multi_ul_fps_delay_b50_ue3}
         \subfigure[$\beta 90$, UE1]
         {\includegraphics[width=0.33\linewidth]{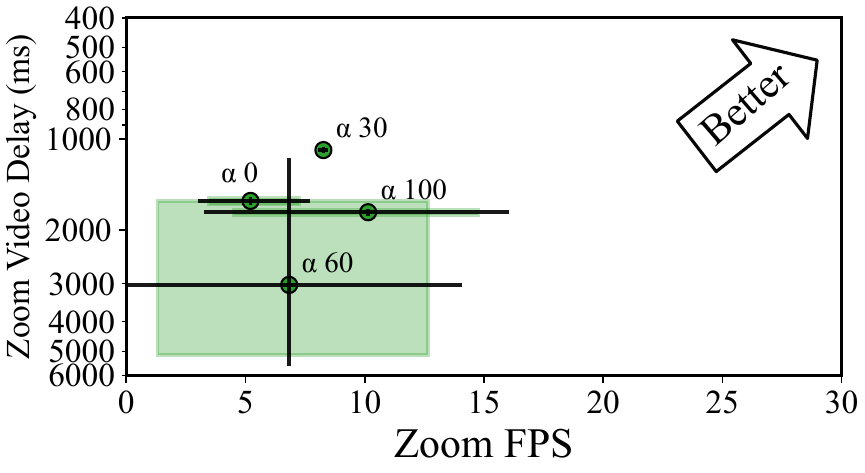}}\label{fig:multi_ul_fps_delay_b90_ue1}
         \subfigure[$\beta 90$, UE2]
         {\includegraphics[width=0.33\linewidth]{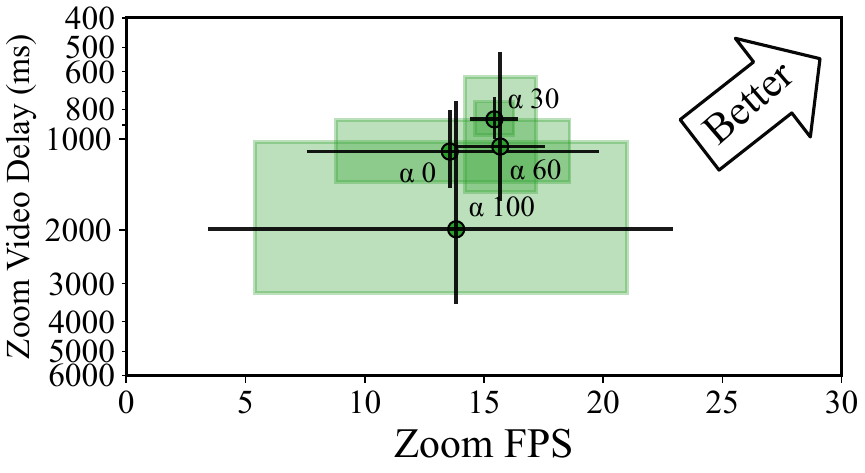}}\label{fig:multi_ul_fps_delay_b90_ue2}
         \subfigure[$\beta 90$, UE3]
         {\includegraphics[width=0.33\linewidth]{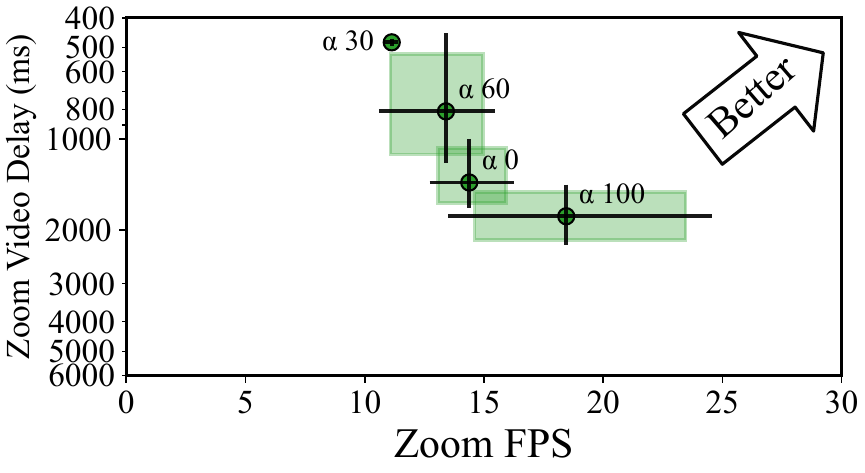}}\label{fig:multi_ul_fps_delay_b90_ue3}
         \subfigure[Optimal $\beta,\alpha$ and baseline, UE1]
         {\includegraphics[width=0.33\linewidth]{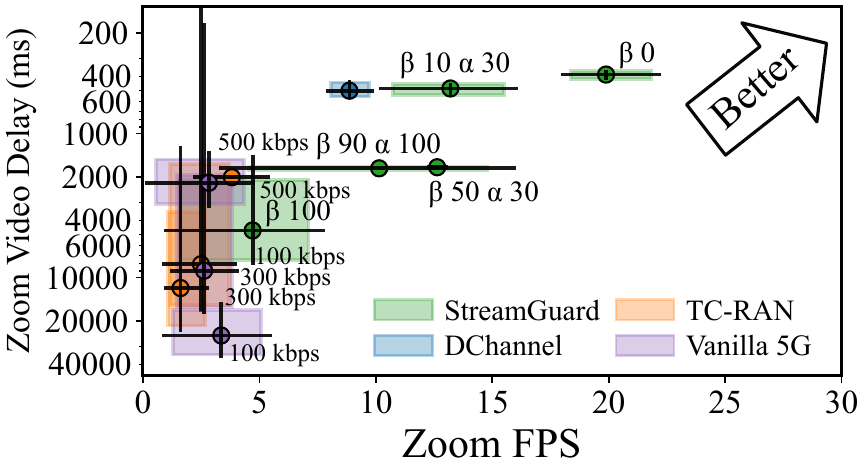}}\label{fig:multi_ul_fps_delay_all_ue1}
         \subfigure[Optimal $\beta,\alpha$ and baseline, UE2]
         {\includegraphics[width=0.33\linewidth]{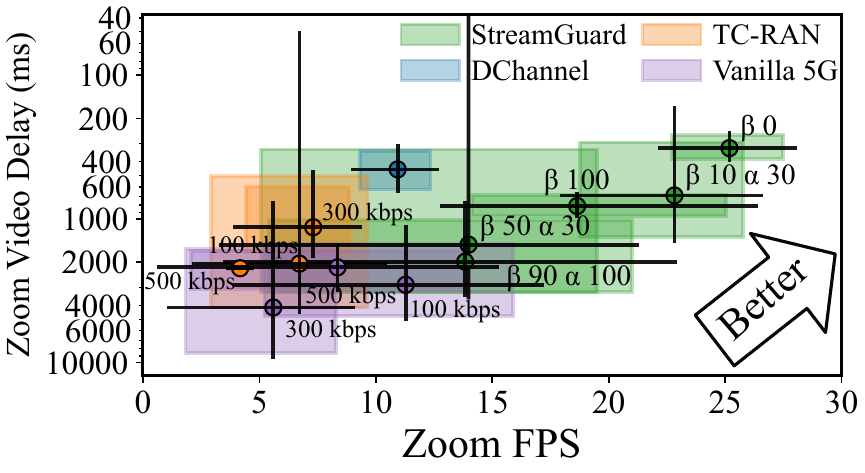}}\label{fig:multi_ul_fps_delay_all_ue2}
         \subfigure[Optimal $\beta,\alpha$ and baseline, UE3]
         {\includegraphics[width=0.33\linewidth]{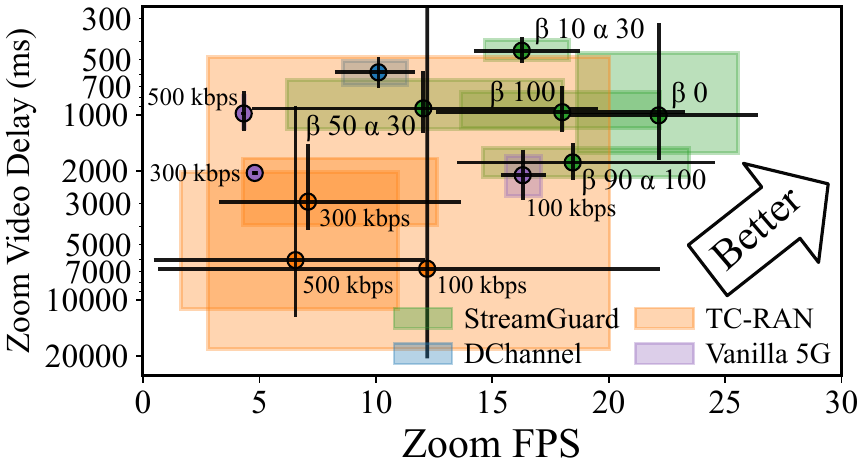}}\label{fig:multi_ul_fps_delay_all_ue3}
        \caption{FPS vs screen-to-camera delay for uplink multi Zoom call experiment.}
        \label{fig:multi_ul_fps_delay}
\end{figure*}

\begin{figure*}
    \centering
         \subfigure[$\beta 10$, UE1]
         {\includegraphics[width=0.33\linewidth]{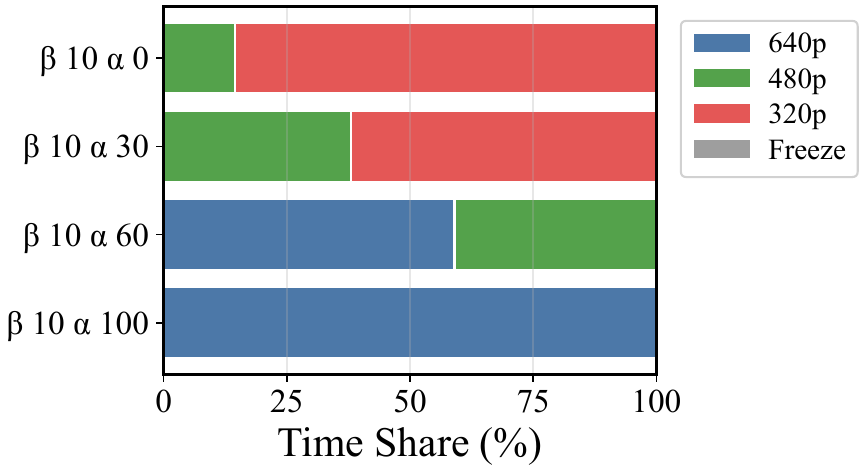}}\label{fig:res_b10_ue1_ul}
         \subfigure[$\beta 10$, UE2]
         {\includegraphics[width=0.33\linewidth]{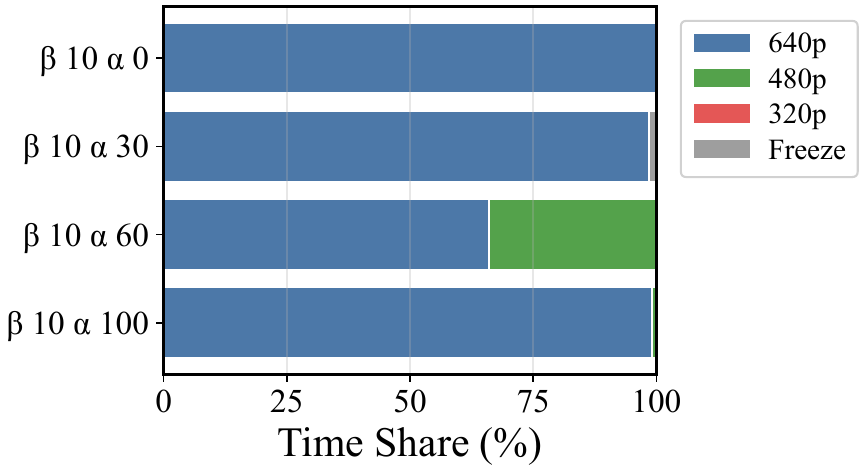}}\label{fig:res_b10_ue2_ul}
         \subfigure[$\beta 10$, UE3]
         {\includegraphics[width=0.33\linewidth]{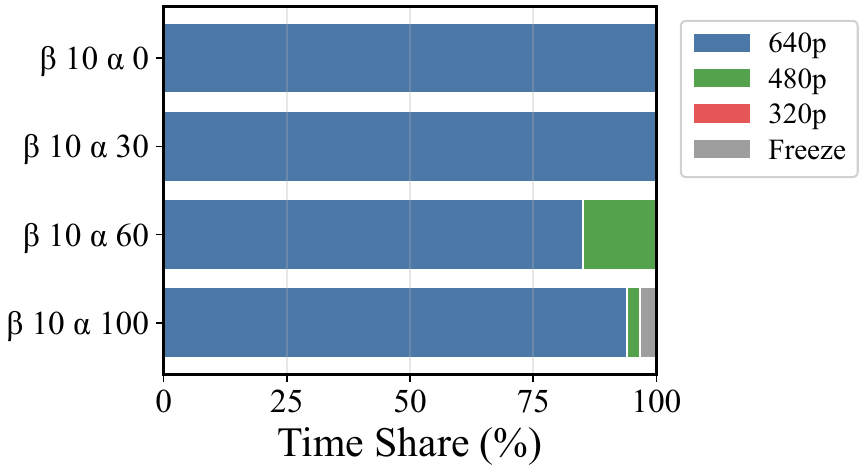}}\label{fig:res_b10_ue3_ul}
         \subfigure[$\beta 50$, UE1]
         {\includegraphics[width=0.33\linewidth]{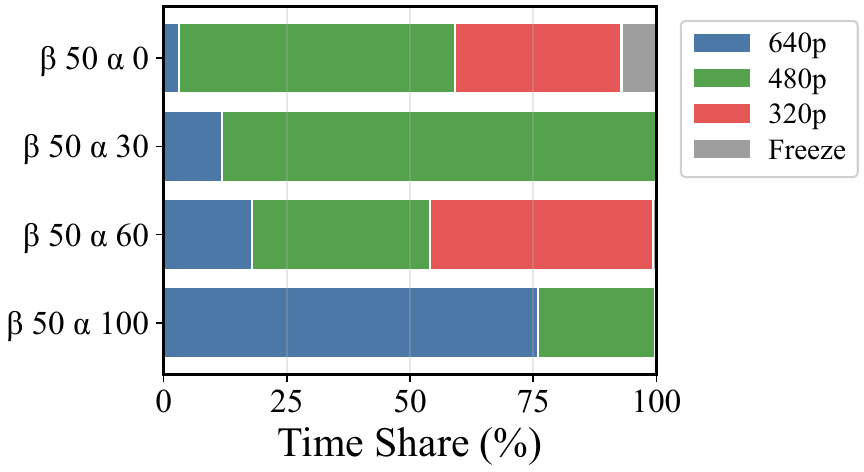}}\label{fig:res_b50_ue1_ul}
         \subfigure[$\beta 50$, UE2]
         {\includegraphics[width=0.33\linewidth]{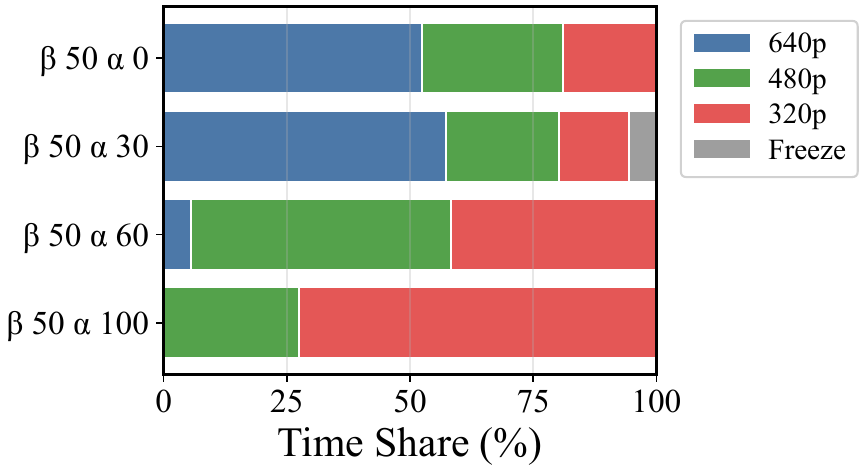}}\label{fig:res_b50_ue2_ul}
         \subfigure[$\beta 50$, UE3]
         {\includegraphics[width=0.33\linewidth]{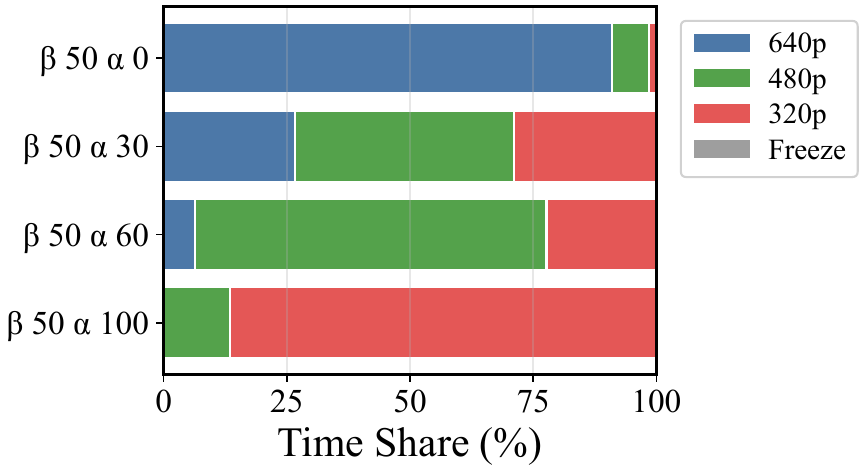}}\label{fig:res_b50_ue3_ul}
         \subfigure[$\beta 90$, UE1]
         {\includegraphics[width=0.33\linewidth]{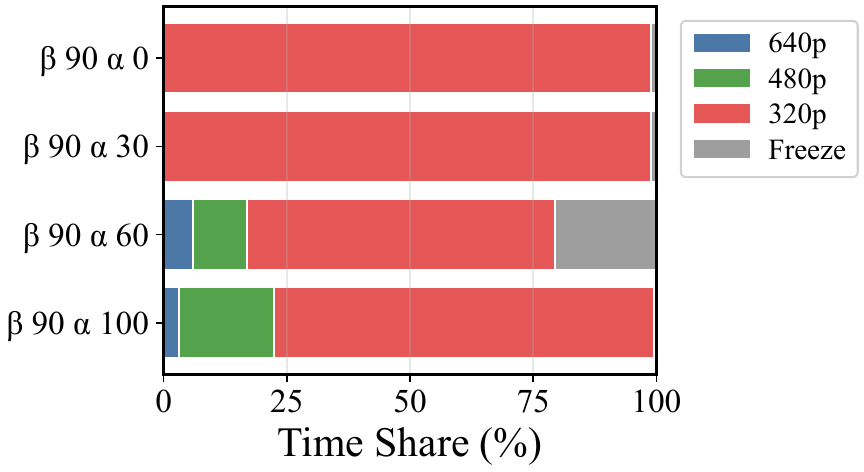}}\label{fig:res_b90_ue1_ul}
         \subfigure[$\beta 90$, UE2]
         {\includegraphics[width=0.33\linewidth]{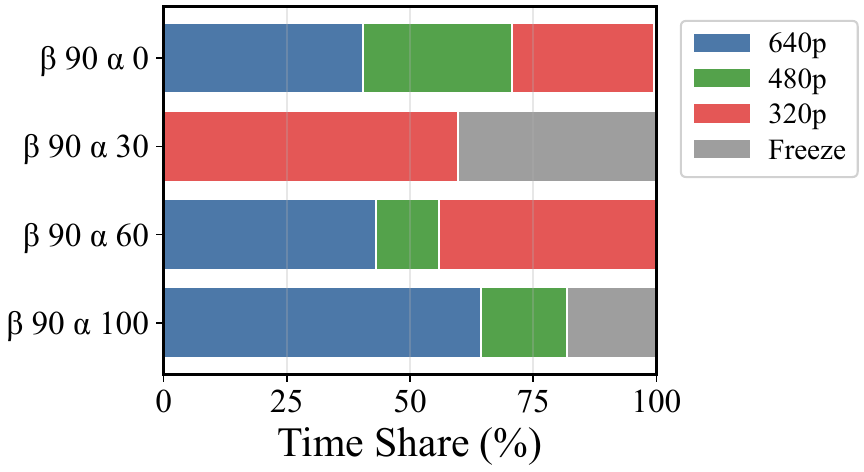}}\label{fig:res_b90_ue2_ul}
         \subfigure[$\beta 90$, UE3]
         {\includegraphics[width=0.33\linewidth]{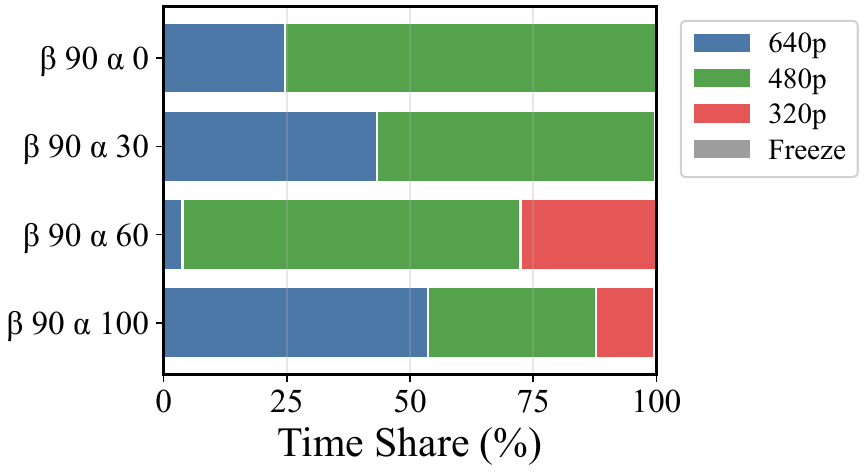}}\label{fig:res_b90_ue3_ul}
         \subfigure[Optimal $\beta,\alpha$ and baseline, UE1]
         {\includegraphics[width=0.33\linewidth]{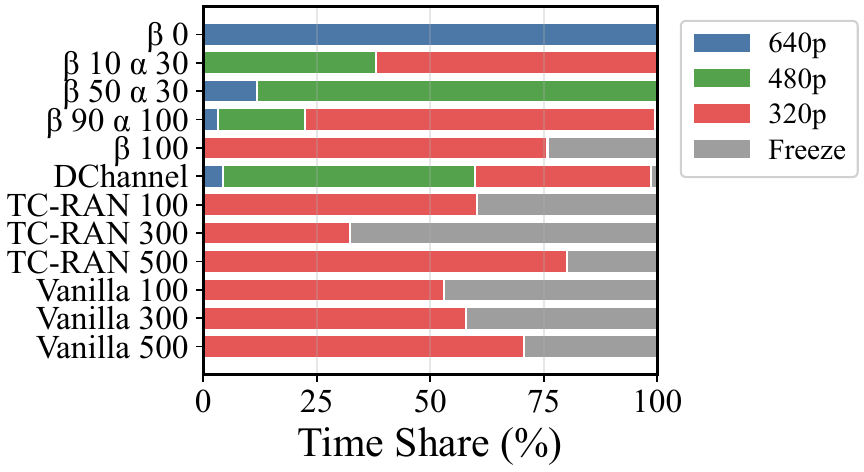}}\label{fig:res_all_ue1_ul}
         \subfigure[Optimal $\beta,\alpha$ and baseline, UE2]
         {\includegraphics[width=0.33\linewidth]{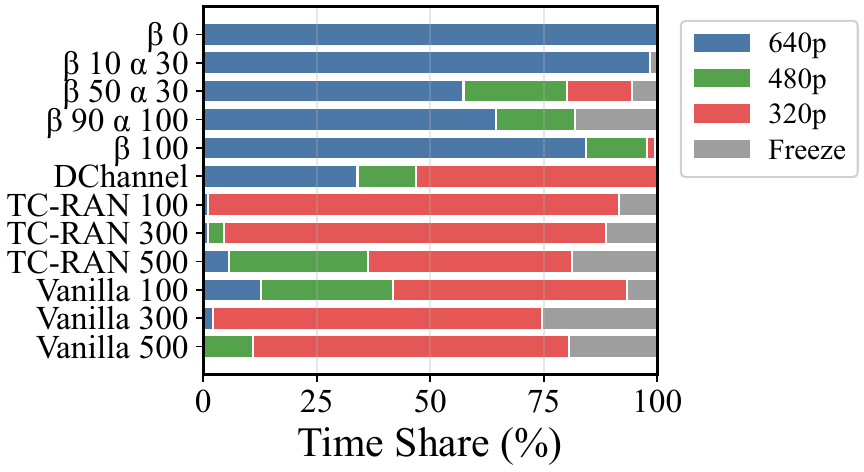}}\label{fig:res_all_ue2_ul}
         \subfigure[Optimal $\beta,\alpha$ and baseline, UE3]
         {\includegraphics[width=0.33\linewidth]{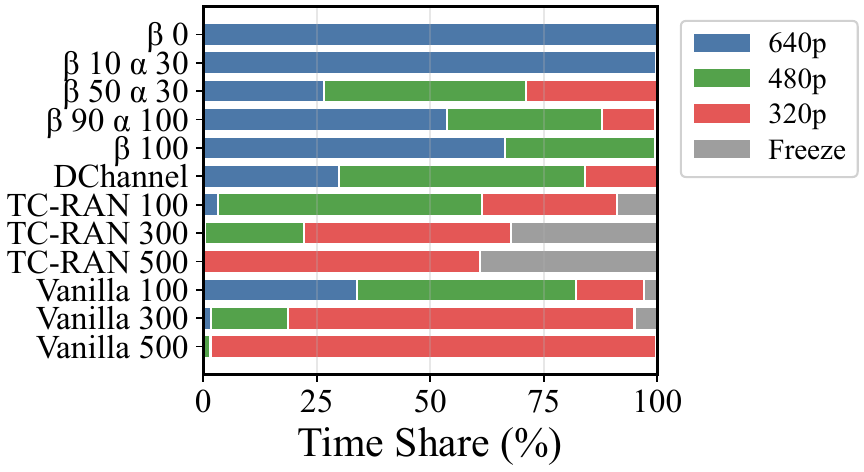}}\label{fig:res_all_ue3_ul}
        \caption{Resolution distributions for uplink multi Zoom call experiment.}
        \label{fig:multi_ul_res}
\end{figure*}

\begin{figure*}
    \centering
         \subfigure[$\beta 10$]
         {\includegraphics[width=0.49\linewidth]{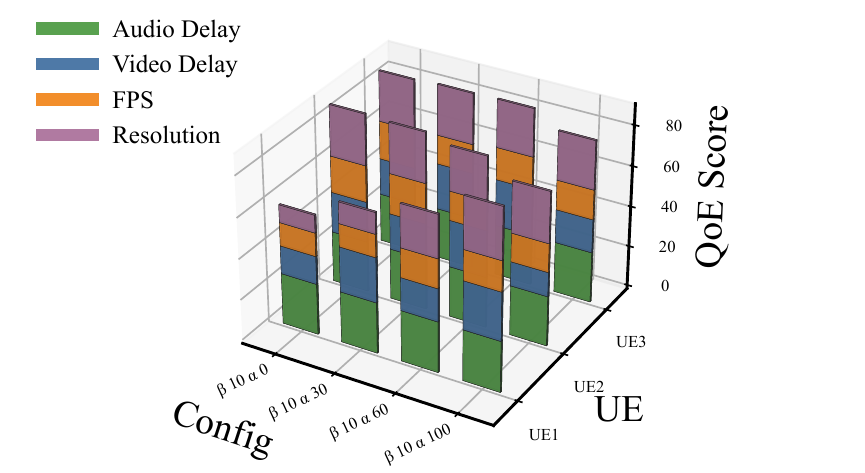}}\label{fig:qoe_breakdown_b10_ul}
         \subfigure[$\beta 50$]
         {\includegraphics[width=0.49\linewidth]{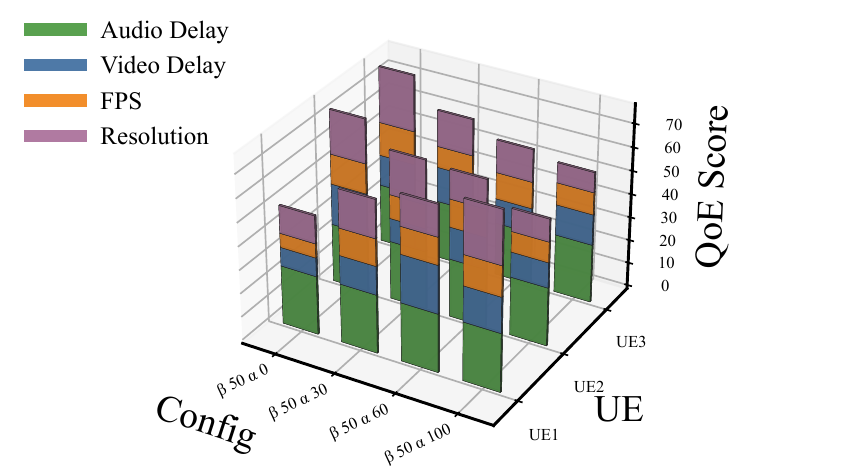}}\label{fig:qoe_breakdown_b50_ul}
         \subfigure[$\beta 90$]
         {\includegraphics[width=0.49\linewidth]{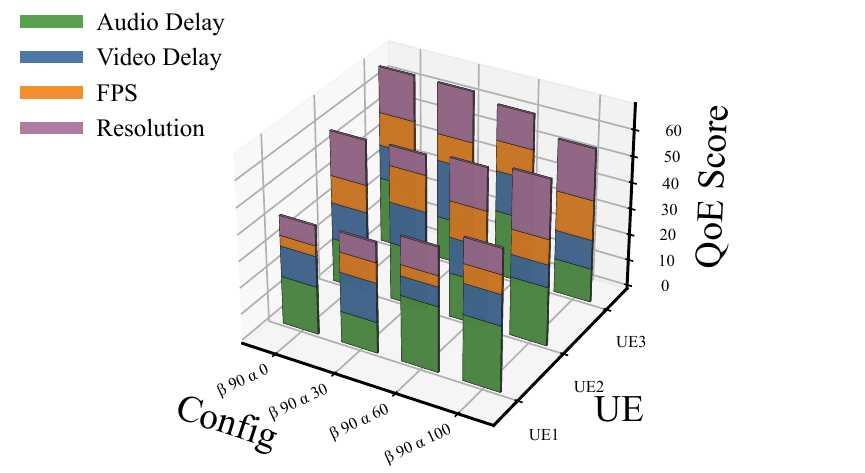}}\label{fig:qoe_breakdown_b90_ul}
         \subfigure[Optimal $\beta, \alpha$ and baseline]
         {\includegraphics[width=0.49\linewidth]{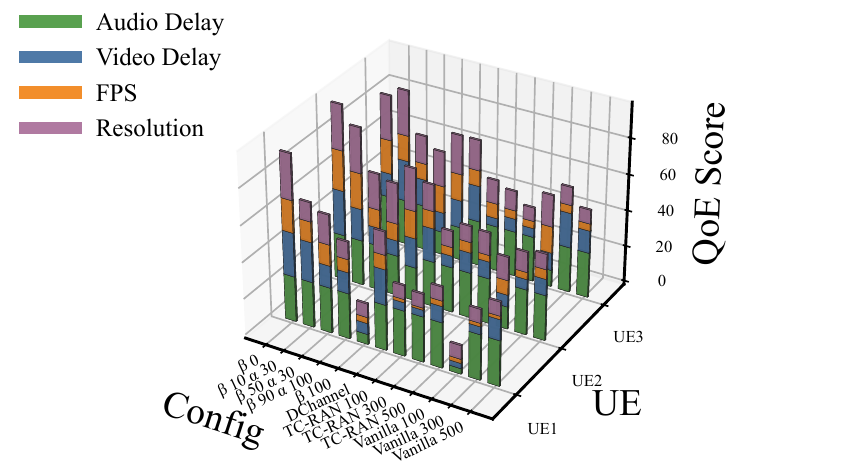}}\label{fig:qoe_breakdown_all_ul}    
        \caption{QoE score breakdowns for uplink multi Zoom call experiment.}
        \label{fig:multi_ul_qoe_breakdown}
\end{figure*}

\noindent\circnum{4} \textbf{Three Uplink Zoom Calls.}
We show the video delay–FPS relationship in~\cref{fig:multi_ul_fps_delay}, the resolution distribution in~\cref{fig:multi_ul_res}, and the QoE score breakdown in~\cref{fig:multi_ul_qoe_breakdown}.

\begin{figure*}
    \centering
         \subfigure[Downlink]
         {\includegraphics[width=0.49\linewidth]{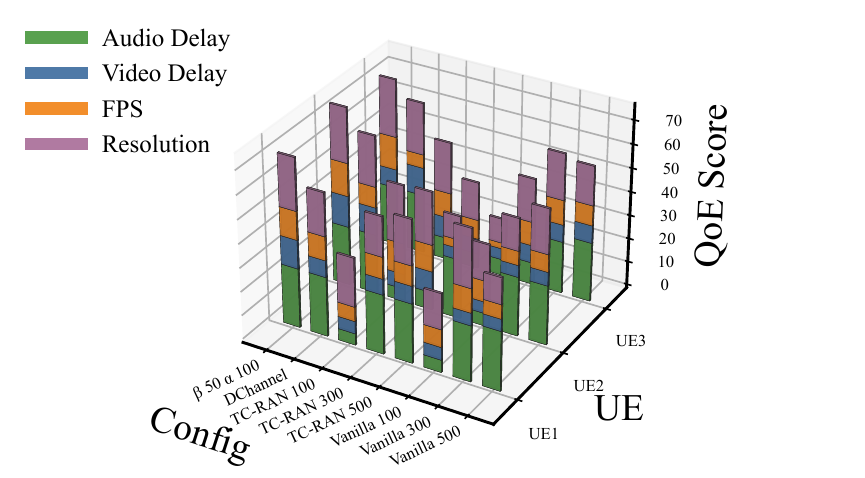}}\label{fig:qoe_breakdown_all_bi_dl} 
         \subfigure[Uplink]
         {\includegraphics[width=0.49\linewidth]{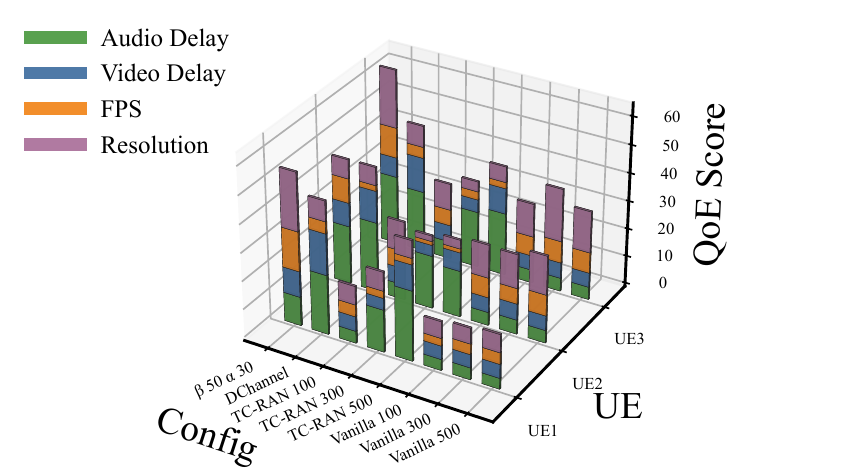}}\label{fig:qoe_breakdown_all_bi_ul}    
        \caption{QoE score breakdowns for bidirectional multi Zoom call experiment.}
        \label{fig:bi_qoe_breakdown}
\end{figure*}

\begin{figure*}
    \centering
         \subfigure[Selected $\beta,\alpha$ and baseline, UE1]
         {\includegraphics[width=0.33\linewidth]{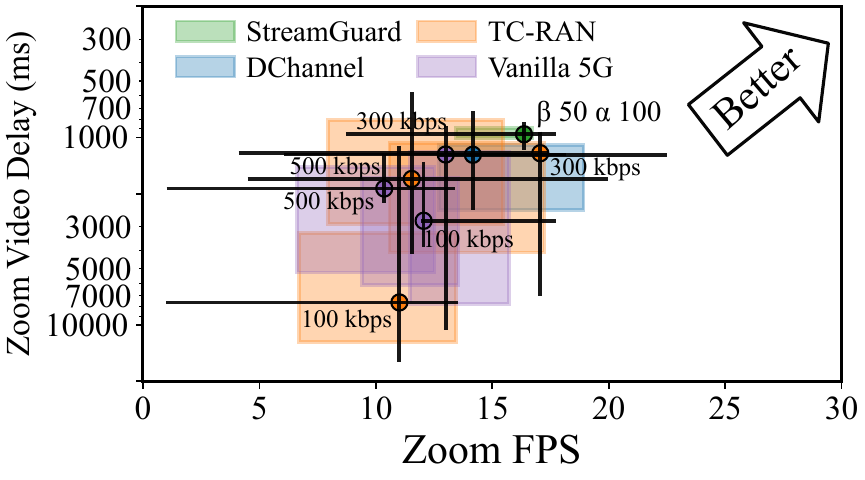}}\label{fig:fps_delay_ue1_bi_dl}
        \subfigure[Selected $\beta,\alpha$ and baseline, UE2]
         {\includegraphics[width=0.33\linewidth]{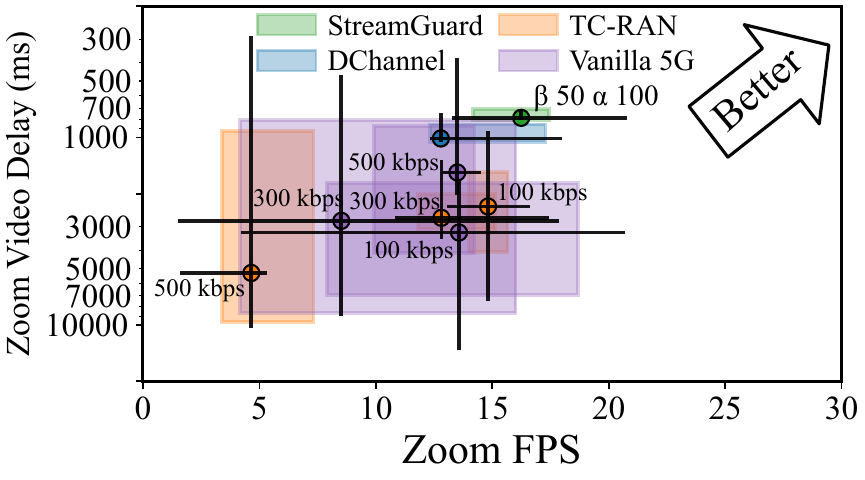}}\label{fig:fps_delay_ue2_bi_dl}
        \subfigure[Selected $\beta,\alpha$ and baseline, UE3]
         {\includegraphics[width=0.33\linewidth]{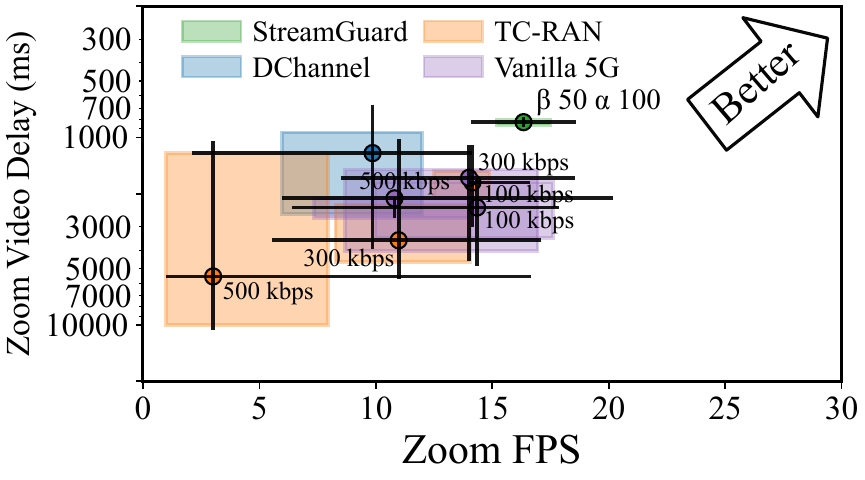}}\label{fig:fps_delay_ue3_bi_dl}
        \caption{Downlink-side FPS vs screen-to-camera delay in bidirectional multi Zoom call experiment.}
        \label{fig:fps_delay_bi_dl}
\end{figure*}

\begin{figure*}
    \centering
         \subfigure[Selected $\beta,\alpha$ and baseline, UE1]
         {\includegraphics[width=0.33\linewidth]{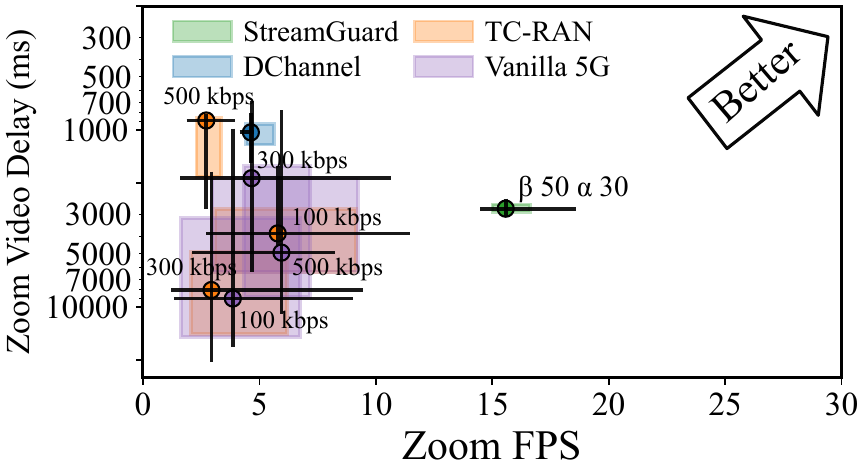}}\label{fig:fps_delay_ue1_bi_ul}
        \subfigure[Selected $\beta,\alpha$ and baseline, UE2]
         {\includegraphics[width=0.33\linewidth]{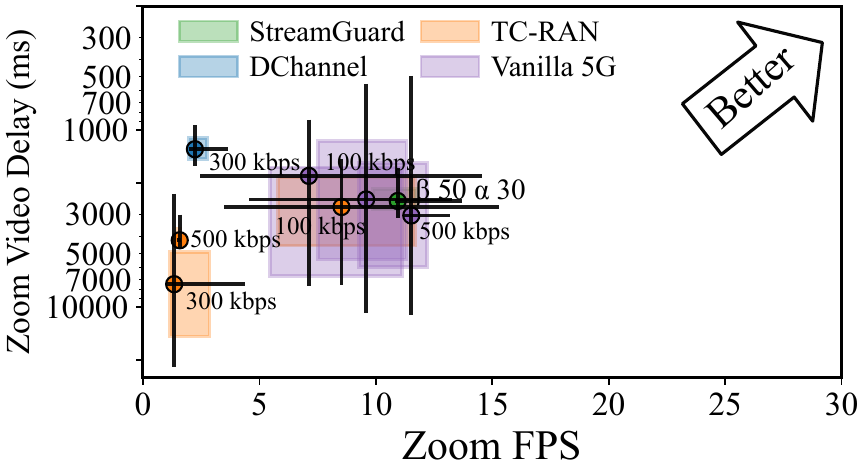}}\label{fig:fps_delay_ue2_bi_ul}
        \subfigure[Selected $\beta,\alpha$ and baseline, UE3]
         {\includegraphics[width=0.33\linewidth]{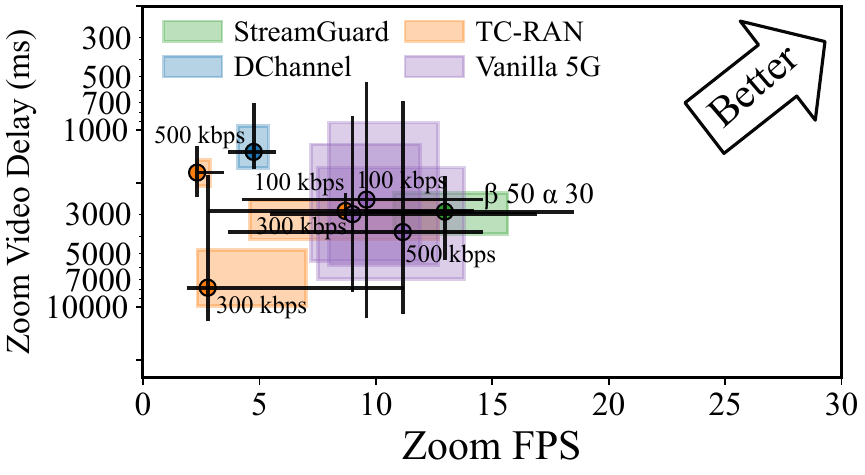}}\label{fig:fps_delay_ue3_bi_ul}
        \caption{Uplink-side FPS vs screen-to-camera delay in bidirectional multi Zoom call experiment.}
        \label{fig:fps_delay_bi_ul}
\end{figure*}

\begin{figure*}
    \centering
         \subfigure[Selected $\beta, \alpha$ and baseline UE1]
         {\includegraphics[width=0.33\linewidth]{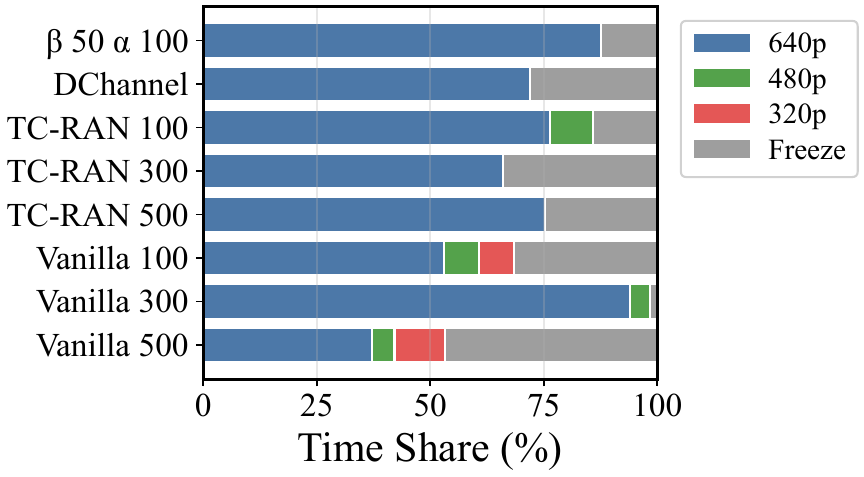}}\label{fig:res_ue1_bi_dl}
         \subfigure[Selected $\beta, \alpha$ and baseline UE2]
         {\includegraphics[width=0.33\linewidth]{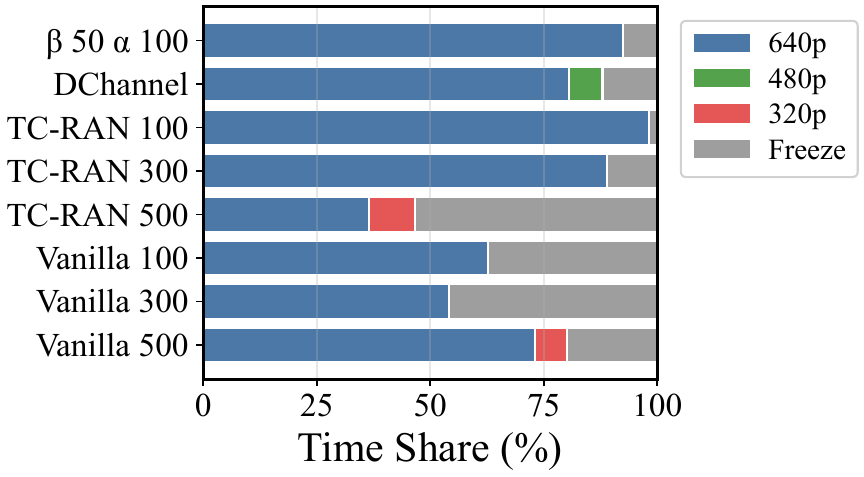}}\label{fig:res_ue2_bi_dl}
         \subfigure[Selected $\beta, \alpha$ and baseline UE3]
         {\includegraphics[width=0.33\linewidth]{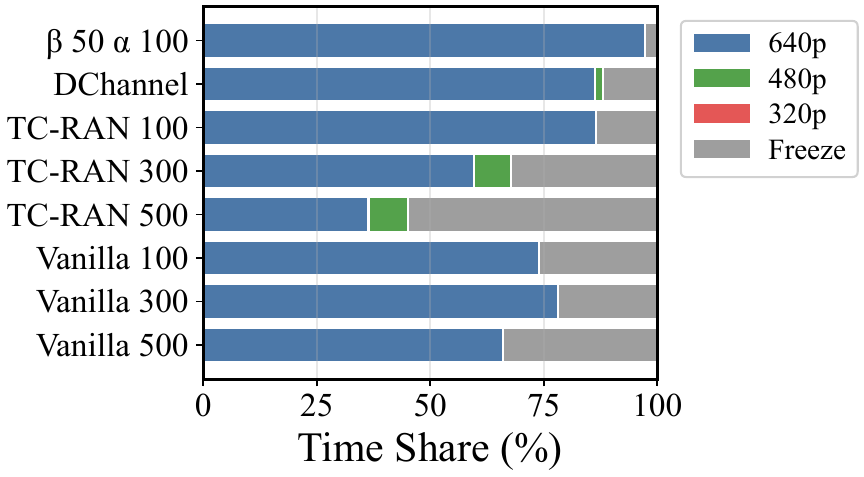}}\label{fig:res_ue3_bi_dl}
        \caption{Downlink-side resolution distributions in bidirectional multi Zoom call experiment.}
        \label{fig:bi_dl_res}
\end{figure*}

\begin{figure*}
    \centering
         \subfigure[Selected $\beta, \alpha$ and baseline UE1]
         {\includegraphics[width=0.33\linewidth]{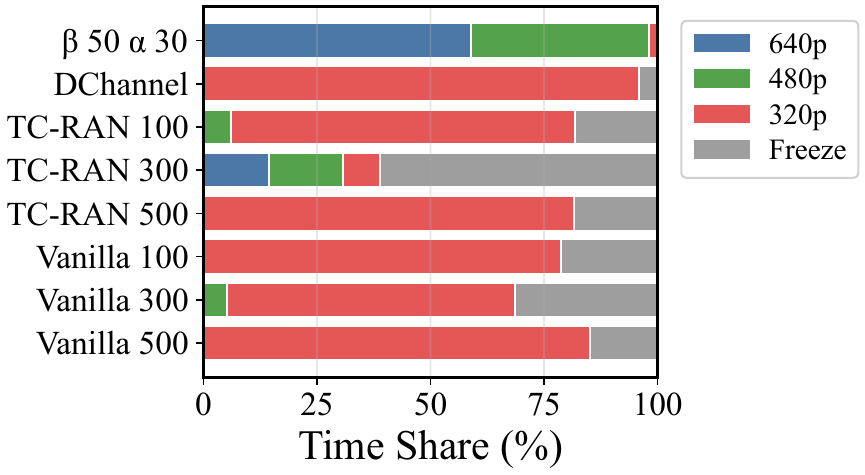}}\label{fig:res_ue1_bi_ul}
         \subfigure[Selected $\beta, \alpha$ and baseline UE2]
         {\includegraphics[width=0.33\linewidth]{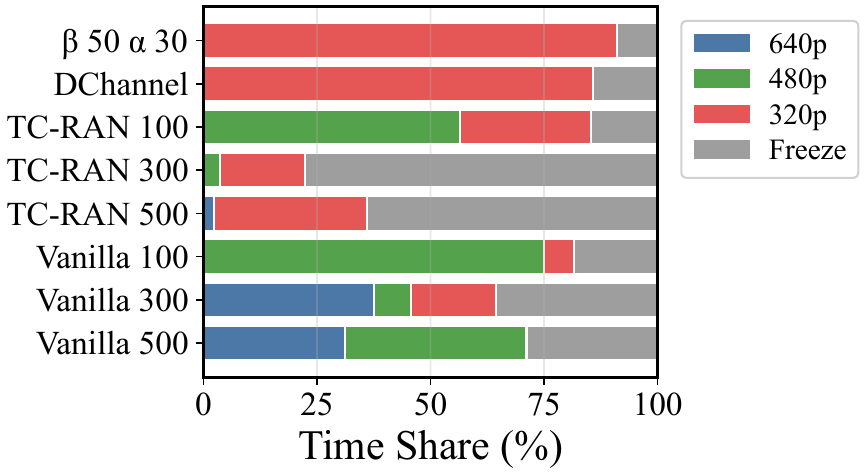}}\label{fig:res_ue2_bi_ul}
         \subfigure[Selected $\beta, \alpha$ and baseline UE3]
         {\includegraphics[width=0.33\linewidth]{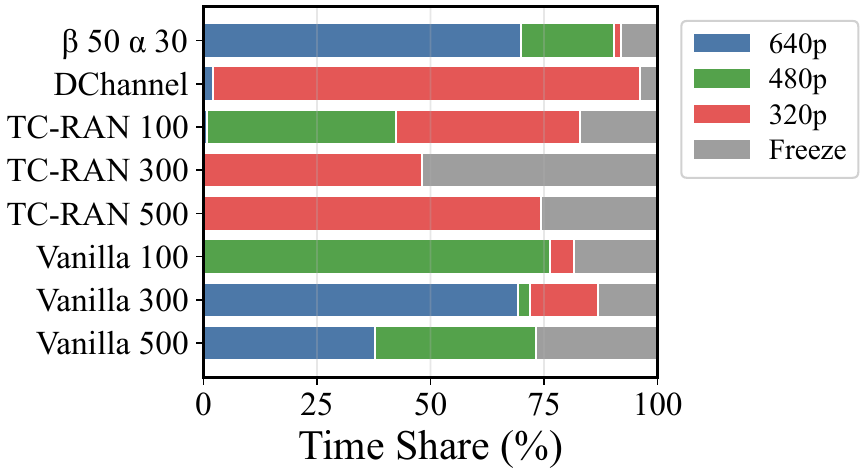}}\label{fig:res_ue3_bi_ul}
        \caption{Uplink-side resolution distributions in bidirectional multi Zoom call experiment.}
        \label{fig:bi_ul_res}
\end{figure*}

\noindent\circnum{5} \textbf{Three Bidirectional Zoom Calls.}
For the downlink (uplink), we show the video delay–FPS relationship in~\cref{fig:fps_delay_bi_dl} (\cref{fig:fps_delay_bi_ul}) and the resolution distribution in~\cref{fig:bi_dl_res} (\cref{fig:bi_ul_res}).
The QoE score breakdown is shown in~\cref{fig:bi_qoe_breakdown}.

For the delay-FPS figures (\cref{fig:single_dl_fps_delay}, \cref{fig:multi_dl_fps_delay}, \cref{fig:single_ul_fps_delay}, \cref{fig:multi_ul_fps_delay}, \cref{fig:fps_delay_bi_dl}, \cref{fig:fps_delay_bi_ul}), the center dot denotes the median FPS and median delay. The box spans the interquartile range (25th to 75th percentiles) of FPS and delay. The bars denote Tukey whiskers, extending to the most extreme non-outlier values within 1.5 times the interquartile range.

\end{appendices}

\end{document}